\documentclass[onecolumn,showpacs,superscriptaddress,floatfix,nofootinbib,prd]{revtex4}

\usepackage{graphicx}
\usepackage{epsf}
\usepackage{epsfig}
\usepackage{amssymb,amsmath}
\usepackage[usenames]{color}
\usepackage{amssymb}
\usepackage{times}
\usepackage{mathrsfs}
\usepackage{hyperref}
\hypersetup{
  colorlinks=true,        % false: boxed links; true: colored links
  linkcolor=blue,         % color of internal links
  citecolor=cyan,         % color of links to bibliography
}
\usepackage{float}

\newcommand{\ie}{{i.e.,}~}
\newcommand{\eg}{{e.g.,}~}

\renewcommand{\u}{\boldsymbol{u}}
\newcommand{\halb}{\frac{1}{2}}

\renewcommand{\d}{\mathrm{d}}

\newcommand{\x}{\boldsymbol{x}}
\newcommand{\Q}{\boldsymbol{Q}}
\renewcommand{\u}{\boldsymbol{u}}
\newcommand{\w}{\boldsymbol{w}}
\newcommand{\q}{\boldsymbol{q}}

\allowdisplaybreaks

\begin{document}
%%%%%%%%%%%%%%%%%%%%%%%%%%%%%%%

%\title[A strongly hyperbolic first--order CCZ4 system]{A strongly hyperbolic first-order reduction of the 
%	conformal and covariant Z4 formulation of the Einstein equations and its solution with discontinuous Galerkin schemes}

\title[A strongly hyperbolic first--order CCZ4 system]{Conformal and covariant Z4 formulation of the Einstein equations: strongly hyperbolic first-order reduction and solution with discontinuous Galerkin schemes}

\date{\today}
\label{firstpage}

\author{Michael Dumbser}
\affiliation{
Laboratory of Applied Mathematics, University of Trento,
Via Mesiano 77, 38123 Trento, Italy
}
\author{Federico Guercilena}
\affiliation{
Institute for Theoretical Physics,
Max-von-Laue Str. 1, 60438 Frankfurt, Germany
}
\author{Sven K\"oppel}
\author{Luciano Rezzolla}
\affiliation{
Institute for Theoretical Physics,
Max-von-Laue Str. 1, 60438 Frankfurt, Germany
}
\affiliation{
Frankfurt Institute for Advanced Studies,
Ruth-Moufang-Str. 1, 60438 Frankfurt, Germany
}
\author{Olindo Zanotti}
\affiliation{
Laboratory of Applied Mathematics, University of Trento,
Via Mesiano 77, 38123 Trento, Italy
}

\begin{abstract}
We present a strongly hyperbolic first-order formulation of the Einstein
equations based on the conformal and covariant Z4 system (CCZ4) with
constraint-violation damping, which we refer to as FO-CCZ4. As CCZ4, this
formulation combines the advantages of a conformal and traceless
formulation, with the suppression of constraint violations given by the
damping terms, but being first order in time and space, it is
particularly suited for a discontinuous Galerkin (DG) implementation. The
strongly hyperbolic first-order formulation has been obtained by making
careful use of first and second-order ordering constraints. A proof of
strong hyperbolicity is given for a selected choice of standard gauges 
via an analytical computation of the entire eigenstructure of the FO-CCZ4
system. The resulting governing partial differential equations system is
written in non-conservative form and requires the evolution of 58
unknowns. A key feature of our formulation is that the first-order CCZ4
system decouples into a set of pure ordinary differential equations and a
reduced hyperbolic system of partial differential equations that contains
only linearly degenerate fields. We implement FO-CCZ4 in a high-order
path-conservative arbitrary-high-order-method-using-derivatives (ADER)-DG
scheme with adaptive mesh refinement and local time-stepping,
supplemented with a third-order ADER-WENO subcell finite-volume limiter
in order to deal with singularities arising with black holes. We validate
the correctness of the formulation through a series of standard tests in
vacuum, performed in one, two and three spatial dimensions, and also
present preliminary results on the evolution of binary black-hole
systems. To the best of our knowledge, these are the first successful
three-dimensional simulations of moving punctures carried out with
high-order DG schemes using a first-order formulation of the Einstein
equations.
\end{abstract}

\maketitle

%%%%%%%%%%%%%%%%%%%%%%%%%%%%%%%%%%%%%%%%%%%%%%%%%%%%%%%%%%%%%%%%%%%%%%%%

\section{Introduction}

%%%%%%%%%%%%%%%%%%%%%%%%%%%%%%%%%%%%%%%%%%%%%%%%%%%%%%%%%%%%%%%%%%%%%%%%

Large scale, fully general-relativistic numerical simulations have
emerged in the last decade as a very powerful tool for the study of
astrophysical systems, following the breakthrough calculations of the
inspiral and merger of binary black holes \cite{Pretorius:2005gq,
  Campanelli06, Baker05a}. The interest for such numerical techniques and
the results they can produce has been only strengthened by the recent
direct detection of gravitational waves
\cite{Abbot2016-GW-detection-prl}, which paves the way for the
forthcoming era of gravitational-wave astronomy.

General-relativistic simulations require (among other issues) stable and
accurate methods for evolving the spacetime, \ie for solving the Einstein
field equations. The development of hyperbolic formulations of the
Einstein equations that allow for long-term simulations of generic
spacetimes, including the ones encompassing the physical singularities
arising in the presence of black holes, has been therefore of great
importance in numerical relativity. The first step in this direction has
been the derivation of the Arnowitt-Deser-Misner (ADM) formulation
(originally introduced in \cite{Arnowitt62unfindable}, but see
\cite{Alcubierre:2008, Bona2009, Baumgarte2010, Gourgoulhon2012,
  Rezzolla_book:2013, Shibata_book:2016} for a more modern perspective).
While this formulation splits time and space and naturally presents
general relativity as an initial boundary-value problem, suitable for
numerical implementation, it is known to be not hyperbolic -- at least
when usual gauge choices are considered (see \cite{Sarbach2012} for a
discussion) -- and therefore unstable in numerical applications.

Subsequently, a lot of effort has been devoted to find hyperbolic
formulations of the Einstein equations. These efforts have lead to the
derivation of the Baumgarte-Shapiro-Shibata-Nakamura-Oohara-Kojima
(BSSNOK) formulation \cite{Shibata95,Baumgarte99,Nakamura87,Brown09},
which achieves hyperbolicity via a conformal transformation of the
3-metric and the promotion of some contractions of the Christoffel
symbols to independently evolved variables and, most importantly, by
inserting the momentum and Hamiltonian constraint expressions in the
evolution system. A general-covariant alternative is the Z4 formulation
of \cite{Bona:2003fj,Bona:2003qn,Alic:2009}, which has been presented
both in first- and second-order form in the spatial derivatives. More
successful have been formulations based on the Z4 one that include a
conformal transformation of the metric. These are the Z4c formulation,
that removes some source terms in the Einstein equations in order to
bring the evolution equations into a form which is closer to the BSSNOK
system \cite{Bernuzzi:2009ex}, and the conformal and covariant CCZ4
formulation \cite{Alic:2011a, Alic2013} (see also \cite{Sanchis2014,
Bezares2017} for some recent and slight variants).  The Z4 family of
formulations also admits mechanisms to damp constraint violations as they
arise during the evolution \cite{Gundlach2005:constraint-damping,
Weyhausen:2011cg,Alic:2011a, Alic2013}.

For completeness, it should be mentioned that the 3+1 formalism is not
the only way to develop a formulation of the Einstein equations suitable for
numerical implementation: alternatives are the generalized-harmonic
formalism \eg \cite{Garfinkle02, Pretorius:2005gq, Lindblom:2005gh,
  Szilagyi:2006qy}; the characteristic-evolution formalism
\cite{Winicour05} the conformal approach \cite{Friedrich:2002xz,
  Husa02b} and fully-constrained formulations \cite{Cordero09}. 
	These approaches, however, are not the subject of the present work.

Parallel to the quest for better formulations of the equations, the
development and implementation of better numerical methods has been a
main priority of ongoing research. While most general-relativistic codes
use finite-differences (\eg \cite{Brown2007b}) or spectral methods (\eg
\cite{Szilagyi:2009qz}) for the spacetime evolution, increasing interest 
is being focused towards DG methods (see, \eg \cite{hesthaven2008} for an
introduction and review). DG methods are very attractive due to their
excellent scalability and wave-propagation properties. The latter allow
the propagation of smooth linear and nonlinear waves over long distances
with little dissipation and dispersion errors, and should thus be in
principle particularly well suited for the solution of the Einstein
equations, where (apart from physical singularities in black holes) the
fields are smooth and high accuracy can be achieved.

So far, however, only a rather limited number of attempts have been made
to solve the Einstein equations with DG methods. Field et al. \cite{field10}
tested a second-order BSSNOK formulation, while Brown et
al. \cite{Brown2012} developed a first-order formulation of BSSNOK,
however both works were limited to spherical symmetry and vacuum
spacetimes. The first DG implementation in non-vacuum spacetimes was
published by Radice \& Rezzolla \cite{Radice2011}, but was still
restricted to spherical symmetry. The first three-dimensional (3D)
implementation, albeit in a fixed spacetime and focused on hydrodynamics
was developed by Bugner et al. \cite{Bugner2015}. More recently, Miller
and Schnetter \cite{Miller2016} proposed an operator-based DG method
suitable also for second-order systems and applied it to the BSSNOK
system, while Kidder et al. \cite{Kidder2016} developed a task based
relativistic magnetohydrodynamics code.

In this work we propose a novel first-order form of the CCZ4 system,
which we refer to as FO-CCZ4. We thoroughly study its eigenstructure and
in particular show that it is strongly hyperbolic for two typical  
choices of gauges, namely zero shift with harmonic lapse and the 
Gamma-driver with 1+log slicing. 
We then implement this formulation in a fully 
three-dimensional code, using an ADER-DG algorithm with adaptive mesh
refinement (AMR) and local time-stepping (LTS), supplemented with a high
order ADER-WENO \cite{eno,Liu1994, Jiang1996} finite-volume subcell
limiter \cite{Dumbser2009a,Dumbser2010,Dumbser2011} to deal with
singularities in black-hole spacetimes. This family of schemes has
already been successfully applied also to the classical and special
relativistic MHD equations (see \cite{Zanotti2015b,Zanotti2015c}).

We test the stability and accuracy of the ADER-DG discretization applied
to our novel FO-CCZ4 formulation in a series of standard tests for
general-relativistic codes \cite{Alcubierre2003:mexico-I, Babiuc:2007vr}.
We also verify that our scheme converges at the expected order of
accuracy and we provide evidence of long-time robustness and stability.
Finally we apply the method to the long-term evolution of single
black-hole spacetimes, showing that we are able to stably evolve a
puncture black-hole spacetime for a time scale of $\sim 1000$ $M$ ($M$
being the mass of the black hole). We also present preliminary results
for the head-on collision of two black holes. To the best of our
knowledge, these are the first simulations of black-hole spacetimes
performed in three spatial dimensions with a high-order DG code.

The structure of the paper is as follows. In Section \ref{sec.FO-CCZ4} we
derive the full set of first-order evolution equations and prove the
strong hyperbolicity for common gauge choices by computing the full
eigenstructure. In Section \ref{sec.ader} we introduce the numerical
scheme intended to solve the partial differential equations (PDE)
system. In Section \ref{sec.tests} we show a number of benchmark results
to demonstrate correctness of both the formulation and the numerical
solver. Finally, the conclusions are summarised in Section
\ref{sec.conclusions}.

We work in a geometrized set of units, in which the speed of light and
the gravitational constant are set to unity, \ie $c=G=1$. Greek indices
run from $0$ to $3$, Latin indices run from $1$ to $3$ and we use the
Einstein summation convention of repeated indices.

%==============================================================================
\section{A first-order strongly hyperbolic CCZ4 system: FO-CCZ4}
\label{sec.FO-CCZ4}

\subsection{The original second-order CCZ4 system}
The second-order CCZ4 system can be derived from the Z4 Lagrangian
$\mathcal{L} = g^{\mu\nu} ( R_{\mu\nu} + 2 \nabla_\mu Z_\nu )$, which
adds terms dependent on the $Z_\mu$ vector to the classical
Einstein-Hilbert Lagrangian (see \cite{Bona2009} for a complete
derivation). The Einstein field equations are recovered by minimizing the
corresponding action and the algebraic constraints $Z_\mu = 0$.
Additional constraint-damping terms can be introduced
\cite{Gundlach2005:constraint-damping}, so that the Einstein equations of
the constraint-damped Z4 system in vacuum read
\begin{equation}
R_{\mu\nu} + \nabla_{(\mu} Z_{\nu)} +
\kappa_1 \left( n_{(\mu} Z_{\nu)}
-(1+\kappa_2) g_{\mu\nu} n_\alpha Z^\alpha \right)
= 0\,,
\end{equation}
where $R_{\mu\nu}$ is the Ricci tensor and $\boldsymbol{n}$ is the unit
vector normal to the spatial hypersurfaces. Here, $\kappa_1, \kappa_2$
are tuning constants related to the characteristic time of the
exponential damping the of constraint violations.

In order to formulate a well-posed Cauchy problem, we apply the 3+1
decomposition of space time (see, \eg \cite{Bona2009,Baumgarte2010}), so
that the line element reads
\begin{equation}
d s^2 = -\alpha^2 d t^2 + \gamma_{ij}
(d x^i + \beta^i d t)
(d x^j + \beta^j d t)\,,
\end{equation}
with lapse $\alpha$, shift $\beta^i$ and 3-metric $\gamma_{ij}$. The 3+1
split leads to evolution equations for $\gamma_{ij}$ as well as the
extrinsic curvature $K_{ij} = -\frac{1}{2}\mathcal{L}_n \gamma_{ij}$,
$\mathcal{L}_n$ being the Lie derivative along $n^\mu$; because of the
gauge freedom of general relativity, the functions $\alpha$ and $\beta$
can be in principle freely specified. The four constraint equations of
the ADM system (generally formulated as an elliptic system, but see, \eg
\cite{Racz2016} for an alternative formulation) become four evolution
equations for the $Z_\mu$ vector.

The CCZ4 formulation, as presented in \cite{Alic:2011a}, introduces the
conformal factor $\phi := (\det(\gamma_{ij}))^{-1/6}$ to define the
conformal 3-metric $\tilde\gamma_{ij}:= \phi^2\gamma_{ij}$, with unit
determinant. As in the BSSNOK system, the extrinsic curvature is
decomposed into its trace $K=K_{ij}\gamma^{ij}$ and a trace-free part
$\tilde{A}_{ij}$, which are promoted to primary evolution variables \ie
\begin{equation}
\tilde A_{ij} := \phi^2\left(K_{ij}-\frac{1}{3}K\gamma_{ij}\right)\,.
\end{equation}

The second-order version of the vacuum CCZ4 equations, including the
evolution equations for the $1+\log$ slicing [Eq. \eqref{1plog}] and
Gamma-driver shift condition
[Eqs. \eqref{gammadriver1}--\eqref{gammadriver2}], is reported here for
clarity, using essentially the same notation as in \cite{Alic:2011a}
\begin{subequations}
\begin{eqnarray}
\partial_t\tilde\gamma_{ij} &=& - 2\alpha \tilde A_{ij}
+ 2\tilde\gamma_{k(i}\partial_{j)}~\beta^k
- \frac{2}{3}\tilde\gamma_{ij}\partial_k~\beta^k
+\beta^k \partial_k \tilde\gamma_{ij} \,,  \label{gamma_eq}\\
\partial_t \tilde A_{ij} &=& \phi^2  \left[-\nabla_i \nabla_j \alpha
+ \alpha \left(R_{ij} + \nabla_i Z_j + \nabla_j Z_i %- 8 \pi S_{ij}
\right)\right]^{\rm TF}
+ \alpha \tilde A_{ij}\left(K- 2\Theta\right) \nonumber \\
&&
- 2\alpha \tilde A_{il}\tilde A^l_j %\nonumber  \\
+ 2\tilde A_{k(i}\partial_{j)}~\beta^k
-\frac{2}{3}\tilde A_{ij}\partial_k~\beta^k + \beta^k \partial_k \tilde A_{ij}  \,,  \label{A_eq} \\
\partial_t\phi &=& \frac{1}{3} \alpha \phi K
- \frac{1}{3} \phi \partial_k \beta^k + \beta^k \partial_k \phi \,, \label{phi_eq} \\
\partial_t K &=& - \nabla^i \nabla_i \alpha + \alpha \left(R + 2
  \nabla_i Z^i + K^2 -2 \Theta K \right)
+ \beta^j \partial_j K - 3 \alpha \kappa_1 \left(1 +
\kappa_2\right) \Theta % + 4 \pi \alpha \left(S - 3 \tau\right)
\,, \\
\partial_t \Theta &=& \frac{1}{2} \alpha \left(R + 2 \nabla_i Z^i - \tilde A_{ij} \tilde
A^{ij} + \frac{2}{3} K^2 - 2 \Theta K\right) - Z^i
\partial_i \alpha+ \beta^k \partial_k \Theta
- \alpha \kappa_1 \left(2 + \kappa_2\right) \Theta % - 8\pi \alpha\,\tau
\,, \\
\partial_t \hat\Gamma^i &=& 2\alpha \left(\tilde\Gamma^i_{jk} \tilde A^{jk}
- 3 \tilde A^{ij} \frac{\partial_j \phi}{\phi} - \frac{2}{3}
\tilde\gamma^{ij} \partial_j K \right)
+2\tilde\gamma^{ki}\left(\alpha \partial_k \Theta - \Theta
\partial_k \alpha
- \frac{2}{3} \alpha K Z_k\right) -  2\tilde A^{ij} \partial_j \alpha + \beta^k \partial_k \hat\Gamma^i \nonumber \\
&& + \tilde\gamma^{kl} \partial_k \partial_l \beta^i
+ \frac{1}{3}\tilde\gamma^{ik}\partial_k\partial_l \beta^l
+ \frac{2}{3} \tilde\Gamma^i \partial_k \beta^k -
\tilde\Gamma^k \partial_k \beta^i
+ 2 \kappa_3 \left(\frac{2}{3} \tilde\gamma^{ij} Z_j \partial_k \beta^k -
 \tilde\gamma^{jk} Z_j \partial_k \beta^i \right)
- 2 \alpha \kappa_1 \tilde \gamma^{ij} Z_j %- 16 \pi \alpha \tilde\gamma^{ij} S_{j}
\,, \label{Gamma_eq}\\
\label{1plog}
\partial_t \alpha &=& - \alpha^2 g(\alpha) \left(K - K_0 - 2\Theta \right) + \beta^k \partial_k \alpha \,, \\
\label{gammadriver1}
\partial_t \beta^i &=& f b^i +\beta^k\partial_k\beta^i \,, \\
\label{gammadriver2}
\partial_t b^i &=& \partial_t \hat\Gamma^i - \beta^k \partial_k \hat\Gamma^i
+ \beta^k\partial_k b^i - \eta b^i \,,
\end{eqnarray}
\end{subequations}
with the contracted Christoffel symbols $\tilde \Gamma^i := \tilde
\gamma^{jk} \tilde \Gamma^i_{jk} = \tilde \gamma^{ij} \tilde \gamma^{kl}
\partial_l \tilde \gamma_{jk}$, the shorthand $\hat \Gamma^i := \tilde
\Gamma^i + 2 \tilde \gamma^{ij} Z_j\,$, and the use of the upper index
${\rm TF}$ to indicate a quantity whose trace has been removed.

We recall that the four-vector $Z^\mu$ is an extra dynamical field
specifically introduced to account for the energy and momentum
constraints of the Einstein equations~\cite{Bona:2003fj, Bona:2003qn,
Bona:2004yp}. Its temporal component is $Z^0 = \Theta/\alpha$ and the
indices of its spatial part may be raised and lowered with the spatial
physical metric $\gamma_{ij}$. Following \cite{Alic:2011a}, the
Hamiltonian constraint $H$ and the momentum constraint $M_i$ of the CCZ4
system read as usual, namely
\begin{equation}
\label{eqn.adm}
H := R - K_{ij} K^{ij} + K^2\,, \qquad M_i := \gamma^{jl} \left( \partial_l
K_{ij} - \partial_i K_{jl} - \Gamma^m_{jl} K_{mi} + \Gamma^m_{ji} K_{ml}
\right)\,,
\end{equation}
where of course $H=0=M_i$ in the continuum limit.

\subsection{Introduction of the auxiliary variables and resulting
ordering constraints}

We introduce the following $33$ auxiliary variables, which involve first
spatial derivatives of the metric terms,
\begin{align}
\label{eq:Auxiliary}
A_i := \partial_i\ln\alpha = \frac{\partial_i \alpha }{\alpha}\,, \qquad
B_k^{i} := \partial_k\beta^i\,,
\qquad
D_{kij} := \frac{1}{2}\partial_k\tilde\gamma_{ij}\,, \qquad
P_i       := \partial_i\ln\phi = \frac{\partial_i \phi}{\phi}\,.
\end{align}

An immediate consequence of \eqref{eq:Auxiliary} and the Schwarz theorem
on the symmetry of second-order derivatives are the following second
order ordering constraints \cite{Gundlach:2005ta}, which read:
\begin{align}
\label{eqn.second.ord.const}
 \mathcal{A}_{ki}   &:= \partial_k A_i - \partial _i A_k        = 0\,, &
 \mathcal{B}_{kl}^i &:= \partial_k B_l^i - \partial_l B_k^i     = 0\,, \nonumber \\
 \mathcal{D}_{klij} &:= \partial_k D_{lij} - \partial_l D_{kij} = 0\,, &
 \mathcal{P}_{ki} &:= \partial_k P_i - \partial _i P_k = 0\,.
\end{align}

Since $\tilde{A}_{ij}$ is by construction trace-free, the following
additional constraint holds: $\tilde{\gamma}^{ij} \tilde{A}_{ij} = 0$,
and thus
\begin{equation}
\label{atf.diff}
\mathcal{T}_k := \partial_k \left( \tilde{\gamma}^{ij} \tilde{A}_{ij} \right) = \partial_k \tilde{\gamma}^{ij} \tilde{A}_{ij} + \tilde{\gamma}^{ij} \partial_k \tilde{A}_{ij} = 0. 
\end{equation}
These relations will be important later in order to derive a
\textit{strongly} hyperbolic system in first-order form.
Furthermore, from the constraint $\det(\tilde{\gamma}_{ij}) =1$ and via
the Jacobi formula $ \partial_k \det(\boldsymbol{A}) = {\rm tr}(
\det(\boldsymbol{A}) \boldsymbol{A}^{-1} \partial_k \boldsymbol{A})$ 
%% or in component form: $\partial_k \det(A) = \det(A) \gamma_i^j A^{il}
%% \partial_k A_{lj}$ 
on the derivatives of the determinant of a matrix, we obtain the
following additional algebraic constraints on the auxiliary variables
$D_{kij}$ (see also \cite{Brown2012})
\begin{equation} 
  \tilde{\gamma}^{ij} D_{kij} = 0\,. 
	\label{eqn.dcons} 
\end{equation} 
From Eq. \eqref{eqn.dcons}, another differential constraint follows,
namely, 
\begin{equation}
\partial_l \tilde{\gamma}^{ij} D_{kij} + \tilde{\gamma}^{ij} \partial_l
D_{kij}=0\,.
\end{equation}
In practical implementations, however, we have not found particular
benefits from making use of this additional constraint in the FO-CCZ4
formulation.

The evolution equations for the auxiliary quantities are obtained by
applying the temporal derivative operator $\partial_t$ to equations
\eqref{eq:Auxiliary}, by subsequently exchanging the spatial and temporal
derivatives on the right-hand side of the resulting equations and by
making use of the PDEs \eqref{gamma_eq}, \eqref{phi_eq}, \eqref{1plog}
and \eqref{gammadriver1}.

Many different first-order formulations of the CCZ4 system are possible,
since any non-purely algebraic term in the original second-order system
can be written as a combination of conservative terms and
non-conservative products (see \cite{Gundlach:2005ta, Hilditch2015} for a
parametric study of such families of systems). In this work, we
considered the two extreme cases: a first one, where as many terms as
possible are written in a conservative flux-divergence form (see, \eg
\cite{Alic:2009}, as an example for the first-order Z4 system) and a
second formulation, similar to the ideas outlined in
\cite{Alcubierre:2008}, making maximum use of the first-order ordering
constraints, so that the variables defining the 4-metric ($\alpha$,
$\beta^i$, $\phi$ and $\tilde{\gamma}_{ij}$) are only evolved by a
nonlinear system of ordinary differential equations (ODEs) and where the
rest of the dynamics is written in terms of non-conservative products.
The coefficients of these non-conservative products are only functions of
$\alpha$, $\beta^i$, $\phi$ and $\tilde{\gamma}_{ij}$ and no differential
terms in these variables appear. The dynamical variables of the FO-CCZ4
system with Gamma-driver shift condition are then: $\tilde{A}_{ij}$, $K$,
$\Theta$, $\hat{\Gamma}^i$, $b^i$ (the $b^i$ vector is an auxiliary field
used to write the Gamma-driver gauge condition
\cite{Alcubierre:2008,Alic:2011a}) and the auxiliary variables $A_k$,
$B_k^i$, $P_k$ and $D_{kij}$. In this paper we will follow the second
approach, \ie the final system of 58 evolution equations will consist of
11 ODEs and 47 PDEs and will have a very special structure discussed
later in Section \ref{sec.hyp}.

%-------------------------------------------------------------------
\subsection{Strongly hyperbolic first-order form of the CCZ4 system}
\label{sec.foccz4}
%-------------------------------------------------------------------

The most natural first-order formulation of the CCZ4 system is
non-conservative and appears in the following form discussed later in
more detail
\begin{equation}
\label{eqn.pde.mat.preview}
\frac{\partial \boldsymbol{Q} }{\partial t} +
\boldsymbol{A}_1(\Q) \frac{\partial \Q}{\partial x_1} +
\boldsymbol{A}_2(\Q) \frac{\partial \Q}{\partial x_2} +
\boldsymbol{A}_3(\Q) \frac{\partial \Q}{\partial x_3}  = \boldsymbol{S}(\boldsymbol{Q}),
\end{equation}
where one has the state vector $\Q$, the system matrices
$\boldsymbol{A}_i$ and the purely algebraic source terms
$\boldsymbol{S}(\Q)$. To obtain a \textit{strongly} hyperbolic
first-order system from the second-order CCZ4 formulation of Alic et
al. \cite{Alic:2011a} given by \eqref{gamma_eq}-\eqref{gammadriver2} we
systematically use the constraints \eqref{eqn.second.ord.const} and
\eqref{atf.diff} and make \textit{maximum possible use} of the auxiliary
variables Eq.~\eqref{eq:Auxiliary}. In other words, our first-order CCZ4
system does \textit{not} contain \textit{any} spatial derivatives of
$\alpha$, $\beta^i$, $\tilde{\gamma}_{ij}$ and $\phi$ any more, but all
these terms have been moved to the purely algebraic source term
$\boldsymbol{S}(\Q)$ by using \eqref{eq:Auxiliary}. This has the
immediate consequence that the evolution equations \eqref{eqn.gamma},
\eqref{eqn.alpha}, \eqref{eqn.beta} and \eqref{eqn.phi} reduce to
\textit{ordinary} differential equations instead of \textit{partial}
differential equations.

\begin{figure}[!htbp]
\begin{center}
   \includegraphics*[width=0.7\textwidth]{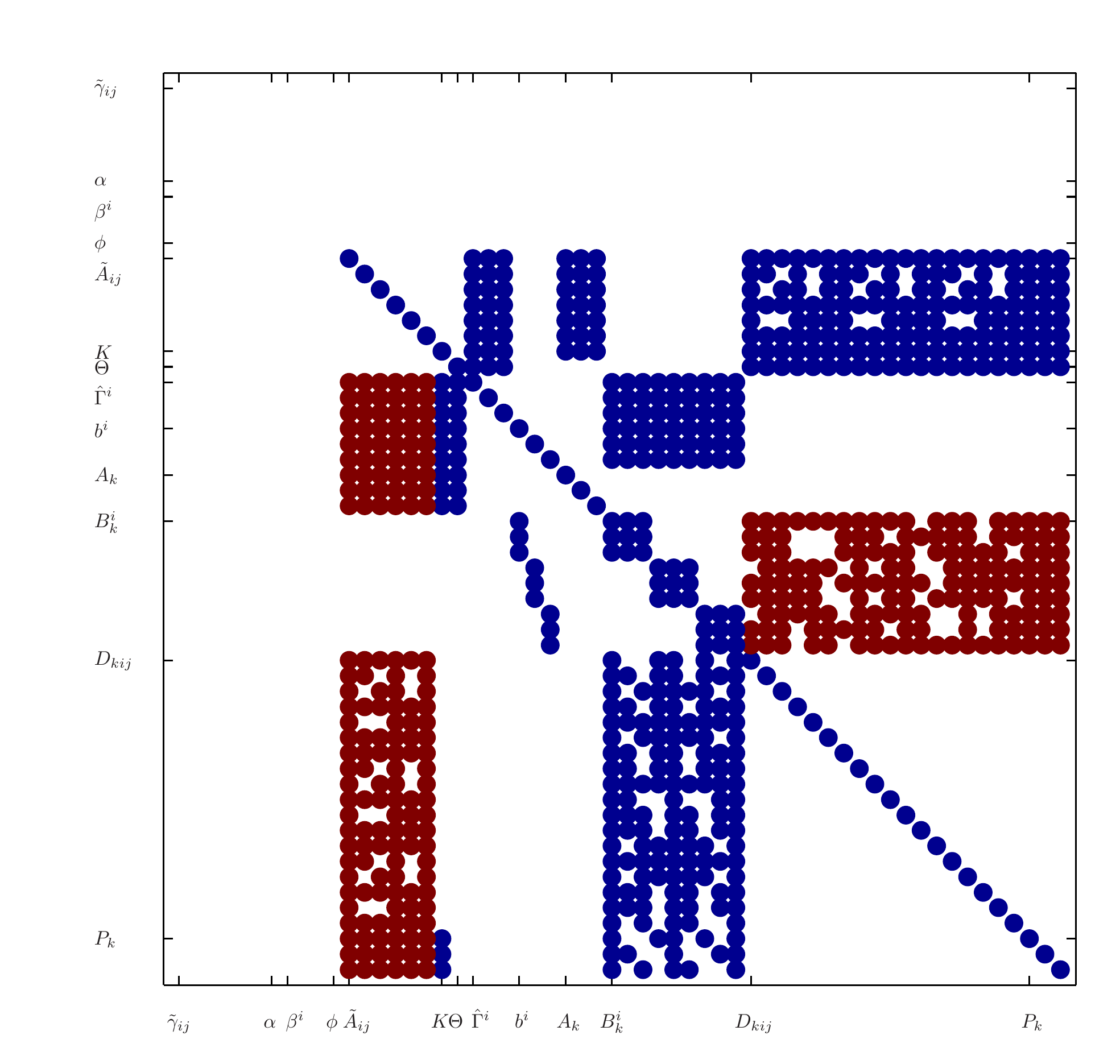}
    \caption{Sparsity pattern of the system matrix $\boldsymbol{A} \cdot
      \boldsymbol{n}$ with $\boldsymbol{n}=(1,1,1) / \sqrt{3}$ for
      randomly perturbed flat Minkowski spacetime using the Gamma-driver
      shift condition ($s=1$) and $1+\log$ slicing
      ($g(\alpha)=2/\alpha$), without the use of the constraints
      \eqref{eqn.second.ord.const} and \eqref{atf.diff} (blue dots) and
      with the use of these constraints (blue \& red dots). The achieved
      \textit{approximate symmetrization} of the sparsity pattern is
      evident. Note also the complete absence of non-zero entries in the
      first 11 lines and columns corresponding to the variables
      $\tilde{\gamma}_{ij}$,$\alpha$, $\beta^i$ and $\phi$, which clearly
      highlights the special structure of our FO-CCZ4 system that can be
      split into a set of pure ODEs and a reduced PDE system, as
      discussed in Section \ref{sec.hyp}.}
    \label{fig.pattern}
\end{center}
\end{figure}

Our final non-conservative first-order CCZ4 system reads as follows:
\begin{subequations}
\begin{eqnarray}
\label{eqn.gamma}
\partial_t\tilde\gamma_{ij}
  & = &  {\beta^k 2 D_{kij} + \tilde\gamma_{ki} B_{j}^k  + \tilde\gamma_{kj} B_{i}^k - \frac{2}{3}\tilde\gamma_{ij} B_k^k }
	- 2\alpha \left( \tilde A_{ij} - {\frac{1}{3} \tilde \gamma_{ij} {\rm tr}{\tilde A} } \right)
  - { \tau^{-1} ( \tilde{\gamma} -1 ) \, \tilde{\gamma}_{ij}},
\\
\label{eqn.alpha}
{ \partial_t \ln{\alpha} }  &=&  { \beta^k A_k } - \alpha g(\alpha) ( K - K_0 - 2\Theta {c} )   \,, \\
\label{eqn.beta}
\partial_t \beta^i   &=& 
 s \beta^k B_k^i +
 s f b^i \\
\label{eqn.phi}
{ \partial_t \ln{\phi} }  & = & { \beta^k P_k } + \frac{1}{3} \left( \alpha K - {B_k^k} \right)  \,,\\
\label{eqn.Aij}
\partial_t\tilde A _{ij} - \beta^k \partial_k\tilde A_{ij} & - & \phi^2 \bigg[  -\nabla_i\nabla_j  \alpha + \alpha \left( R_{ij}+ \nabla_i Z_j + \nabla_j Z_i \right) \bigg]
+ \phi^2 \frac{1}{3} \frac{\tilde\gamma_{ij}}{\phi^2} \bigg[ -\nabla^k \nabla_k \alpha + \alpha (R +  2 \nabla_k Z^k ) \bigg]    \\
\nonumber
& = & { \tilde A_{ki} B_j^k + \tilde A_{kj} B_i^k - \frac{2}{3}\tilde A_{ij} B_k^k }
+ \alpha \tilde A_{ij}(K - 2 \Theta {c} ) - 2 \alpha\tilde A_{il} \tilde\gamma^{lm} \tilde A_{mj}  - { \tau^{-1} \, \tilde{\gamma}_{ij} \, {\rm tr}{\tilde A}   } \,, \\
\label{eqn.K}
\partial_t K - \beta^k \partial_k K & + & \nabla^i \nabla_i \alpha - \alpha( R + 2 \nabla_i Z^i) =
\alpha K (K - 2\Theta {c} ) - 3\alpha\kappa_1(1+\kappa_2)\Theta \\
\label{eqn.theta}
\partial_t \Theta  - \beta^k\partial_k\Theta & - & \frac{1}{2}\alpha {e^2} ( R + 2 \nabla_i Z^i)
  =  \frac{1}{2} \alpha {e^2} \left( \frac{2}{3} K^2 - \tilde{A}_{ij} \tilde{A}^{ij} \right) - \alpha \Theta K {c} - {Z^i \alpha A_i}
				- \alpha\kappa_1(2+ \kappa_2)\Theta  \,,   \\
\label{eqn.Ghat}
\partial_t \hat\Gamma^i - \beta^k \partial_k \hat \Gamma^i &+& \frac{4}{3} \alpha \tilde{\gamma}^{ij} \partial_j K  - 2 \alpha \tilde{\gamma}^{ki} \partial_k \Theta
- s \tilde{\gamma}^{kl} \partial_{(k} B_{l)}^i
- s \frac{1}{3} \tilde{\gamma}^{ik}  \partial_{(k} B_{l)}^l - \textcolor{red}{ s 2 \alpha \tilde{\gamma}^{ik}  \tilde{\gamma}^{nm} \partial_k \tilde{A}_{nm}   }
   \\
 & = & { \frac{2}{3} \tilde{\Gamma}^i B_k^k - \tilde{\Gamma}^k B_k^i  } +
       2 \alpha \left( \tilde{\Gamma}^i_{jk} \tilde{A}^{jk} - 3 \tilde{A}^{ij} P_j \right) -
       2 \alpha \tilde{\gamma}^{ki} \left( \Theta A_k + \frac{2}{3} K Z_k \right) -
			 2 \alpha \tilde{A}^{ij} A_j \nonumber \\
	&&   - \textcolor{red}{ 4 s \, \alpha \tilde{\gamma}^{ik} D_k^{~\,nm} \tilde{A}_{nm} } + 2\kappa_3 \left( \frac{2}{3} \tilde{\gamma}^{ij} Z_j B_k^k - \tilde{\gamma}^{jk} Z_j B_k^i \right) - 2 \alpha \kappa_1 \tilde{\gamma}^{ij} Z_j \nonumber \\
\label{eqn.b}
\partial_t b^i  - s \beta^k \partial_k b^i & = & s \left(  \partial_t \hat\Gamma^i - \beta^k \partial_k \hat \Gamma^i - \eta b^i \right),
% \beta^k\partial_k \left( b^i - \hat\Gamma^i \right) + ({\cal NCP}^i)_{\hat\Gamma}  = -\eta b^i
% -\frac{2}{3}\alpha K \hat\Gamma^i-\alpha\kappa_1 \hat\Gamma^i\,\\
\end{eqnarray}
with the PDEs for the auxiliary variables given by:
\begin{eqnarray}
\label{eqn.A}
\partial_t A_{k} - {\beta^l \partial_l A_k} &+& \alpha g(\alpha) \left( \partial_k K - \partial _k K_0 - 2c \partial_k \Theta \right)
+ \textcolor{red}{s \alpha g(\alpha) \tilde{\gamma}^{nm} \partial_k \tilde{A}_{nm} }\\ \nonumber &=&  
+ \textcolor{red}{2s\, \alpha g(\alpha) D_k^{~\,nm} \tilde{A}_{nm} }
-\alpha A_k \left( K - K_0 - 2 \Theta c \right) \left( g(\alpha) + \alpha g'(\alpha)  \right) + B_k^l ~A_{l} \,,
\\
\label{eqn.B}
\partial_t B_k^i - s\beta^l \partial_l B_k^i &-& s\left(  f \partial_k b^i + \textcolor{red}{ \alpha^2 \mu \, \tilde{\gamma}^{ij} \left( \partial_k P_j - \partial_j P_k \right) } 
- \textcolor{red}{\alpha^2 \mu \, \tilde{\gamma}^{ij} \tilde{\gamma}^{nl} \left( \partial_k D_{ljn} - \partial_l D_{kjn} \right) } \right)
   % \textcolor{red}{\beta^l \partial_l B_k^i} +
	\\ \nonumber &=& %0,
		s B^l_k~B^i_l \,,
\\
\label{eqn.D}
 \partial_t D_{kij} - {\beta^l \partial_l D_{kij}} &+& s \left(
         - \frac{1}{2} \tilde{\gamma}_{mi} \partial_{(k} {B}_{j)}^m
         - \frac{1}{2} \tilde{\gamma}_{mj} \partial_{(k} {B}_{i)}^m
				 + \frac{1}{3} \tilde{\gamma}_{ij} \partial_{(k} {B}_{m)}^m  \right)
				 +  \alpha \partial_k \tilde{A}_{ij}
				 -  \textcolor{red}{ \alpha \frac{1}{3} \tilde{\gamma}_{ij} \tilde{\gamma}^{nm} \partial_k \tilde{A}_{nm} }  \\ \nonumber
  &=&   B_k^l D_{lij} + B_j^l D_{kli} + B_i^l D_{klj} - \frac{2}{3} B_l^l D_{kij} - \textcolor{red}{  \alpha \frac{2}{3}  \tilde{\gamma}_{ij} D_k^{~\,nm} \tilde{A}_{nm} } - \alpha A_k \left( \tilde{A}_{ij} - \frac{1}{3} \tilde{\gamma}_{ij} {\rm tr} \tilde{A} \right), 
\\
\label{eqn.P}
 \partial_t P_{k} - \beta^l \partial_l P_{k} &-& \frac{1}{3} \alpha \partial_k K
+ s \frac{1}{3} \partial_{(k} {B}_{i)}^i   - \textcolor{red}{s \frac{1}{3} \alpha \tilde{\gamma}^{nm} \partial_k \tilde{A}_{nm} } \\  \nonumber &=&
\frac{1}{3} \alpha A_k K + B_k^l P_l - \textcolor{red}{s \frac{2}{3} \alpha \, D_k^{~\,nm} \tilde{A}_{nm} }. 
%% && \partial_t K_0  =  0.
\end{eqnarray}
\end{subequations}
Indicated in red in the equations above are those terms that have been
added to the PDE to obtain an approximate symmetrization of the sparsity
pattern of the system matrices (see discussion in Sec. \ref{sec.hyp} and
Fig. \ref{fig.pattern}).

A few remarks should be made now. First, the function $g(\alpha)$ in the
PDE for the lapse $\alpha$ controls the slicing condition, where
$g(\alpha)=1$ leads to harmonic slicing and $g(\alpha)=2/\alpha$ leads to
the so-called $1+\log$ slicing condition, see \cite{Bona95b}. Second, in
order to obtain the advective terms along the shift vector in the
evolution equations of the auxiliary variables, we have used the
identities \eqref{eqn.second.ord.const}. We stress that it is important
to use the second-order ordering constraints \eqref{eqn.second.ord.const}
in an appropriate way to guarantee strong hyperbolicity, since a naive
first-order formulation of the second-order CCZ4 system that just uses
the auxiliary variables in order to remove the second-order spatial
derivatives will only lead to a weakly hyperbolic system (see
\cite{Gundlach:2005ta} for a detailed discussion on the use of
second-order ordering constraints in second order in space first order in
time hyperbolic systems). Third, we have found that the use of first and 
second-order ordering constraints alone is \textit{not enough}, but that
one must also literally derive the PDE \eqref{eqn.D} for $D_{kij}$ from
\eqref{gamma_eq} by explicitly exploiting the fact that $\tilde{A}_{ij}$
is trace-free via the use of the constraint $\mathcal{T}_k$ by adding Eq.
\eqref{atf.diff} to Eq. \eqref{eqn.D}. Without the use of
$\mathcal{T}_k$ in Eq. \eqref{eqn.D}, the system immediately loses its 
strong hyperbolicity 
(see also \cite{Cao:2012} for a similar observation in the Z4c system).
Once again, these important additional terms in the FO-CCZ4 system
related to the constraints \eqref{eqn.second.ord.const} and
\eqref{atf.diff} have been highlighted in red in Eqs.
\eqref{eqn.gamma}-\eqref{eqn.P}.

We also have introduced several additional constants compared to the
original second-order CCZ4 system. In particular:
\begin{itemize}
\item the constant $\tau$ is a relaxation time to enforce the algebraic
  constraints on the determinant of $\tilde{\gamma}_{ij}$ and on the
  trace of $\tilde{A}_{ij}$ \textit{``weakly''} (see the discussion in
  \cite{Alic:2011a}).
\item the constant $e$ is a \textit{cleaning speed} for the Hamiltonian
  constraint, following the ideas of the generalized Lagrangian
  multiplier (GLM) approach of Dedner et al. \cite{Dedner:2002}. As the
  cleaning is a non-physical process, $e > 1$ is in principle allowed;
  this leads to faster constraint transport and thus can be used to
  obtain a better satisfaction of the constraints for \textit{purely
    numerical} purposes, but $e \neq 1$ breaks the covariance of the
  FO-CCZ4 system.

\item the constant $\mu>0$ appears in Eq. \eqref{eqn.B} and allows one to
  adjust the contribution of second-order ordering constraints.

\item the constant $s$ contributes to the evolution equations for $b^i$,
  $\beta^i$ and $B^i_k$ and allows to turn on or off the evolution of the
  shift. For $s=0$ we have the simple gauge condition $\partial_t \beta^i
  = 0$, while for $s=1$ the usual Gamma-driver gauge condition is
  obtained.

\item the constant $c$ (not to be confused with the speed of light, which
  is set to unity) allows to remove some of the algebraic source terms of
  the Z4 system, but its default value is $c=1$, see \cite{Alic:2011a}. 

\item instead of evolving the lapse $\alpha$ and the conformal factor
  $\phi$, we evolve their \textit{logarithms}, \ie $\ln(\alpha)$ and
  $\ln(\phi)$. While not a standard choice, this is a very simple method
  to preserve the \textit{positivity} of the lapse and the conformal
  factor also at the discrete level. Note also that when treating black
  holes as punctures, the lapse would vanish at the puncture location and
  its logarithm diverge. We therefore impose a positive lower limit in
  our numerical implementation. Since we employ a DG scheme where the
  solution in every element is represented by an interpolating polynomial
  (see section \ref{sec.ader}), in an element surrounding the puncture
  the polynomial might actually reach values lower than the limit due to
  the Runge phenomenon; even in this case, however, the logarithm would
  not diverge.

\end{itemize}

Furthermore, we have the following expressions and identities for various
terms appearing in the evolution equations:
\begin{eqnarray}
{\rm tr} \tilde{A} & = &  \tilde{\gamma}^{ij} \tilde{A}_{ij}, \qquad \textnormal{ and } \qquad \tilde{\gamma} = \textnormal{det}( \tilde{\gamma}_{ij} ), \\
\partial_k \tilde{\gamma}^{ij} & = &  - 2 \tilde{\gamma}^{in} \tilde{\gamma}^{mj} D_{knm} := -2 D_k^{~\,ij}, 
% for the ones which don't find this obvious:
\qquad\text{(derivative of the inverse matrix)} \\
\tilde{\Gamma}_{ij}^k &=& \tilde{\gamma}^{kl} \left( D_{ijl} + D_{jil} - D_{lij} \right), \\
\label{eqn.dchr}
\partial_k \tilde{\Gamma}_{ij}^m & = & -2 D_k^{ml} \left( D_{ijl} + D_{jil} - D_{lij} \right)
												             + \tilde{\gamma}^{ml} \left( \partial_{(k} {D}_{i)jl} + \partial_{(k} {D}_{j)il} - \partial_{(k} {D}_{l)ij} \right) , \\ %
\Gamma_{ij}^k &=& \tilde{\gamma}^{kl} \left( D_{ijl} + D_{jil} - D_{lij} \right) - \tilde{\gamma}^{kl} \left( \tilde{\gamma}_{jl} P_i + \tilde{\gamma}_{il} P_j - \tilde{\gamma}_{ij} P_l \right)
= \tilde{\Gamma}_{ij}^k - \tilde{\gamma}^{kl} \left( \tilde{\gamma}_{jl} P_i + \tilde{\gamma}_{il} P_j - \tilde{\gamma}_{ij} P_l \right),  \\
\partial_k \Gamma_{ij}^m &=& -2 D_k^{ml} \left( D_{ijl} + D_{jil} - D_{lij} \right) +
2 D_k^{ml} \left( \tilde{\gamma}_{jl} P_i + \tilde{\gamma}_{il} P_j  - \tilde{\gamma}_{ij} P_l \right)
                                       - 2 \tilde{\gamma}^{ml} \left(  D_{kjl} P_i + D_{kil} P_j  - D_{kij} P_l \right)
																			  \nonumber \\
												&&	   + \tilde{\gamma}^{ml} \left( \partial_{(k} {D}_{i)jl} + \partial_{(k} {D}_{j)il} - \partial_{(k} {D}_{l)ij} \right)  -
															\tilde{\gamma}^{ml} \left( \tilde{\gamma}_{jl} \partial_{(k} {P}_{i)} + \tilde{\gamma}_{il} \partial_{(k} {P}_{j)}  - \tilde{\gamma}_{ij} \partial_{(k} {P}_{l)} \right) , \\
R^m_{ikj} & = & \partial_k \Gamma^m_{ij} - \partial_j \Gamma^m_{ik} + \Gamma^l_{ij} \Gamma^m_{lk} - \Gamma^l_{ik} \Gamma^m_{lj}, \\
   R_{ij} & = &  R^m_{imj}, \\
\nabla_i \nabla_j \alpha &=& \alpha A_i A_j - \alpha \Gamma^k_{ij} A_k + \alpha \partial_{(i} {A}_{j)}, \\
\nabla^i \nabla_i \alpha &=& \phi^2 \tilde{\gamma}^{ij} \left( \nabla_i \nabla_j \alpha \right), \\
\tilde{\Gamma}^i & = &  \tilde{\gamma}^{jl} \tilde{\Gamma}^i_{jl},  \\
\partial_k \tilde{\Gamma}^i & = & -2 D_k^{jl} \, \tilde{\Gamma}^i_{jl} + \tilde{\gamma}^{jl} \, \partial_k \tilde{\Gamma}^i_{jl}, \\
Z_i &=& \frac{1}{2} \tilde\gamma_{ij} \left(\hat{\Gamma}^j-\tilde{\Gamma}^j \right), \qquad  Z^i = \frac{1}{2} \phi^2 (\hat\Gamma^i-\tilde\Gamma^i), \\
\nabla_i Z_j &=& D_{ijl} \left(\hat{\Gamma}^l-\tilde{\Gamma}^l \right) + \frac{1}{2} \tilde\gamma_{jl} \left( \partial_i \hat{\Gamma}^l - \partial_i \tilde{\Gamma}^l \right) - \Gamma^l_{ij} Z_l, \\
R + 2 \nabla_k Z^k & = & \phi^2 \tilde{\gamma}^{ij} \left( R_{ij} +
    \nabla_i Z_j + \nabla_j Z_i \right)\,.
%\\
%
%(R_{ij}+\nabla_i Z_j + \nabla_j Z_i) &=& -\tilde\gamma^{lm}\partial_l D_{mij} + \frac{1}{2}\tilde\gamma^{lm}\tilde\gamma^{kn}\bigg[(2D_{lin} + D_{iln} - 2D_{nli})\partial_m\tilde\gamma_{jk} + %(D_{kim} + D_{mik} - D_{ikm})\partial_j \tilde\gamma_{ln}   \\
%\nonumber
%&& + D_{nlj}\partial_i\tilde\gamma_{km}\bigg] + \tilde\gamma_{k(i}\partial_{j)}\hat\Gamma^k + \frac{1}{2}\hat\Gamma^m\partial_m \tilde\gamma_{ij} + %2\frac{1}{\phi}\hat\Gamma^m\tilde\gamma_{m(i}\partial_{j)}\phi - \gij \hat\Gamma^m \frac{\partial_m\phi}{\phi} - 2\tilde\gamma^{pq}D_{qip}\frac{\partial_j\phi}{\phi} \\
%\nonumber
%&&
%- 2\tilde\gamma^{pq}D_{qjp}\frac{\partial_i\phi}{\phi} + P_i \frac{\partial_j\phi}{\phi} + \partial_i P_j - \frac{1}{2}\tilde\gamma^{km}P_k(\partial_i\tilde\gamma_{jm} +
%\partial_j\tilde\gamma_{im} - \partial_m\tilde\gamma_{ij}) - \gij\tilde\gamma^{lk}P_k \frac{\partial_l\phi}{\phi} \\
%\nonumber
%&&  + \gij\tilde\gamma^{lk}\partial_k P_l\,\\
%
%( R + 2\nabla_i Z^i)&=&{{\gamma^{ij}(R_{ij}+\nabla_i Z_j + \nabla_j Z_i)=\phi^2\tilde\gamma^{ij}(R_{ij}+\nabla_i Z_j + \nabla_j Z_i)}} \,,\\
%
%
\end{eqnarray}

Here, we have again made use of the second-order ordering
constraints \eqref{eqn.second.ord.const} by \textit{symmetrizing}
the spatial derivatives of the auxiliary variables as follows:

\begin{equation}
 \partial_{(k} {A}_{i)}     := \frac{ \partial_k A_i + \partial_i A_k       }{2}, \quad
 \partial_{(k} {P}_{i)}     := \frac{ \partial_k P_i + \partial_i P_k       }{2}, \quad
%\end{equation*}
%\begin{equation}
 \partial_{(k} {B}^i_{j)}   := \frac{ \partial_k B^i_j + \partial_j B^i_k     }{2}, \quad
 \partial_{(k} {D}_{l)ij}  := \frac{ \partial_k D_{lij} + \partial_l D_{kij} }{2}.
\label{eqn.symm.aux}
\end{equation}
Many of the above terms will contribute to the purely algebraic source
term, as well as to the non-conservative product. For example, in the
spatial derivatives of the Christoffel symbols of the conformal metric
\eqref{eqn.dchr}, the first bracket contributes only to the purely
algebraic source term, while the second bracket is a non-conservative
product.

In a practical implementation, it is therefore necessary to carefully
separate each contribution. We also stress that in our FO-CCZ4
formulation, the Ricci tensor $R_{ij}$ is directly calculated from the
Riemann tensor $R^m_{ikj}$ and the Christoffel symbols and their
derivatives \textit{ab definitionem}, without making use of the typical
splitting of the Ricci tensor as \eg used in \cite{Alic:2011a}. We also
compute the contracted Christoffel symbols $\tilde{\Gamma}^i$ directly
from their definition, without making use of the fact that the
determinant of $\tilde{\gamma}_{ij}$ is unity, since in general this
cannot be guaranteed to hold exactly at the discrete level, unless the
algebraic constraints are rigorously enforced.

From a more formal and mathematical point of view, the additional use of
the second-order ordering constraints \eqref{eqn.second.ord.const} and
the constraint $\mathcal{T}_k$ (the terms colored in red) can be
motivated by looking at the structure of the sparsity pattern of the
system matrix $\boldsymbol{A} \cdot \boldsymbol{n} = \boldsymbol{A}_1 n_1
+ \boldsymbol{A}_2 n_2 + \boldsymbol{A}_3 n_3$ with and without the use
of these constraints. In Fig. \ref{fig.pattern} we report the sparsity
pattern of the system matrix in the normal direction $\boldsymbol{n} =
1/\sqrt{3} (1,1,1)$ for the Gamma-driver shift condition and the $1+\log$
slicing condition for a randomly perturbed flat Minkowski spacetime,
neglecting all matrix entries whose absolute value is below a threshold
of $10^{-7}$. The blue dots represent the original sparsity structure
\textit{without} the use of the second-order ordering constraints
\eqref{eqn.second.ord.const} and without using the constraint
\eqref{atf.diff}, while the combination of the blue and the red dots
shows the sparsity pattern after the terms colored in red have been added
to the PDE system. Our approach for finding a suitable form of the
ordering constraints to be added is based on \textit{approximate
symmetrization} of the sparsity pattern of the system matrix, in order
to avoid \textit{Jordan blocks}, which cannot be diagonalized. Such
Jordan blocks are evident in the sparsity pattern given by the blue dots
alone in Fig. \ref{fig.pattern}.

We are not aware of works in which the constraint $\mathcal{T}_k$ has
been used in conformal first-order hyperbolic formulations of the 3+1
Einstein equations, but its effect becomes rather clear from Fig.
\ref{fig.pattern}. It is also directly evident from Fig.
\ref{fig.pattern} that the first 11 quantities $\tilde{\gamma}_{ij}$,
$\alpha$, $\beta^i$ and $\phi$ are only evolved by ODEs and that the
entire system does not depend on spatial derivatives of these variables,
since all entries in the first 11 rows and columns of the system matrix
are zero. However, we explicitly stress here that our FO-CCZ4 system is
\textit{not} symmetric hyperbolic in the sense of Friedrichs
\cite{Friedrichs1954}, like for example the family of symmetric
hyperbolic and thermodynamically compatible systems of Godunov and
Romenski \cite{GodunovRomenski72, Godunov:1995a, Rom1998}. Further work
in this direction will be necessary to try and achieve a symmetric
hyperbolic form of FO-CCZ4 with a convex extension.

In summary and as an aid to the reader, we list below the key ideas that
have been used in order to obtain the \textit{strongly hyperbolic} FO-CCZ4
system proposed in this paper:

\begin{enumerate}
 \item maximum use of the first-order ordering constraints
   \eqref{eq:Auxiliary} in order to \textit{split} the complete system
   into 11 pure ODEs \eqref{eqn.ode} for the evolution of the quantities
   defining the 4-metric ($\alpha$, $\beta^i$, $\tilde{\gamma}_{ij}$ and
   $\phi$), and with \textit{no spatial derivatives} of these quantities
   appearing in the remaining PDE system \eqref{eqn.pde.red}. However, if
   we want to keep this very particular split structure of the PDE
   system, it is \textit{not} possible to add damping terms proportional
   to the first-order ordering constraints \eqref{eq:Auxiliary} to the
   system, since this would make spatial derivatives of $\alpha$,
   $\beta^i$, $\tilde{\gamma}_{ij}$, $\phi$ appear again and may
   eventually lead to Jordan blocks which cannot be diagonalized. We
   therefore explicitly refrain from adding these terms, in contrast to
   what has been done in \cite{Brown2012}. Following the philosophy
   above, also writing the system in a flux-conservative form like in
   \cite{Bona97a,Alic:2009} is not possible, since the fluxes will in
   general depend on the 4-metric and thus, after application of the
   chain rule, spatial derivatives of $\alpha$, $\beta^i$,
   $\tilde{\gamma}_{ij}$ and $\phi$ would appear again in the
   quasi-linear form. We note that not adding any damping terms
   proportional to the first-order ordering constraints
   \eqref{eq:Auxiliary} may lead to a rapid growth of these constraints
   on the discrete level (see also \cite{Lindblom:2005gh}). This effect,
   however, may be reduced by a periodic reinitialization of the
   auxiliary variables with appropriate discrete versions of
   Eq. \eqref{eq:Auxiliary}, either after a certain number of timesteps,
   or if a large growth of the first-order constraint violations is
   detected. However, in this paper this option has not been further
   investigated.

 \item \textit{approximate symmetrization} of the sparsity pattern of the
   system matrix $\boldsymbol{A} \cdot \boldsymbol{n}$ by appropriate use
   of the second-order ordering constraints \eqref{eqn.second.ord.const}
   and the constraint \eqref{atf.diff}, \ie by adding the terms
   highlighted in red in PDEs \eqref{eqn.gamma}-\eqref{eqn.P}.
   Symmetrization of the first derivatives of the auxiliary variables by
   using \eqref{eqn.symm.aux}, apart from the advective terms along the
   shift vector.

 \item introduction of an \textit{independent} constraint propagation
   speed $e$ for the Hamiltonian constraint $H$ in the PDE
   \eqref{eqn.theta} for the variable $\Theta$, following the ideas of
   the GLM approach of Dedner et al. \cite{Dedner:2002}.

 \item use of the \textit{logarithms} of $\alpha$ and $\phi$ as evolution
   variables, in order to guarantee positivity for $\alpha$ and $\phi$ in
   a simple and natural way. These evolution quantities are consistent
   with the definitions of the auxiliary variables $A_k$ and $P_k$.
\end{enumerate}

\subsection{Strong hyperbolicity}
\label{sec.hyp}

As already shown briefly above, the FO-CCZ4 system
\eqref{eqn.gamma}-\eqref{eqn.P} can be written in compact matrix-vector
form \eqref{eqn.pde.mat.preview}, where the complete state vector is
given by ${\boldsymbol Q}^T := \left( \tilde\gamma_{ij}, \ln{\alpha},
\beta^i, \ln{\phi}, \tilde A_{ij}, K, \Theta, \hat\Gamma^i, b^i, A_k,
B^i_k, D_{kij}, P_k \right)$, containing a total of 58 variables that
have to be evolved in time. For clarity, we show the full sequential form
of all 58 variables in vector $\Q$ in Appendix \ref{sec:eigenappendix}.
The vector $\Q$ can be split as $\Q^T = ( \boldsymbol{V}^T,
\boldsymbol{U}^T )$ into a vector $\boldsymbol{V}$ of the 11 quantities
that define the 4-metric, $\boldsymbol{V}^T :=(\tilde\gamma_{ij},
\ln{\alpha}, \beta^i, \ln{\phi})$, and a vector $\boldsymbol{U}$ of the
remaining 47 dynamic variables $\boldsymbol{U}^T :=\left( \tilde A_{ij},
K, \Theta, \hat\Gamma^i, b^i, A_k, B^i_k, D_{kij}, P_k \right)$. From
\eqref{eqn.gamma}-\eqref{eqn.P} and Fig. \ref{fig.pattern} it is obvious
that the vector $\boldsymbol{V}$ is evolved in time only via ODEs of the
type
\begin{equation}
\label{eqn.ode}
  \frac{\partial \boldsymbol{V}}{\partial t} = \boldsymbol{S}'(\Q),
\end{equation}
where $\boldsymbol{S}'(\Q)$ contains the first 11 elements of the vector
of purely algebraic source terms $\boldsymbol{S}(\Q)$. Therefore, the
eigenvalues associated with the ODE subsystem for $\boldsymbol{V}$ are
trivially zero. Since in our formulation of the FO-CCZ4 system we have
made maximum use of the first-order ordering constraints, Eqs.
\eqref{eqn.gamma}--\eqref{eqn.P} do not contain \textit{any} spatial
derivative of the quantities in $\boldsymbol{V}$, so that the columns in
the matrices of the related eigenvectors are trivially the unit vectors.
The remaining reduced system that needs to be analyzed contains the
vector $\boldsymbol{U}$ of the dynamic quantities and has the very
particular structure
\begin{equation}
 \label{eqn.pde.red}
\frac{\partial \boldsymbol{U} }{\partial t} +
\boldsymbol{B}_1 (\boldsymbol{V}) \frac{\partial \boldsymbol{U}}{\partial x_1} +
\boldsymbol{B}_2 (\boldsymbol{V}) \frac{\partial \boldsymbol{U}}{\partial x_2} +
\boldsymbol{B}_3 (\boldsymbol{V}) \frac{\partial \boldsymbol{U}}{\partial x_3}  =
\boldsymbol{S}''(\Q)\,,
\end{equation}
where the source term $\boldsymbol{S}''(\Q)$ contains the remaining
elements of the source vector $\boldsymbol{S}(\Q)$ and where the system
matrices $\boldsymbol{B}_i$ depend only on the vector $\boldsymbol{V}$
defining the 4-metric and do \textit{not} depend on the vector
$\boldsymbol{U}$. The non-trivial eigenvectors of the complete system
\eqref{eqn.pde.mat.preview} can thus be obtained from those of the reduced system
\eqref{eqn.pde.red} by simply adding zeros corresponding to the
quantities contained in $\boldsymbol{V}$.

An immediate consequence of the very particular splitting of
\eqref{eqn.pde.mat.preview} into the ODEs \eqref{eqn.ode} and the reduced
PDEs \eqref{eqn.pde.red} is that all waves appearing in the system
\eqref{eqn.pde.red} and thus in \eqref{eqn.pde.mat.preview} are
\textit{linearly degenerate} (see \cite{Toro09} for a detailed
discussion), since the eigenvalues $\lambda_i$ depend only on
$\boldsymbol{V}$ and not on $\boldsymbol{U}$ and hence $\partial
\lambda_i / \partial \Q \cdot \boldsymbol{r}_{i} = 0, \, \forall\,
\lambda_i$. This also means that the FO-CCZ4 system cannot generate shock
waves, since the formation of classical shock waves requires the
compression of characteristics and thus the presence of genuinely
nonlinear fields \cite{Toro09, Rezzolla_book:2013}.

In order to prove strong hyperbolicity of the FO-CCZ4 system proposed in
this paper, we compute the \textit{entire} eigenstructure of the system
matrix $\boldsymbol{B}_1$ in the $x_1$ direction for two standard gauge
choices: i) zero shift $\beta^i=0$ (hence $s=0$) with harmonic slicing,
\ie $g(\alpha)=1$ and ii) the gamma driver shift condition ($s=1$) with
1+log slicing, i.e. $g(\alpha)=2/\alpha$.  Note that, in principle, the
eigenstructure of the principal symbol of the system should be computed
for every normal direction $\mathbf{n} \neq 0$ in space.  However, this
is not necessary in this case, since the Einstein equations are isotropic
(see \cite{Sarbach2012}).

For the first shift condition, there is no need to evolve the  
quantities $b^i$ and $B_k^i$, whose corresponding PDEs can therefore be 
neglected in the following analysis (the associated eigenvalues are 
simply zero and the eigenvectors are the unit vectors).
For zero shift the vector $\boldsymbol{U}$ can thus be furthermore reduced to only 
35 remaining dynamic quantities $\boldsymbol{U}^T = (\tilde{A}_{ij}, K, 
\Theta, \hat{\Gamma}^i, A_k, D_{kij}, P_k)$. In this case the 35 eigenvalues of 
matrix $\boldsymbol{B}_1$ in $x_1$ direction are 
\begin{subequations}
\begin{align}
\lambda_{1,2,\cdots,21} &= 0\,, &
\lambda_{22,23} &= \pm \sqrt{\tilde{\gamma}^{11}} \phi \, \alpha \,e\,, \\
\lambda_{24,25,\cdots,29}&=+\sqrt{\tilde{\gamma}^{11}} \phi \, \alpha\,, &
\lambda_{30,31,\cdots,35}&=-\sqrt{\tilde{\gamma}^{11}} \phi \, \alpha\,.
\end{align}
\end{subequations}
The associated complete set of 35 right eigenvectors defining the right 
eigenvector matrix $\boldsymbol{R}$ as well as the inverse right 
eigenvector matrix ($\boldsymbol{L} = \boldsymbol{R}^{-1}$) that defines 
the so-called left eigenvectors are given in the first section of the 
Appendix \ref{sec:eigenappendix}. 

The fact that the FO-CCZ4 system has only real eigenvalues and a complete
set of linearly independent eigenvectors (where the matrix of
eigenvectors is uniformly bounded) is a necessary and sufficient
condition for strong hyperbolicity. Note that for harmonic lapse the
eigenvectors $\boldsymbol{r}_{22,23}$ are only linearly independent of
$\boldsymbol{r}_{24,\cdots35}$ if $c=1, \, \forall e>0$ or for $e \neq 1,
\, \forall c \geq 0$.  The choice $c=1$ and $e=1$ corresponds to the
standard setting typically used for second order Z4 and CCZ4 systems, and
the importance of using $c=1$ has already been shown in the hyperbolicity
analysis for the first and second order Z4 system carried out in
\cite{Bona:2003qn, Bona:2004yp}, i.e. our results on the FO-CCZ4 system
confirm previous findings made in the literature.
For the gamma driver shift condition, the hyperbolicity analysis is much
more complex and requires the computation of all 47 eigenvectors of the
reduced dynamical system \eqref{eqn.pde.red}, this time including also
the quantities $b^i$ and $B^i_k$. After tedious calculations it was
possible to obtain analytical expressions for the eigenvalues and all 47
eigenvectors also in this case. The results are reported in the second
section of the Appendix \ref{sec:eigenappendix}.
To the best of our knowledge, this is the first time that a hyperbolicity
analysis of a first-order reduction of the CCZ4 system including the
gamma driver shift condition has been carried out. An analysis of the
FO-CCZ4 system with other shift conditions, such as the generalized
harmonic shift \cite{Bona:2004yp, Bona05a}, is left to future work.

At this point, we would like to add the following clarifying remark.  The
hyperbolicity analysis has been carried out for the FO-CCZ4 evolution
system \eqref{eqn.gamma}-\eqref{eqn.P}, which in principle admits
violations of the algebraic constraints $\det(\tilde{\gamma}_{ij})=1$,
$\tilde{\gamma}^{ij} \tilde{A}_{ij} = 0$ and $\tilde{\gamma}^{ij} D_{kij}
= 0$. Hence, compared to the original Z4 system
\cite{Bona:2003fj,Bona:2003qn,Alic:2009}, it has an augmented solution
space.  Since our hyperbolicity analysis has been made without enforcing
the algebraic constraints, it is valid for the FO-CCZ4 system with the
augmented solution space, but should not be regarded as an analysis of
the original Z4 system. However, if the initial data satisfies the
algebraic constraints, a direct consequence of the system
\eqref{eqn.gamma}-\eqref{eqn.P} is that the constraints will remain
satisfied for all times, so that our hyperbolicity analysis also covers
solutions that satisfy the algebraic constraints.

%--------------------------------------------------------
\section{Path-conservative ADER Discontinuous Galerkin schemes with
ADER-WENO subcell finite-volume limiter}
\label{sec.ader}
\subsubsection{Unlimited ADER-DG scheme and Riemann solvers}

As mentioned above, the FO-CCZ4 system \eqref{eqn.gamma}-\eqref{eqn.P}
above can be written as a non-conservative first-order hyperbolic system
of the symbolic form given by Eq. \eqref{eqn.pde.mat.preview} (see also
\cite{Dumbser2009a,Dumbser2010,Dumbser2011}), where the matrices
$\boldsymbol{A}_i$ are the system matrices in the coordinate direction
$x_i$ and their eigenstructure has been analyzed in the previous section.

When solving numerically the system \eqref{eqn.pde.mat.preview}, the
computational domain $\Omega$ is covered by a set of non-overlapping
Cartesian tensor-product elements $\Omega_{i} = [x_i - \halb \Delta x_i,
  x_i + \halb \Delta x_i] \times [y_i - \halb \Delta y_i, y_i + \halb
  \Delta y_i] \times [z_i - \halb \Delta z_i, z_i + \halb \Delta z_i] $
where $\boldsymbol{x}_i = (x_i, y_i, z_i)$ indicates the barycenter of
cell $\Omega_i$ and $\Delta \boldsymbol{x}_i = (\Delta x_i,\Delta
y_i,\Delta z_i)$ defines the size of $\Omega_i$ in each spatial
coordinate direction. Furthermore, the domain $\Omega$ is the union of
all elements, \ie $\Omega = \bigcup \Omega_i$. Adaptive mesh refinement
(AMR) has been implemented in a cell-by-cell framework based on a tree
structure \cite{Khokhlov1998}, together with time-accurate local
time-stepping (LTS; see Refs. \cite{Dumbser2013, Dumbser2014,
  Zanotti2015, Zanotti2015c, Zanotti2015b} for details). In the DG
finite-element framework, the discrete solution of the PDE system
\eqref{eqn.pde.mat.preview} is denoted by $\boldsymbol{u}_h$ in the
following and is defined in the space of tensor products of piecewise
polynomials of degree $N$ in each spatial direction, denoted
$\mathcal{U}_h$ in the following. At time $t^n$, in each element
$\Omega_i$ the discrete solution is written in terms of some spatial
basis functions $\Phi_l(\boldsymbol{x})$ and some unknown degrees of
freedom $\hat{\boldsymbol{u}}_{i,l}^n$ as follows,
\begin{equation}
\boldsymbol{u}_h(\boldsymbol{x},t^n) = \sum \limits_l \hat{\boldsymbol{u}}_{i,l}
\Phi_l(\boldsymbol{x}) := \hat{\boldsymbol{u}}_{i,l}^n \Phi_l(\boldsymbol{x})\,,
\label{eqn.ansatz.uh}
\end{equation}
where $l:=(l_1,l_2,l_3)$ is a multi-index and the spatial basis functions
$\Phi_l(\boldsymbol{x}) = \varphi_{l_1}(\xi) \varphi_{l_2}(\eta)
\varphi_{l_3}(\zeta)$ are generated via the tensor product of
one-dimensional basis functions $\varphi_{k}(\xi)$ on the reference
element $[0,1]$. The mapping from physical coordinates $\boldsymbol{x}
\in \Omega_i$ to reference coordinates $\boldsymbol{\xi}=\left(
\xi,\eta,\zeta \right) \in [0,1]^3$ is simply given by $\boldsymbol{x} =
\boldsymbol{x}_i - \halb \Delta \boldsymbol{x}_i + (\xi \Delta x_i, \eta
\Delta y_i, \zeta \Delta z_i)^T$. For the one-dimensional basis functions
$\varphi_k(\xi)$ we use the Lagrange interpolation polynomials passing
through the Gauss-Legendre quadrature nodes $\xi_j$ of an $N+1$ point
Gauss quadrature formula (see Fig. \ref{fig.subcellgrid}). Hence, the
basis polynomials satisfy the interpolation property $\varphi_k(\xi_j) =
\delta_{kj}$, where $\delta_{kj}$ is the usual Kronecker symbol. Due to
this particular choice of a \textit{nodal} tensor-product basis, the
entire scheme can be written in a dimension-by-dimension fashion, where
all integral operators can be decomposed into a sequence of
one-dimensional operators acting only on the $N+1$ degrees of freedom in
the respective dimension.

To derive the ADER-DG method, we first multiply the governing equations
\eqref{eqn.pde.mat.preview} by a test function $\Phi_k \in
\mathcal{U}_h$, identical to the spatial basis functions of
Eq.~\eqref{eqn.ansatz.uh}. After that, we integrate over the spacetime
control volume $\Omega_i \times [t^n;t^{n+1}]$ and obtain
\begin{equation}
\label{eqn.pde.nc.gw1}
\int \limits_{t^n}^{t^{n+1}} \int \limits_{\Omega_i}
\Phi_k \frac{\partial \Q}{\partial t} d \x \, d t
+\int \limits_{t^n}^{t^{n+1}} \int \limits_{\Omega_i}
\Phi_k \left( \boldsymbol{A}(\Q) \cdot \nabla \Q  \right) d \x \, d t
= \int \limits_{t^n}^{t^{n+1}} \int \limits_{\Omega_i}   \Phi_k
\boldsymbol{S}(\Q) d \x \, d t\,,
\end{equation}
with $d \boldsymbol{x} = d x \, d y \,d z$, \ie we integrate over
\textit{coordinate volumes} rather than over physical volumes. Since the
solution is discontinuous across element interfaces, the resulting jump
terms have to be taken properly into account. This is done in our
numerical scheme with the aid of the path-conservative approach, first
developed by Castro and Par\'es in the finite-volume framework
\cite{Castro2006, Pares2006} and later extended also to the DG
finite-element framework in \cite{Rhebergen2008, Dumbser2009a,
  Dumbser2010}. In the ADER-DG framework, higher order in time is
achieved with the use of an element-local spacetime predictor, denoted by
$\boldsymbol{q}_h(\boldsymbol{x},t)$, and which will be discussed
later. Using \eqref{eqn.ansatz.uh}, integrating the first term by parts
in time, taking into account the jumps between elements and making use of
the local predictor solution $\boldsymbol{q}_h$ instead of $\Q$, the weak
formulation \eqref{eqn.pde.nc.gw1} can be rewritten as
\begin{equation}
\label{eqn.pde.nc.gw2}
\left( \int \limits_{\Omega_i}  \Phi_k \Phi_l d \x \right)
\left( \hat{\boldsymbol{u}}^{n+1}_{i,l} - \hat{\boldsymbol{u}}^{n}_{i,l}  \right)
+ \int \limits_{t^n}^{t^{n+1}} \! \! \int \limits_{\Omega_i^\circ}
\Phi_k \left( \boldsymbol{A}(\q_h) \cdot \nabla \q_h  \right)  d \x \, d t
+ \int \limits_{t^n}^{t^{n+1}} \! \! \int \limits_{\partial \Omega_i}
\Phi_k \mathcal{D}^-\left( \q_h^-, \q_h^+ \right) \cdot \boldsymbol{n} \,d S d t
= \int \limits_{t^n}^{t^{n+1}} \! \! \int \limits_{\Omega_i}
\Phi_k \boldsymbol{S}(\q_h)  d \x \, d t\,.
\end{equation}
In \eqref{eqn.pde.nc.gw2}, the first integral leads to the so-called
``element mass matrix'', which is diagonal for our choice of the basis
and test functions, the second integral accounts for the smooth part of
the discrete solution in the interior $\Omega_i^\circ = \Omega_i
\backslash \partial \Omega_i$ of the element $\Omega_i$, the boundary
integral accounts for the jumps across the element interfaces and the
term on the right-hand side accounts for the presence of the purely
algebraic source terms $\boldsymbol{S}$. Following the path-conservative
approach \cite{Pares2006,Castro2006,Dumbser2011}, the jump terms are
defined via a path-integral in phase space between the boundary
extrapolated states at the left $\boldsymbol{q}_h^-$ and at the right
$\q_h^+$ of the interface as follows:
\begin{equation}
\label{eqn.pc.scheme}
\mathcal{D}^-\left( \q_h^-, \q_h^+ \right) \cdot \boldsymbol{n} =
\frac{1}{2} \left( \, \int \limits_{0}^1
\boldsymbol{A}(\boldsymbol{\psi}) \cdot \boldsymbol{n} \, \d s \right)
\left( \q_h^+ - \q_h^- \right) - \frac{1}{2} \boldsymbol{\Theta} \left(
\q_h^+ - \q_h^- \right),
\end{equation}
with $\boldsymbol{A} \cdot \boldsymbol{n} = \boldsymbol{A}_1 n_1 + \boldsymbol{A}_2 n_2 +
\boldsymbol{A}_3 n_3$ the system matrix in normal direction and where we have
used the simple segment path
\begin{equation}
\boldsymbol{\psi} = \boldsymbol{\psi}(\q_h^-, \q_h^+, s) =
\q_h^- + s \left( \q_h^+ - \q_h^- \right),
\qquad 0 \leq s \leq 1\,.
\end{equation}
In Eq. \eqref{eqn.pc.scheme} $\boldsymbol{\Theta} > 0$ denotes an
appropriate numerical viscosity matrix. According to
\cite{Dumbser2009a,Dumbser2010,Dumbser2011}, the path integral appearing
in \eqref{eqn.pc.scheme} is simply computed numerically via a
Gauss-Legendre quadrature formula of sufficient order of accuracy. In
this paper, we use one to three Gaussian quadrature points to approximate
the path integral above. For a simple path-conservative Rusanov-type
method \cite{Dumbser2009a,Castro2010}, the viscosity matrix reads
\begin{equation}
\label{eqn.rusanov}
\boldsymbol{\Theta}_{\textnormal{Rus}} = s_{\max}
\boldsymbol{I}_{58\times58}, \qquad \textnormal{with} \qquad s_{\max} =
\max \left( \left| \Lambda(\q_h^-) \right|, \left| \Lambda(\q_h^+)
\right| \right)\,,
\end{equation}
and where $s_{\max}$ denotes the maximum wave speed found at the
interface. Furthermore, $\boldsymbol{I}_{p \times q} = \delta_{ij}$ with
$i,j \in \mathbb{N}$, $1 \leq i \leq p$, $1 \leq j \leq q$ is the $p
\times q$ identity matrix. In order to reduce numerical dissipation for
the quantities evolved via ODEs, \ie for $\alpha$, $\beta^i$,
$\tilde{\gamma}_{ij}$ and $\phi$, in alternative to the Rusanov scheme we
also employ the recently-proposed path-conservative Harten-Lax-van Leer-Einfeldt-Munz 
(HLLEM) method \cite{Dumbser2015}, which is a  
generalization of the original HLLEM method \cite{Harten83,Einfeldt1991}, and for
which the jump terms on the element boundary read 
\begin{equation}
\label{eqn.hllem.scheme}
\mathcal{D}^-\left( \q_h^-, \q_h^+ \right) \cdot \boldsymbol{n} =
-\frac{s_L}{s_R-s_L} \left( \, \int \limits_{0}^1 \boldsymbol{A}(\boldsymbol{\psi})
\cdot \boldsymbol{n} \, \d s \right) \left( \q_h^+ - \q_h^- \right)
+ \frac{s_L s_R}{s_R-s_L} \left( \q_h^+ - \q_h^- \right)
- \frac{s_L s_R}{s_R-s_L} \boldsymbol{R}_* \boldsymbol{\delta}_* \boldsymbol{L}_*
\left( \q_h^+ - \q_h^- \right)\,,
\end{equation}
with
\begin{equation}
\boldsymbol{\delta}_* = \boldsymbol{I}_{11\times11} -
\frac{\boldsymbol{\Lambda}_*^-}{s_L} - \frac{\boldsymbol{\Lambda}_*^+}{s_R},
\qquad
\textnormal{and} \qquad \boldsymbol{\Lambda}_*^{\pm} = \halb
\left( \boldsymbol{\Lambda}_* \pm \left| \boldsymbol{\Lambda}_* \right|
\right)\,.
\end{equation}

Here, $\boldsymbol{R}_*$ and $\boldsymbol{L}_*$ are the rectangular
matrices containing only the right and left eigenvectors of the internal
waves associated with the eigenvalues $\boldsymbol{\Lambda}_*$ that one
wants to resolve exactly in the HLLEM Riemann solver. In our case, these
internal waves are exactly the 11 stationary contact waves associated
with the 11 ODEs for $\alpha$, $\beta^i$, $\phi$ and
$\tilde{\gamma}_{ij}$, hence their wave speed is zero and thus
$\boldsymbol{\Lambda}_* = 0$ and $\boldsymbol{\delta}_*=
\boldsymbol{I}_{11\times11}$. The associated 11 eigenvectors are the unit
vectors, hence $\boldsymbol{R}_* = \boldsymbol{I}_{58 \times 11}$ and
$\boldsymbol{L}_* = \boldsymbol{I}_{11 \times 58}$. With the left and
right signal speeds simply chosen as $s_L = -s_{\max}$ and $s_R =
+s_{\max}$ and with $s_{\max}$ computed as in Eq. \eqref{eqn.rusanov},
the HLLEM scheme takes the same form of Eq. \eqref{eqn.pc.scheme} with
the viscosity matrix given by
\begin{equation}
\label{eqn.hllem.visc}
\boldsymbol{\Theta}_{\textnormal{HLLEM}} = s_{\max}
\left( \boldsymbol{I}_{58\times58} - \boldsymbol{I}_{58\times11}
\boldsymbol{I}_{11\times58} \right)\,.
\end{equation}
The choice of the approximate Riemann solver closes the description of
the numerical scheme \eqref{eqn.pde.nc.gw2}. Next, we will briefly
describe the computation of the local spacetime predictor solution
$\q_h$ needed in Eq. \eqref{eqn.pde.nc.gw2} and \eqref{eqn.pc.scheme}.

\subsubsection{Spacetime predictor}
The element-local spacetime predictor solution $\q_h(\boldsymbol{x},t)$
is computed from the known discrete solution $\u_h(\boldsymbol{x},t^n)$
at time $t^n$ using a solution of the Cauchy problem \textit{``in the
  small''}, \ie without considering the interaction with the neighbors,
according to the terminology introduced by Harten et al. in \cite{eno}.
In the ENO scheme of Harten et al. \cite{eno}, and in the original ADER
approach of Toro and Titarev \cite{Titarev2002,Titarev2005,Toro2006}, the
so-called Cauchy-Kovalewski procedure was used. This procedure is very
cumbersome and is based on local Taylor series expansions in space and
time and where time derivatives are replaced by the known space
derivatives at time $t^n$ by successively differentiating the governing
PDE system with respect to space and time and inserting the resulting
terms into the Taylor series. For an explicit example of the
Cauchy-Kovalewski procedure applied to the three-dimensional Euler
equations of compressible gas dynamics, see \cite{Dumbser2007}. However,
it is obvious that a highly complex PDE system as the FO-CCZ4 model
\eqref{eqn.gamma}-\eqref{eqn.P} is not amenable to such an approach,
which heavily relies on analytical manipulations of the PDE
system. Therefore, we use an alternative local spacetime DG predictor
\cite{Dumbser2008,Dumbser2009,Dumbser2009a}, which only requires the
pointwise computation of source terms and non-conservative products. The
solution $\q_h$ is expanded into a local spacetime basis
\begin{equation}
\label{eqn.spacetime}
\q_h(\boldsymbol{x},t) = \sum \limits_l \theta_l(\boldsymbol{x},t)
\hat{\boldsymbol{q}}_{i,l} := \theta_l(\boldsymbol{x},t) \hat{\boldsymbol{q}}_{i,l}\,,
\end{equation}
with the multi-index $l=(l_0,l_1,l_2,l_3)$ and where the spacetime basis
functions $\theta_l(\boldsymbol{x},t) = \varphi_{l_0}(\tau)
\varphi_{l_1}(\xi) \varphi_{l_2}(\eta) \varphi_{l_3}(\zeta) $ are again
generated from the same one-dimensional nodal basis functions
$\varphi_{k}(\xi)$ as before, \ie the Lagrange interpolation polynomials
of degree $N$ passing through $N+1$ Gauss-Legendre quadrature nodes. The
spatial mapping $\boldsymbol{x} = \boldsymbol{x}(\boldsymbol{\xi})$ is
also the same as before and the coordinate time is mapped to the
reference time $\tau \in [0,1]$ via $t = t^n + \tau \Delta
t$. Multiplication of the PDE system \eqref{eqn.pde.mat.preview} with a
test function $\theta_k$ and integration over $\Omega_i \times
[t^n,t^{n+1}]$ yields the following weak form in space and time, which is
\textit{different} from \eqref{eqn.pde.nc.gw1}, since now the test and
basis functions are also time dependent:
\begin{equation}
\label{eqn.pde.st1}
\int \limits_{t^n}^{t^{n+1}} \! \! \int \limits_{\Omega_i}
\theta_k(\x,t) \frac{\partial \q_h}{\partial t} d \x \, d t
+\int \limits_{t^n}^{t^{n+1}} \! \! \int \limits_{\Omega_i}
\theta_k(\x,t) \left( \boldsymbol{A}(\q_h) \cdot \nabla \q_h  \right) d \x \, d t
= \int \limits_{t^n}^{t^{n+1}} \! \! \int \limits_{\Omega_i}
\theta_k(\x,t) \boldsymbol{S}(\q_h) d \x \, d t\,.
\end{equation}
Since we are now interested only in an element local predictor solution,
without interaction with the neighbors, at this stage we do \textit{not}
account for the jumps in $\q_h$ across the element interfaces yet, since
this will be done later in the final corrector step of the ADER-DG scheme
\eqref{eqn.pde.nc.gw2}. Instead, we have to introduce the known discrete
solution $\u_h(\x,t^n)$ at time $t^n$. For this purpose, the first term is
integrated by parts in time and leads to
\begin{eqnarray}
\label{eqn.pde.st2}
\int \limits_{\Omega_i}   \theta_k(\x,t^{n+1}) \q_h(\x,t^{n+1}) d \x  -
\int \limits_{\Omega_i}   \theta_k(\x,t^{n}) \u_h(\x,t^{n}) d \x
- \int \limits_{t^n}^{t^{n+1}} \! \! \int \limits_{\Omega_i}
\frac{\partial \theta_k(\x,t)}{\partial t}  \q_h(\x,t)
d \x \, d t = && \nonumber \\
\int \limits_{t^n}^{t^{n+1}} \! \! \int \limits_{\Omega_i}  \theta_k(\x,t)
\boldsymbol{S}(\q_h)  d \x \, d t
- \int \limits_{t^n}^{t^{n+1}} \! \! \int \limits_{\Omega_i} \theta_k(\x,t)
\left( \boldsymbol{A}(\q_h) \cdot \nabla \q_h  \right)  d \x \, d t\,. &&
\end{eqnarray}
Using the local spacetime ansatz \eqref{eqn.spacetime}, Eq.
\eqref{eqn.pde.st2} becomes an element-local nonlinear system for the
unknown degrees of freedom $\hat{\boldsymbol{q}}_{i,l}$ of the spacetime
polynomials $\q_h$. The solution of \eqref{eqn.pde.st2} can be easily
found via a simple and fast converging, fixed-point iteration detailed
\eg in \cite{Dumbser2009a,Toro2009b,Dumbser2014}. This completes the
description of the unlimited ADER-DG scheme.

\subsubsection{ADER-WENO finite-volume subcell limiter}
\label{sec.aderweno}

In general the spatial metric is smooth, which justifies the use of the
high-order unlimited ADER-DG scheme discussed in the previous sections.
However, in the presence of black holes physical singularities appear,
which can lead to numerical instabilities or can even lead to a failure
of the computation. Following the ideas outlined in
Refs. \cite{Dumbser2014, Zanotti2015c, Zanotti2015b}, we therefore
supplement our ADER-DG scheme with a high-order ADER-WENO subcell
finite-volume limiter, which is much more robust than the unlimited DG
scheme, but which is at the same time still high-order accurate in space
and time.

\begin{figure}
\includegraphics[width=\textwidth]{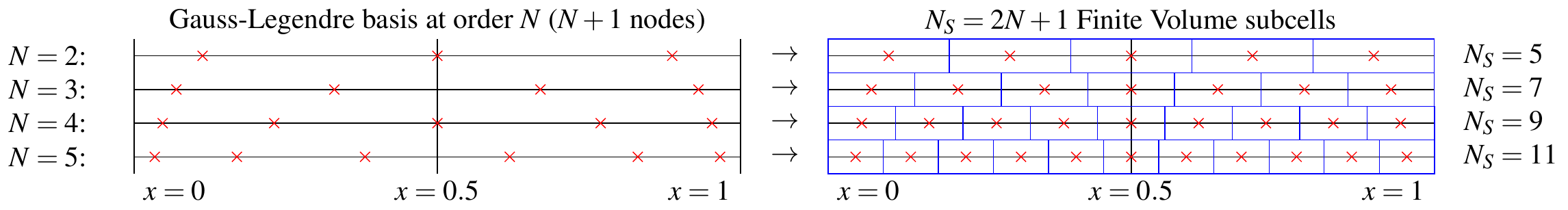}
\caption{\emph{Left}: Node locations in the reference one-dimensional
  cell of the ADER-DG scheme. The coordinate $x\in\{0,1\}$ covers the
  computational cell holding $N$ degrees of freedom, where $N+1$ is the
  order of the method.  \emph{Right}: representation of the finite-volume
  subcell structure, with each cell being divided in a set of $2N+1$
  subcells.}
\label{fig.subcellgrid}
\end{figure}

While in Refs. \cite{Dumbser2014, Zanotti2015c, Zanotti2015b} a
sophisticated \textit{a posteriori} limiting strategy has been proposed,
in this paper for simplicity we fix the limited cells \textit{a priori}
for the entire duration of the simulation, since we assume to know the
location of the black holes. Future simulations with moving black holes
will require a dynamic adjustment of the limiter, as suggested in
\cite{Dumbser2014,Zanotti2015c,Zanotti2015b}. In practice, each
computational cell $\Omega_i$ that has been marked for limiting is split
into $(2N+1)^3$ finite-volume subcells, which are denoted by
$\Omega_{i,s}$ and that satisfy $\Omega_i = \bigcup_s \Omega_{i,s}$ (see
Fig. \ref{fig.subcellgrid}). Note that this very fine division of a DG
element into finite-volume subcells does \textit{not} reduce the timestep
of the overall ADER-DG scheme, since the Courant-Friedrichs-Lewy (CFL)
coefficient of explicit DG schemes scales with $1/(2N+1)$, while the CFL
of finite-volume schemes (used on the subgrid) is of the order of
unity. The discrete solution in the subcells $\Omega_{i,s}$ is
represented at time $t^n$ in terms of \textit{piecewise constant} subcell
averages $\bar{\u}^n_{i,s}$, \ie
\begin{equation}
\label{eqn.subcellaverage}
\bar{\u}^n_{i,s} := \frac{1}{|\Omega_{i,s}|} \int \limits_{\Omega_{i,s}}
\Q(\x,t^n) d \x\,.
\end{equation}
These subcell averages are now evolved in time with a third-order
ADER-WENO finite-volume scheme that looks very similar to the ADER-DG
scheme \eqref{eqn.pde.nc.gw2}, namely
\begin{equation}
\label{eqn.pde.nc.fv}
\left( \bar{\boldsymbol{u}}^{n+1}_{i,s} - \bar{\boldsymbol{u}}^{n}_{i,s}  \right)
+ \int \limits_{t^n}^{t^{n+1}} \! \! \int \limits_{\Omega_{i,s}^\circ}
\left( \boldsymbol{A}(\q_h) \cdot \nabla \q_h  \right)  d \x \, d t
+ \int \limits_{t^n}^{t^{n+1}} \! \! \int \limits_{\partial \Omega_{i,s}}
\mathcal{D}^-\left( \q_h^-, \q_h^+ \right) \cdot \boldsymbol{n} \, d S d t
= \int \limits_{t^n}^{t^{n+1}} \! \! \int \limits_{\Omega_{i,s}}
\boldsymbol{S}(\q_h)  d \x \, d t\,.
\end{equation}
Here, we use again a spacetime predictor solution $\q_h$, which is now
computed from an initial condition given by a WENO \cite{Jiang1996}
reconstruction polynomial $\w_h(\x,t^n)$ computed from the cell averages
$\bar{\u}^n_{i,s}$ via a multi-dimensional WENO reconstruction operator
detailed in \cite{Dumbser2007b,Dumbser2013}. The values at the cell
interfaces $\q_h^-$ and $\q_h^+$ are again computed as the boundary
extrapolated values from the left and the right subcell adjacent to the
interface.

To summarize our nonlinear WENO reconstruction briefly: for each subcell
$\Omega_{i,s}$ we compute several reconstruction polynomials
$\w_h^k(\x,t^n)$ requiring integral conservation of $\w_h^k$ on a set of
different reconstruction stencils $\mathcal{S}^k_{i,s}$, \ie
\begin{equation}
\frac{1}{|\Omega_{i,j}|} \int \limits_{\Omega_{i,j}} \w^k_h(\x,t^n)
d \x = \bar{\u}^n_{i,j} \qquad \forall \, \Omega_{i,j} \in
\mathcal{S}^k_{i,s}\,.
\end{equation}
This system is solved via a constrained least-squares algorithm requiring
at least exact conservation in the cell $\Omega_{i,s}$ itself (see
\cite{Dumbser2007b} for details). From the set of reconstruction
polynomials $\w_h^k$, the final WENO reconstruction polynomial $\w_h$ is
obtained by using a classical nonlinear weighted combination of the
polynomials $\w_h^k$ (see \cite{Jiang1996,Dumbser2007b}) as follows:
\begin{equation}
\label{eqn.weno}
\w_h(\x,t^n) = \sum \limits_k \omega_k \w_h^k(\x,t^n),
\qquad \textnormal{with} \qquad \omega_k = \frac{\tilde{\omega}_k}
{\sum \limits_l \tilde{\omega}_l}
\qquad \textnormal{and}  \qquad \tilde{\omega}_k = \frac{\lambda_k}
{(\sigma_k + \epsilon)^r}\,,
\end{equation}
where the oscillation indicators $\sigma_k$ are computed as usual from
\begin{equation}
\sigma_k := \sum \limits_{l\geq 1} \int \limits_{\Omega_{i,s}}
\Delta \x_{i,s}^{2l -1} \left( \frac{\partial^l}{\partial \x^l}
\w_h^k(\x,t^n) \right)^2 d \x\,.
\end{equation}

The small parameter $\epsilon$ in \eqref{eqn.weno}, which is only
needed to avoid division by zero, is set to $\epsilon=10^{-14}$ and the
exponent in the denominator is chosen as $r=8$. The linear weights are
$\lambda_1 = 10^5$ for the central stencil (\ie $k=1$), while all other
stencils (\ie $k>1$) have linear weight $\lambda_k=1$. This choice
corresponds also to the one made in \cite{Dumbser2007b}.

In a practical implementation it is very convenient to write also the
WENO reconstruction polynomials in terms of some reconstruction basis
functions $\psi_l(\x)$ as $\w_h(\x,t^n) = \Psi_l(\x) \hat{\w}_l^n$. In
this paper, following \cite{Dumbser2013}, the basis functions $\Psi_l$
are defined exactly as the $\Phi_l$, \ie as tensor products of Lagrange
interpolation polynomials through the Gauss-Legendre quadrature
nodes. For the limiter, we only use a piecewise quadratic reconstruction,
leading to a nominally third-order accurate scheme. As already mentioned
before, the predictor is computed according to \eqref{eqn.pde.st2}, where
the initial data $\u_h(\x,t^n)$ is replaced by $\w_h(\x,t^n)$ and the
spatial control volumes $\Omega_i$ are replaced by the subcells
$\Omega_{i,s}$.

Once all subcell averages $\bar{\boldsymbol{u}}^{n+1}_{i,s}$ inside a cell
$\Omega_i$ have been computed according to \eqref{eqn.pde.nc.fv}, the
limited DG polynomial $\u'_h(\x,t^{n+1})$ at the next time level is
obtained again via a classical constrained least squares reconstruction
procedure requiring
\begin{equation}
\frac{1}{|\Omega_{i,s}|} \int \limits_{\Omega_{i,s}} \u'_h(\x,t^{n+1}) d
\x = \bar{\u}^{n+1}_{i,s} \qquad \forall \, \Omega_{i,s} \in \Omega_i,
\end{equation}
and
\begin{equation}
\label{eqn.constr.Omegai}
\int \limits_{\Omega_{i}}
\u'_h(\x,t^{n+1}) d \x = \sum \limits_{\Omega_{i,s} \in \Omega_i}
|\Omega_{i,s}| \bar{\u}^{n+1}_{i,s}\,,
\end{equation}
where \eqref{eqn.constr.Omegai} is a constraint and imposes conservation at
the level of the control volume $\Omega_i$.

This completes the brief description of the ADER-WENO subcell
finite-volume limiter used in this paper. We should emphasize that all
our attempts to simulate puncture black holes with the unlimited ADER-DG
scheme described in Sec. \ref{sec.single.punctures} have failed after
only a few timesteps, and only with the aid of the limiter described in
this section we were able to carry out stable long-time evolutions of
puncture black holes.

\subsubsection{Summary of the path conservative ADER-DG method with
subcell ADER-WENO finite volume limiter}

For the sake of clarity, in this section we briefly summarize the
description of the ADER-DG method with subcell finite-volume limiter
outlined in the previous sections. In particular, to illustrate the
algorithm flow and its practical implementation, we list below the
various stages of the algorithm over a full timestep.

For each element $\Omega_i$, the algorithm to obtain the solution at time
$t^{n+1}$ from the data at time $t^n$ proceeds as follows:
\begin{enumerate}
    \item at all Gauss-Legendre quadrature nodes in each element, the
      data at time $t^n$ for all evolved variables are present, as
      computed from the previous timestep of the scheme (or from the
      initial data); if the cell $\Omega_i$ is flagged as troubled (the
      subcell limiter is active), also all the subcell finite volume
      averages are present;
    \item the discrete data are modified to strictly enforce the
      algebraic constraints of the CCZ4 system (see section
      \ref{sec.tests});
    \item the degrees of freedom $\hat{\boldsymbol{q}}_{i,l}$ of the
      spacetime predictor are computed from \eqref{eqn.pde.st2} using a
      fixed-point iteration method; the spatial derivatives $\nabla
      {\boldsymbol{q}}_h$ of the solution needed in the predictor are
      computed by differentiating the DG polynomial (a so-called
      pseudo-spectral derivative). In unlimited cells, the initial data
      in \eqref{eqn.pde.st2} are given by the DG polynomials $\u_h$, in
      limited cells the initial data are given by the reconstruction
      polynomials $\w_h$ obtained via a nonlinear WENO reconstruction
      operator acting on the subcell averages; inside the spacetime
      predictor, no information from the neighbouring elements is
      necessary;
    \item the predictor is used to compute the volume and boundary
      integrals of the explicit fully discrete corrector stage,
      i.e. Eq. \eqref{eqn.pde.nc.gw2} for the ADER-DG scheme and Eq.
      \eqref{eqn.pde.nc.fv} for the ADER finite volume limiter; in both
      cases the jump terms at the element interfaces are evaluated using
      the path-conservative approach of Par\'es and Castro
      \cite{Pares2006,Castro2006}, see Eq. \eqref{eqn.pc.scheme};
    \item the volume and boundary integrals are used to compute the
      solution at time $t^{n+1}$ in a fully discrete one-step
      time-update, reminiscent of the forward Euler method;
    \item finally, if a cell is flagged for limiting, the finite volume
      subcell averages are used to reconstruct the limited DG polynomial
      $\u_h'(\x,t^{n+1})$.
\end{enumerate}

The algorithm is simplified by the assumption that the cells to be
limited and evolved via the ADER-WENO subcell finite-volume scheme rather
than the ADER-DG scheme, are fixed a-priori throughout the entire
simulation. Otherwise, additional steps would be present in order to
determine if a particular cell is developing pathologies in the numerical
solution and the limiting procedure should be activated, or
not. Furthermore, in this work we do not employ a dynamic adaptive
mesh-refinement (AMR) approach; instead, for simplicity the refined grid
structure is fixed at the beginning of the simulation. Note finally that
the spacetime predictor is computed ``in the small'', disregarding
contributions from the neighbouring cells. This means in particular that
also boundary conditions are not supplied to the spacetime predictor,
but only to the corrector scheme which carries out the fully-discrete
time update of each cell. Combined with the compact stencil of the
ADER-DG method at any order of accuracy, which involves only the cell and
its direct neighbors, this potentially allows for a very efficient
parallel implementation of the algorithm.

%================================================================
\section{Numerical tests}\label{sec.tests}

In the following we present a battery of standard tests that explore the
ability of our formulation to carry out long-term stable evolutions of a
number of different spacetimes with increasing degree of curvature. If
not stated otherwise, in all of the tests we set initially $\Theta = 0$,
$\hat{\Gamma}^i = \tilde{\Gamma}^i$ and $b^i = 0$ and the HLLEM method is
used, \ie Eq. \eqref{eqn.pc.scheme} with the viscosity matrix
\eqref{eqn.hllem.visc}.  In all tests the algebraic constraints on the
unit determinant of $\tilde{\gamma}_{ij}$, the zero trace of
$\tilde{A}_{ij}$ as well as the constraint $\tilde{\gamma}^{ij} D_{kij} =
0$ (which is a consequence of $|\tilde{\gamma}_{ij}|=1$) have all been
\textit{rigorously enforced} in the discrete solution
$\u_h(\boldsymbol{x},t^n)$ at the beginning of each timestep, but they
have \textit{not} been enforced during the computation of the spacetime
predictor $\q_h$. Note that the predictor $\q_h$ is only an auxiliary
quantity that is overwritten after each timestep and which has a role
similar to the evolution stage to the half timelevel in second-order 
MUSCL-Hancock type TVD finite-volume schemes.
%Note that in the very common case of a semidiscrete scheme evolved in
%time with a high-order Runge-Kutta (RK) time integrator, our choice
%corresponds to enforcing the constraints after a full timestep, but not
%in all the intermediate RK stages.
We therefore set $\tau \to \infty$ and thus neglect the corresponding
source terms. In tests involving black holes, the lower limit on the
lapse is set to be $\ln(\alpha) \geq -20$. We will use the notation $P_N$
to indicate an ADER-DG scheme using piecewise polynomials of degree $N$
to represent $\u_h$.

\subsection{Linearized gravitational-wave test}

The first test problem is a simple one-dimensional wave-propagation test
problem in the linearized regime. The computational setup follows the one
suggested by Alcubierre et al. in Ref. \cite{Alcubierre:2003pc}. The
computational domain is $\Omega = [-0.5,0.5]$ with periodic boundary
conditions in the $x$ direction and two simulations are run until a final
time of $t=1000$: \textit{(i)} a first one using 4 ADER-DG $P_5$ elements
(\ie a total number of 24 degrees of freedom) and \textit{(ii)} a second
one using only 2 ADER-DG $P_9$ elements (\ie only 20 degrees of
freedom). This test is run with the unlimited version of the ADER-DG
scheme. The exact solution of the metric of the problem is given by
\begin{equation}
\label{eqn.lw.metric}
d s^2 = - d t^2 + d x^2 + (1+h) d y^2 + (1-h) d z^2\,,
\qquad \textnormal{with} \qquad
h := \epsilon \sin \left( 2 \pi (x-t) \right)\,,
\end{equation}
and the wave amplitude $\epsilon = 10^{-8}$ is chosen small enough in
order to stay in the linear regime, so that terms
$\mathcal{O}(\epsilon^2)$ can be neglected. Since the shift is zero in
the metric \eqref{eqn.lw.metric} ($\beta^i = 0$), we set $s=0$ in our
FO-CCZ4 system and furthermore harmonic slicing is used, \ie
$g(\alpha)=1$. We also set $K_0=0$, $c=0$, $e=2$ and use the 
\textit{undamped} version of the system, setting $\kappa_1 = \kappa_2 =
\kappa_3 = \eta = 0$. 
%We note that when setting $s=0$ and adopting a harmonic slicing it is
%necessary to choose $e > 1$ in order to get a strongly hyperbolic system,
%hence we use $e=2$. 
Using the metric \eqref{eqn.lw.metric}, the 
definition of the extrinsic curvature reduces to $K_{ij} = -
\tfrac{1}{2}\partial_t \gamma_{ij}/(\alpha)$, so that the various
components are given by $K_{xx} = K_{xy} = K_{xz} = K_{yz} = 0$, $K_{yy}
= -\halb \partial_t h$ and $K_{zz} = +\halb \partial_t h$. From this
information, the conformal factor $\phi$, the conformal spatial metric
$\tilde{\gamma}_{ij}$, the traceless conformal extrinsic curvature
$\tilde{A}_{ij}$ and all auxiliary variables can be computed by a direct
calculation according to their definitions.

\begin{figure}[!htbp]
\begin{center}
    \begin{tabular}{cc}
        \includegraphics[width=0.40\textwidth]{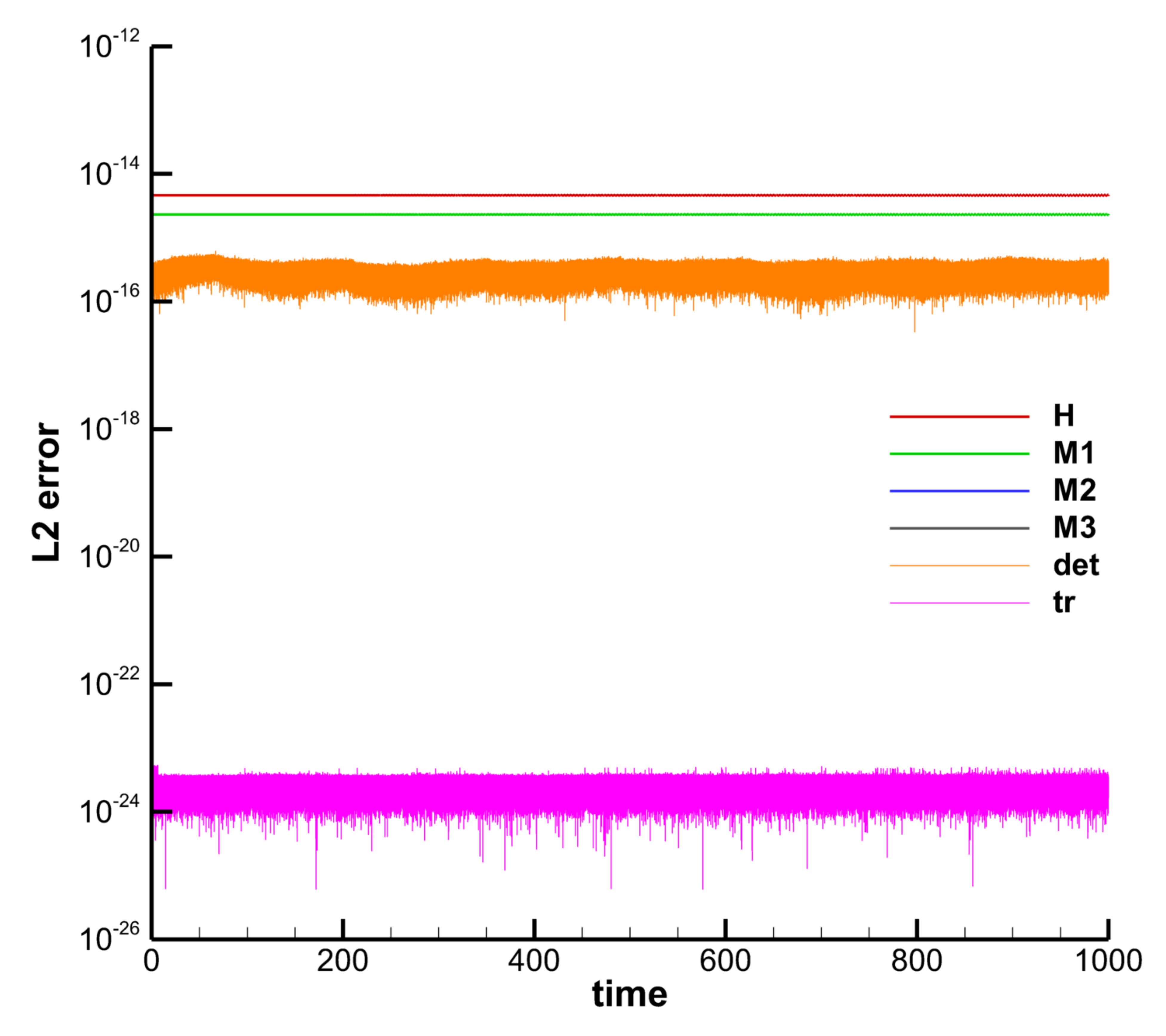}   &
        \includegraphics[width=0.40\textwidth]{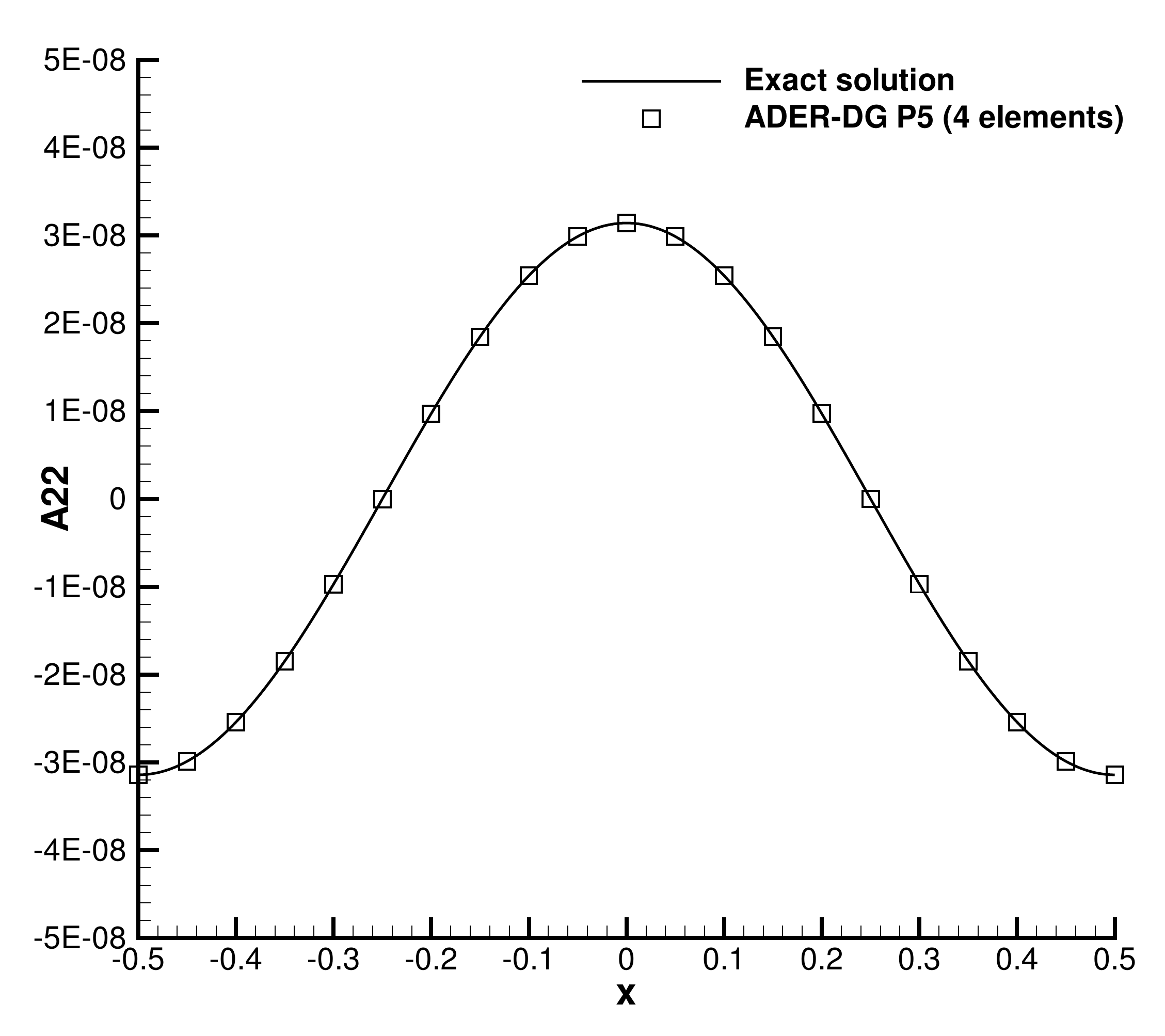} \\
        \includegraphics[width=0.40\textwidth]{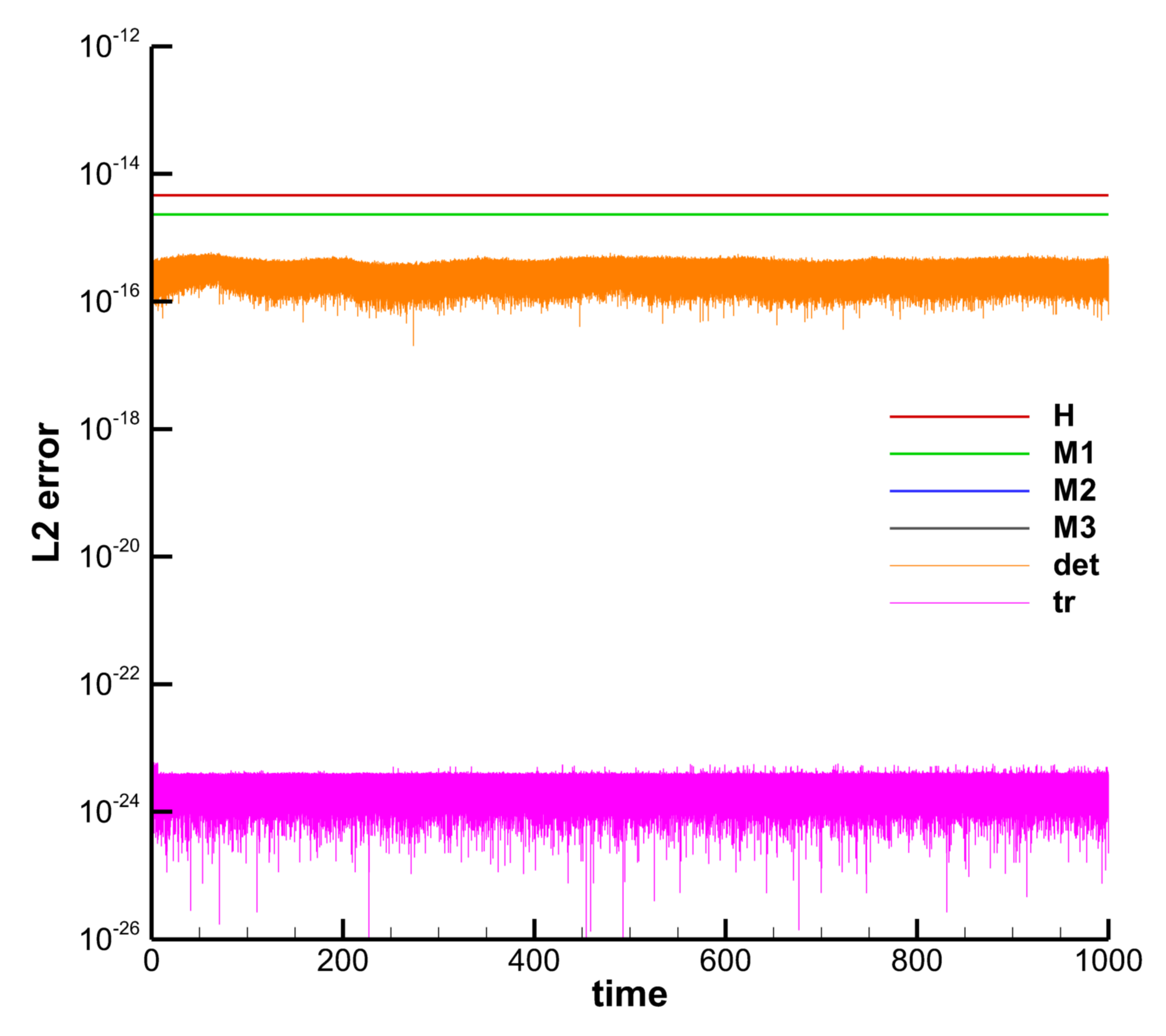}     &
	\includegraphics[width=0.40\textwidth]{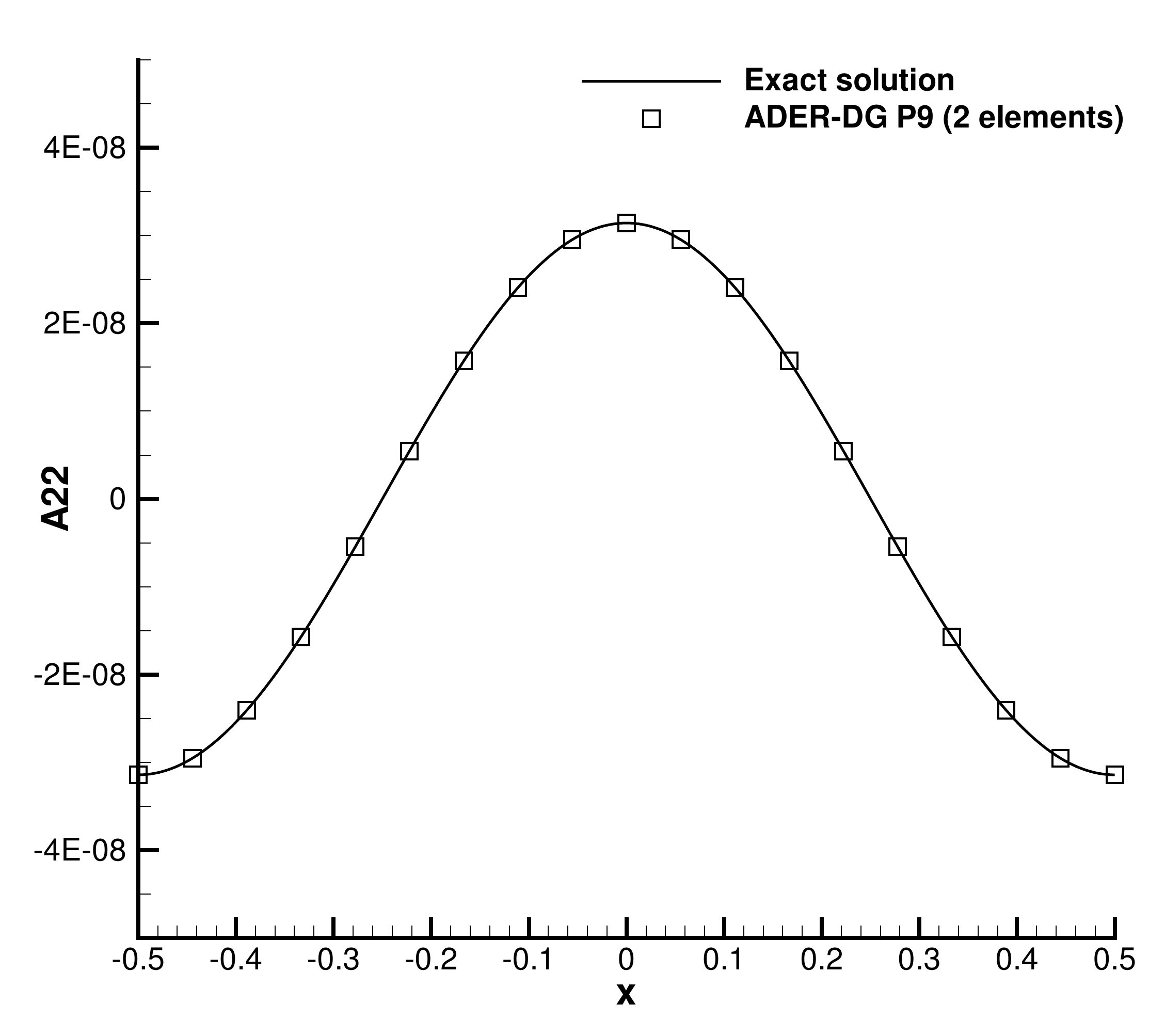}
    \end{tabular}
    \caption{Linearized gravitational-wave test using an ADER-DG $P_5$
      scheme with 4 elements (top row) and an ADER-DG $P_9$ scheme with
      only 2 elements (bottom row). The temporal evolution of the
      constraints (left column) is shown together with the waveform for
      the component $\tilde{A}_{22}$ of the traceless conformal extrinsic
      curvature after 1000 crossing times at time $t=1000$ (right
      column).}
    \label{fig.linearwave}
\end{center}
\end{figure}

In Fig. \ref{fig.linearwave} we report the temporal evolution of all ADM
constraints (Hamiltonian and momentum constraints) as well as the errors
of the algebraic constraints on the determinant of the conformal metric
and the error in the trace of $\tilde{A}_{ij}$ in both cases, \ie using
the ADER-DG $P_5$ and $P_9$ scheme. A comparison of the
extrinsic-curvature component $\tilde{A}_{22}$ with the exact solution is
also provided at the final time $t=1000$, showing overall an excellent
agreement between numerical and exact solution. The quality of the
results obtained with the ADER-DG schemes used in this paper, which are
uniformly high-order accurate in both space and time, is significantly
superior to the results shown in Ref. \cite{Alcubierre:2003pc} for the
same test problem using a finite difference scheme with much more grid
points (between 50 and 200) compared to the very coarse mesh containing
only 20 to 24 degrees of freedom used in our simulations. Note that a
fair comparison between high order finite-difference and DG schemes must
be made in terms of points per wavelength for finite-difference methods
and in degrees of freedom per wavelength for DG schemes.

\subsection{Gauge-wave test}

Also this classical test problem has been taken from the collection of
standard tests of Ref. \cite{Alcubierre:2003pc}. The metric in this case
is given by
\begin{equation}
\label{eqn.gw.metric}
\d s^2 = - H(x,t) d t^2 + H(x,t) \d x^2 + \d y^2 + \d z^2\,,
\qquad \textnormal{with} \qquad H(x,t) := 1-A\,\sin \left( 2\pi(x-t)
\right)\,.
\end{equation}
The metric \eqref{eqn.gw.metric} implies zero shift ($\beta^i = 0$),
hence we use once more $s=0$ together with harmonic slicing
$g(\alpha)=1$. Also for this test we employ the \textit{undamped} version
of the FO-CCZ4 system, setting $\kappa_1 = \kappa_2 = \kappa_3 = \eta =
0$. The computational domain in this case is two-dimensional, with $\Omega = [-0.5,0.5] \times
[-0.05, 0.05]$ with periodic boundary conditions in all directions. Since
$\beta^i = 0$, the extrinsic curvature is again given by $K_{ij} =
-\partial_t \gamma_{ij} / (2\alpha)$, \ie $K_{yy} = K_{zz} = K_{xy} =
K_{xz} = K_{yz} = 0$ and the remaining primary variables are
\begin{equation}
\phi^2 = H^{-1/3}, \qquad  \alpha = \sqrt{H}, \qquad
K_{xx} = - \pi A\frac{\cos\left(2\pi(x-t)\right)}{\sqrt{1 - A\sin\left(2\pi(x-t)\right)}}.
\end{equation}
We furthermore set $K_0=0$. The auxiliary variables can be
obtained from their definition via a straightforward calculation.

\begin{figure}[!htbp]
\begin{center}
    \begin{tabular}{cc}
        \includegraphics[width=0.40\textwidth]{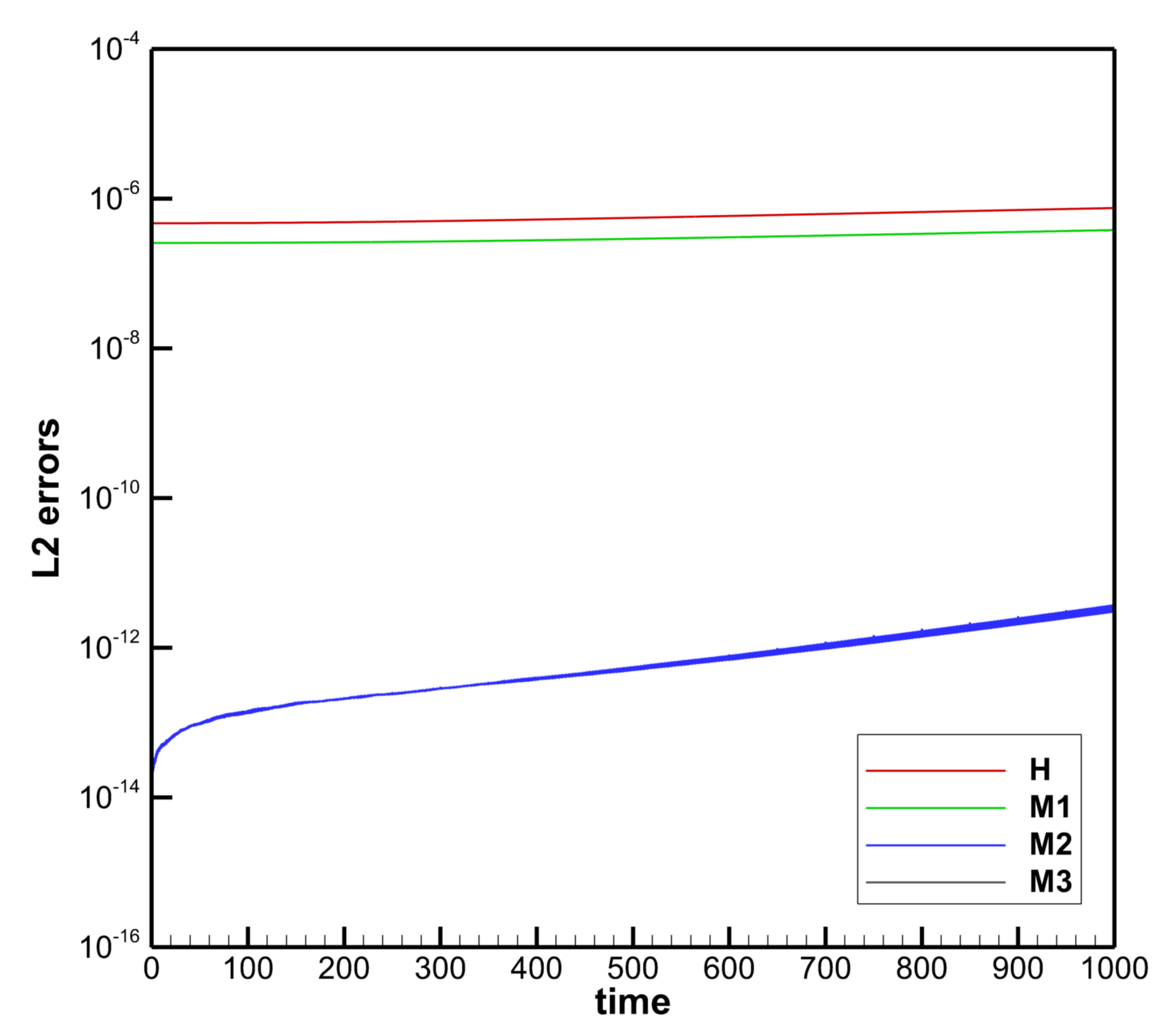} &
				\includegraphics[width=0.40\textwidth]{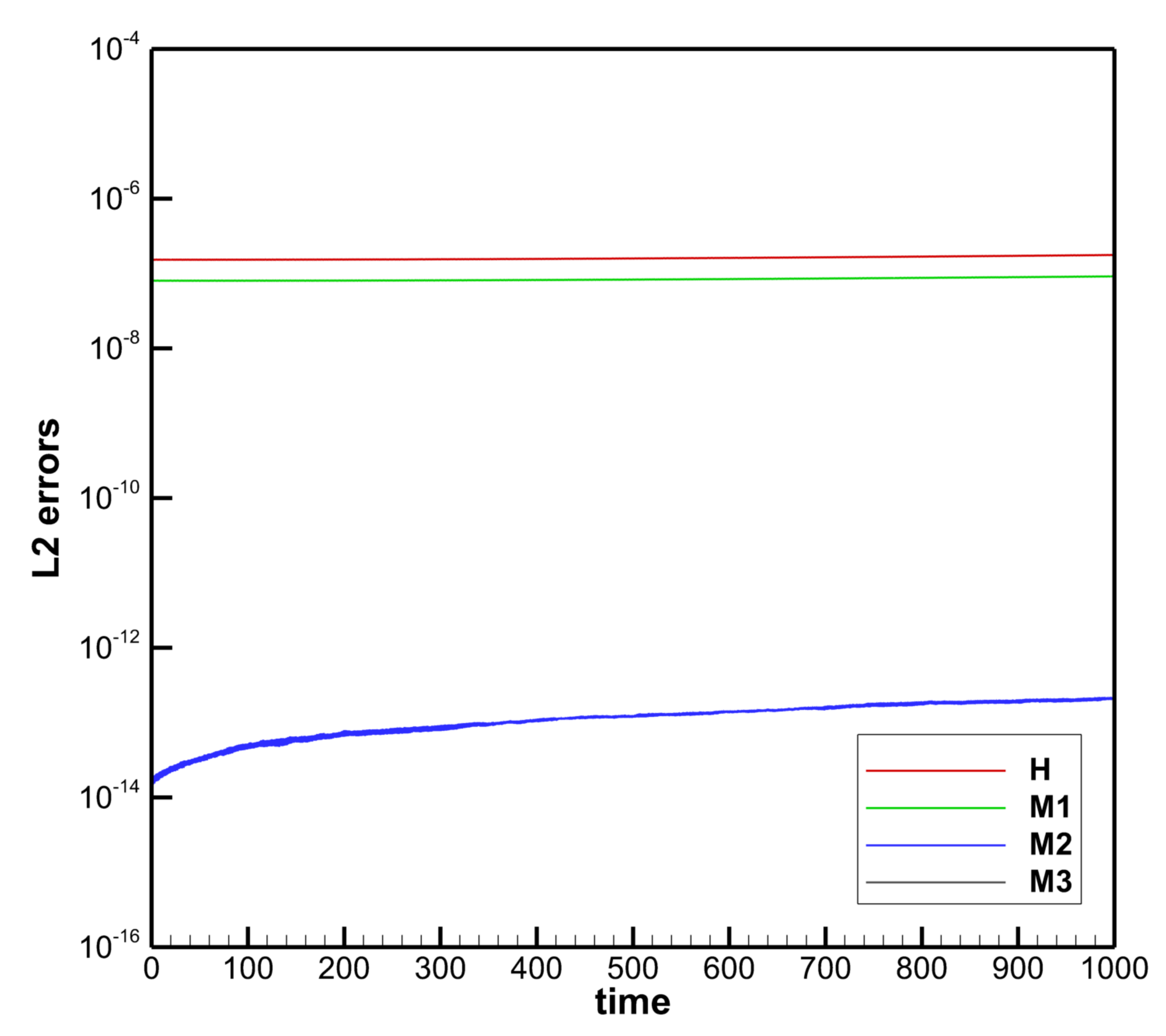} \\ 		
	\includegraphics[width=0.40\textwidth]{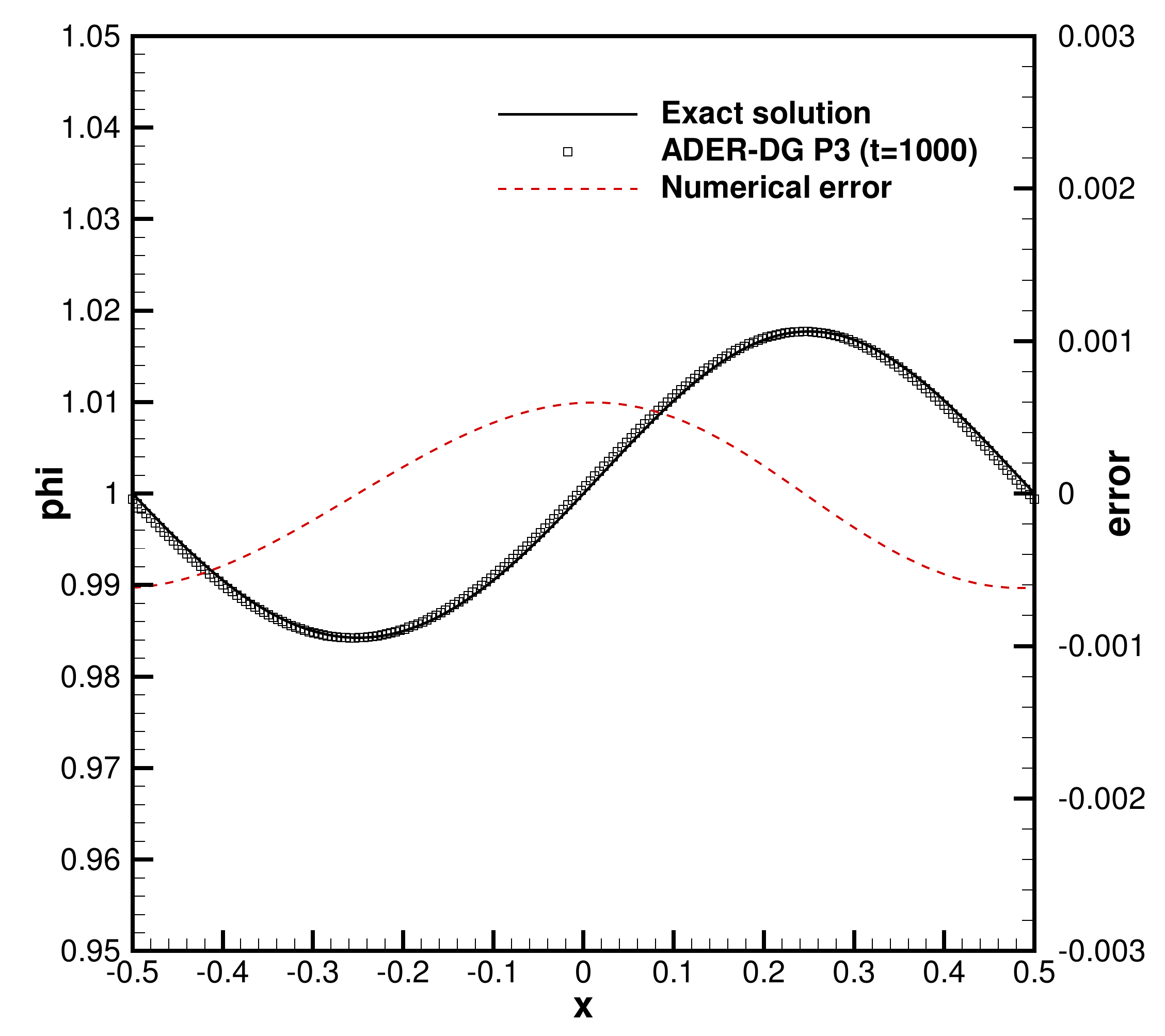} & 
	\includegraphics[width=0.40\textwidth]{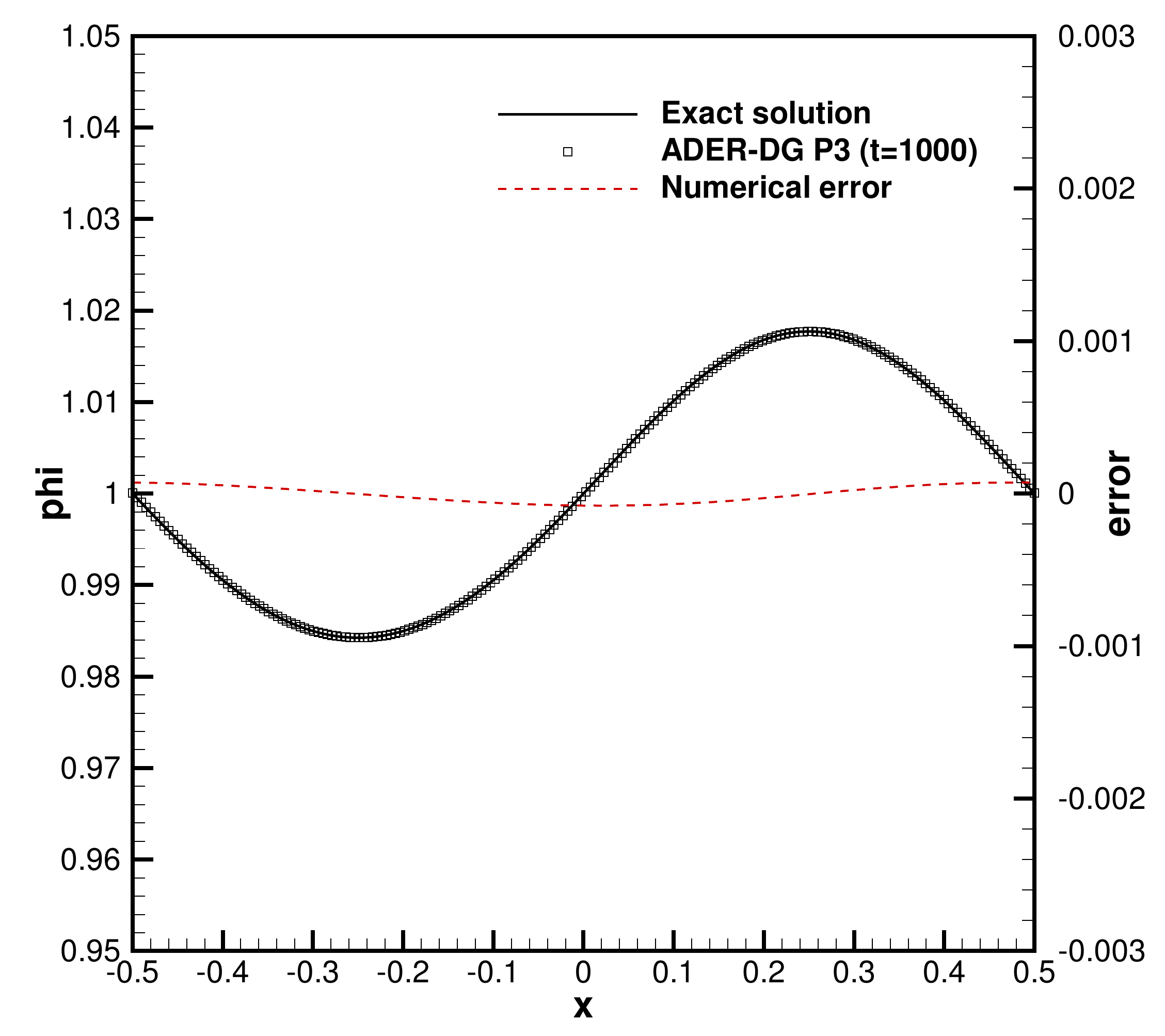} \\
	\includegraphics[width=0.40\textwidth]{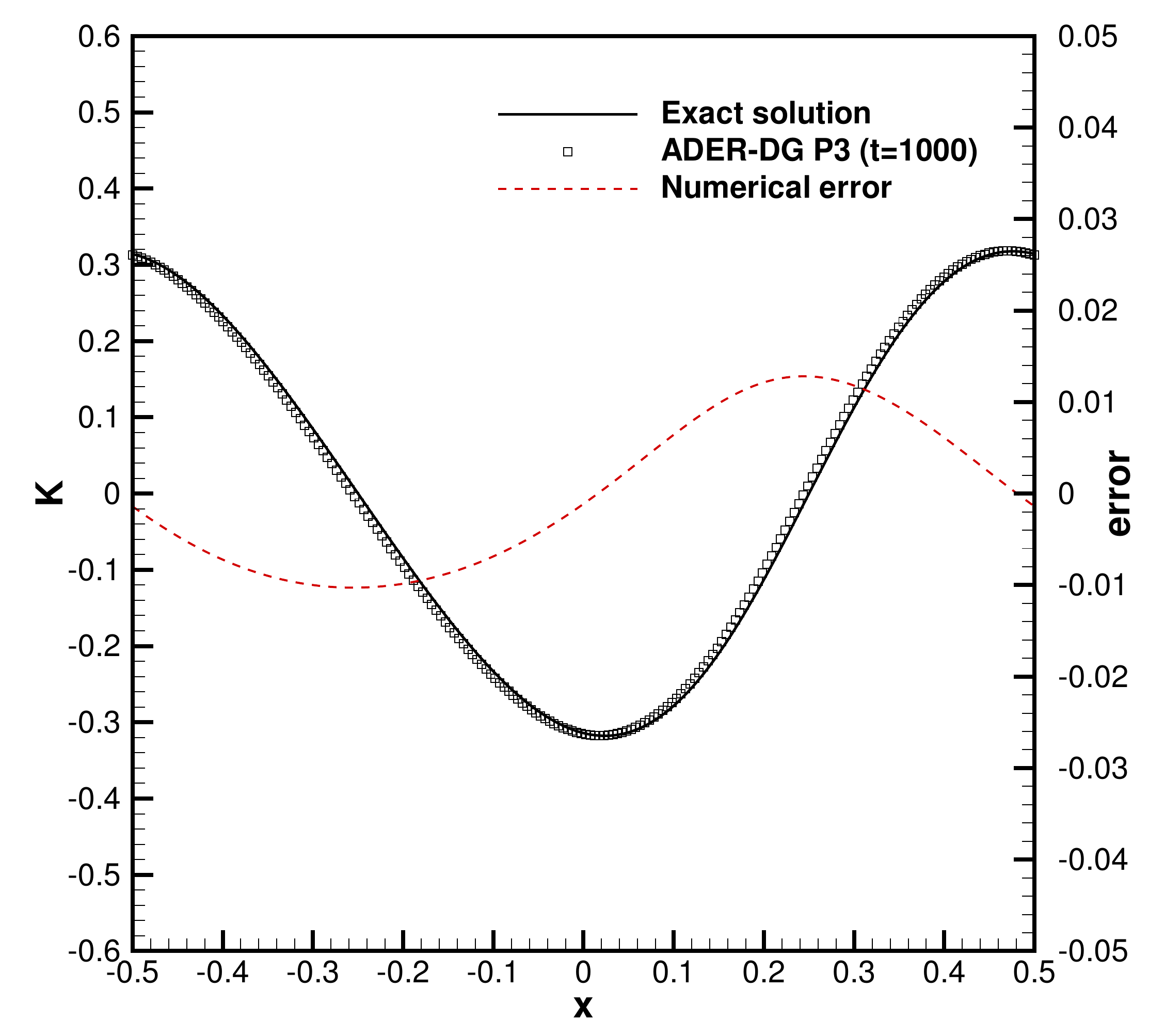} &
	\includegraphics[width=0.40\textwidth]{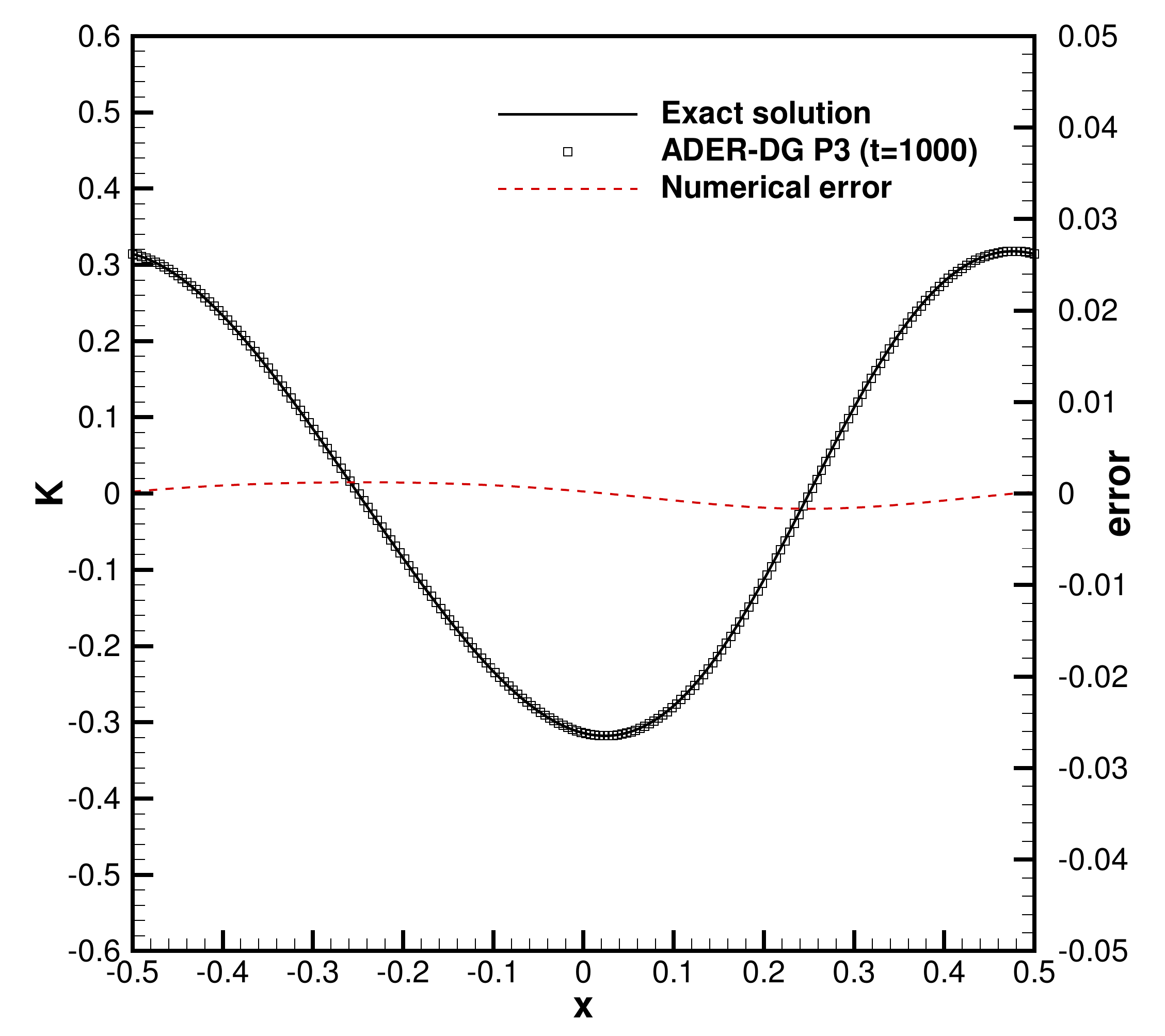}  
    \end{tabular}
    \caption{Gauge-wave test case with amplitude $A=0.1$ using the undamped FO-CCZ4 system ($\kappa_1=\kappa_2=\kappa_3=0$). 
					   Comparison of the default setting $e=1$ and $c=1$ (left) with the modified setting $e=2$ and $c=0$ 
						 leading to an improved cleaning of the Hamiltonian constraint  (right). 
		         Temporal evolution of the Hamiltonian and momentum constraints (top). 
						 Comparison of the wave form of the conformal factor $\phi$ (center) and the trace of the 
						 extrinsic curvature $K$ (bottom) with the exact solution at time $t=1000$. Notice the larger constraint violations and the  
						 slight phase shift in the waveforms present in the results in the left column, corresponding to the default setting $e=c=1$. 
						 Since in this test the momentum constraint $M_3=0$, it is not plotted when using a logarithmic axis.}
    \label{fig.gaugewave}
\end{center}
\end{figure}

We first simulate this test problem with a perturbation amplitude of
$A=0.1$ until $t=1000$ with an unlimited ADER-DG $P_3$ scheme and using
$100 \times 10$ elements to cover the domain $\Omega$. We run this
physical setup twice, once with the default parameters $e=c=1$, according
to the original second order CCZ4 system \cite{Alic:2011a} and a
\textit{modified} setting with $e=2$ and $c=0$ to obtain an improved
cleaning of the Hamiltonian constraint. In both cases the system is
strongly hyperbolic.  The time evolution of the ADM constraints is
reported in Fig. \ref{fig.gaugewave}, showing only a very moderate growth
of the constraint $M_2$ that is sublinear in time and close to machine
precision. The other constraints $H$ and $M_1$ remain essentially
constant during the entire simulation.  We emphasize that we have used
the \textit{undamped} version of the FO-CCZ4 system, and nevertheless
obtain stable results, while the original second-order CCZ4 formulation
was reported to fail for this test problem in the undamped version, and
only the damped CCZ4 system was stable (see \cite{Alic:2011a} for
details). It is also worth recalling that both the first- and the
second-order formulation of the BSSNOK system fail for this test case
after a rather short time (see \cite{Alic:2011a, Brown2012}). In
Fig. \ref{fig.gaugewave} we also provide a direct comparison of the
solution after 1000 crossing times for the conformal factor $\phi$ as
well as for the trace of the extrinsic curvature $K$. Note the overall 
very good agreement between the numerical solution and the exact
one. For the sake of clarity, in the plots of the waveforms we also
report the numerical error computed as the difference between the 
numerical solution and the exact solution at the final time $t=1000$. 
It can be clearly noticed from the computational results shown in Fig. 
\ref{fig.gaugewave} that the constraints and the phase errors in the 
waveforms are significantly smaller for the 
modified setting $e=2$, which may justify the use of a faster cleaning 
speed of the Hamiltonian constraint $e>1$ for \textit{purely numerical} 
purposes. In any case, our FO-CCZ4 system behaves well also with the 
default setting $e=c=1$, which is typically used in the standard 
second order CCZ4 system \cite{Alic:2011a}.

Since the gauge-wave test has a smooth nontrivial exact analytical
solution and is also valid in the nonlinear regime of the equations, we
can use it in order to perform a numerical convergence study. For this
purpose, we run the test again with different unlimited ADER-DG $P_N$
schemes on a sequence of successively refined meshes. To make the test
more difficult, we choose a very large perturbation amplitude of $A=0.9$,
which takes the system in the highly nonlinear regime, although in the
end the test consists only in a nonlinear re-parametrization of the flat
Minkowski spacetime. For thise case we use $c=0$ and $e=2$. We set the 
final simulation time to $t=10$ and continue using the \textit{undamped} 
version of the FO-CCZ4 system. 

\begin{figure}[!htbp]
\begin{center}
    \begin{tabular}{cc}
        \includegraphics[width=0.40\textwidth]{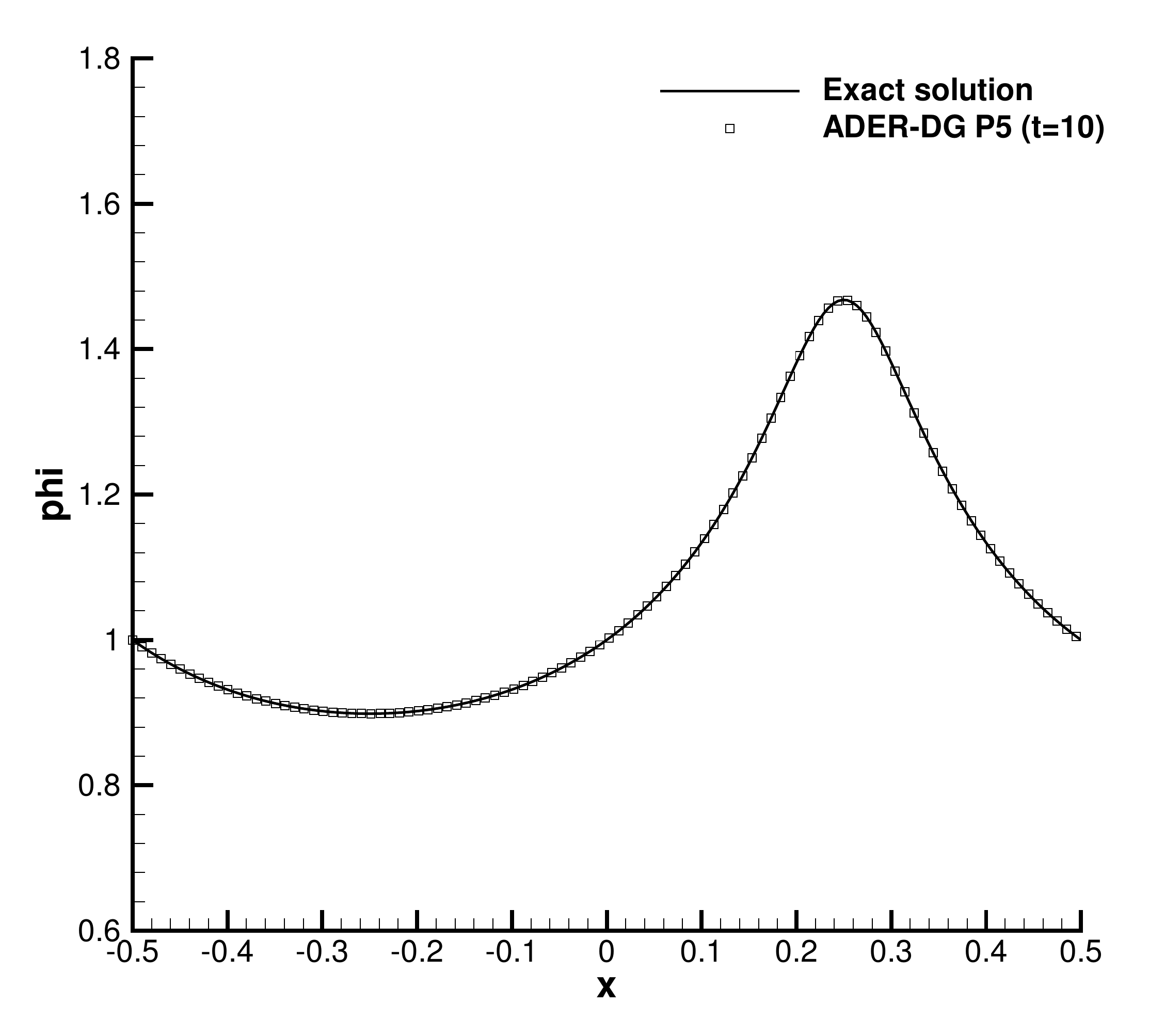}   &
	\includegraphics[width=0.40\textwidth]{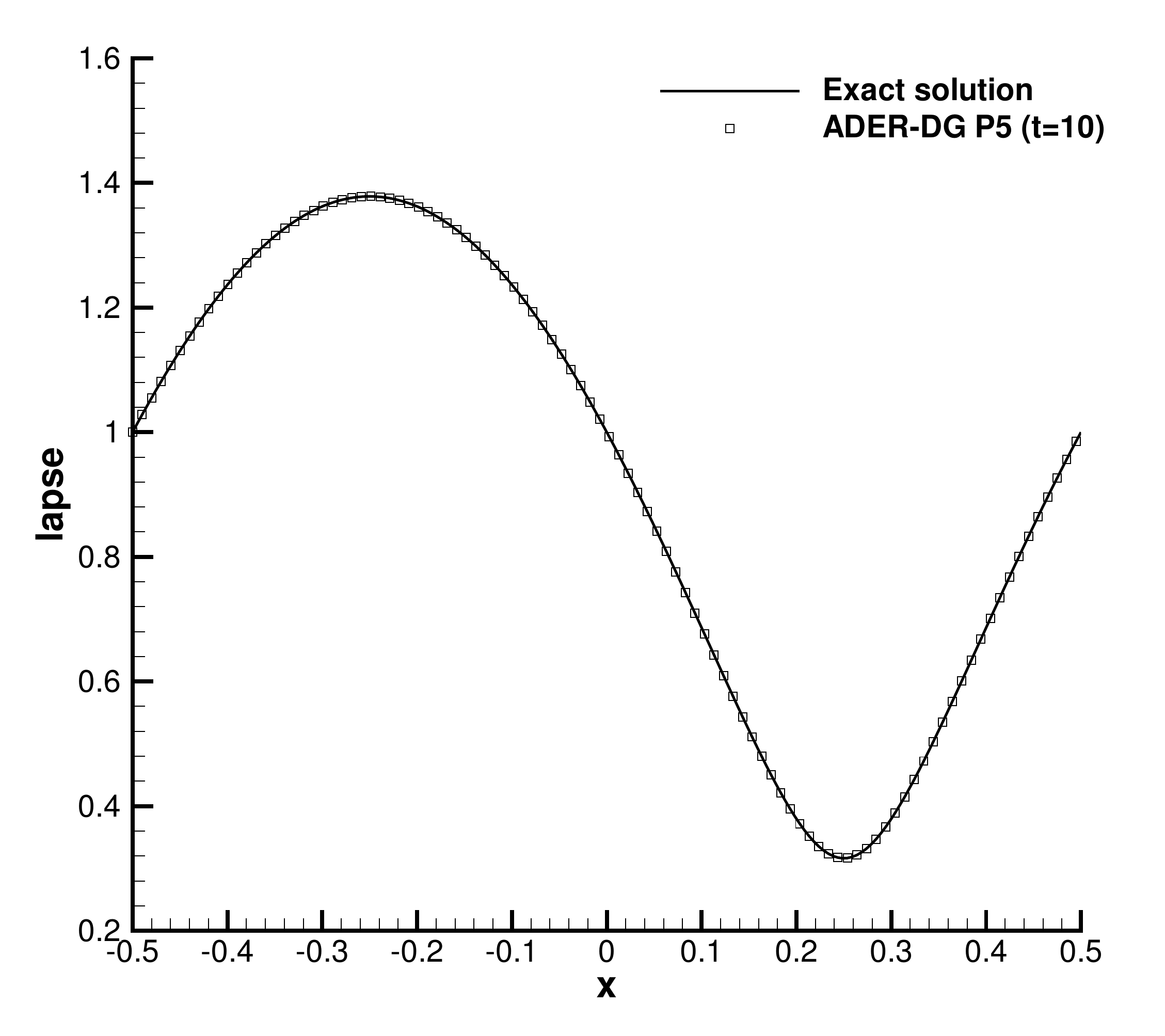} \\
        \includegraphics[width=0.40\textwidth]{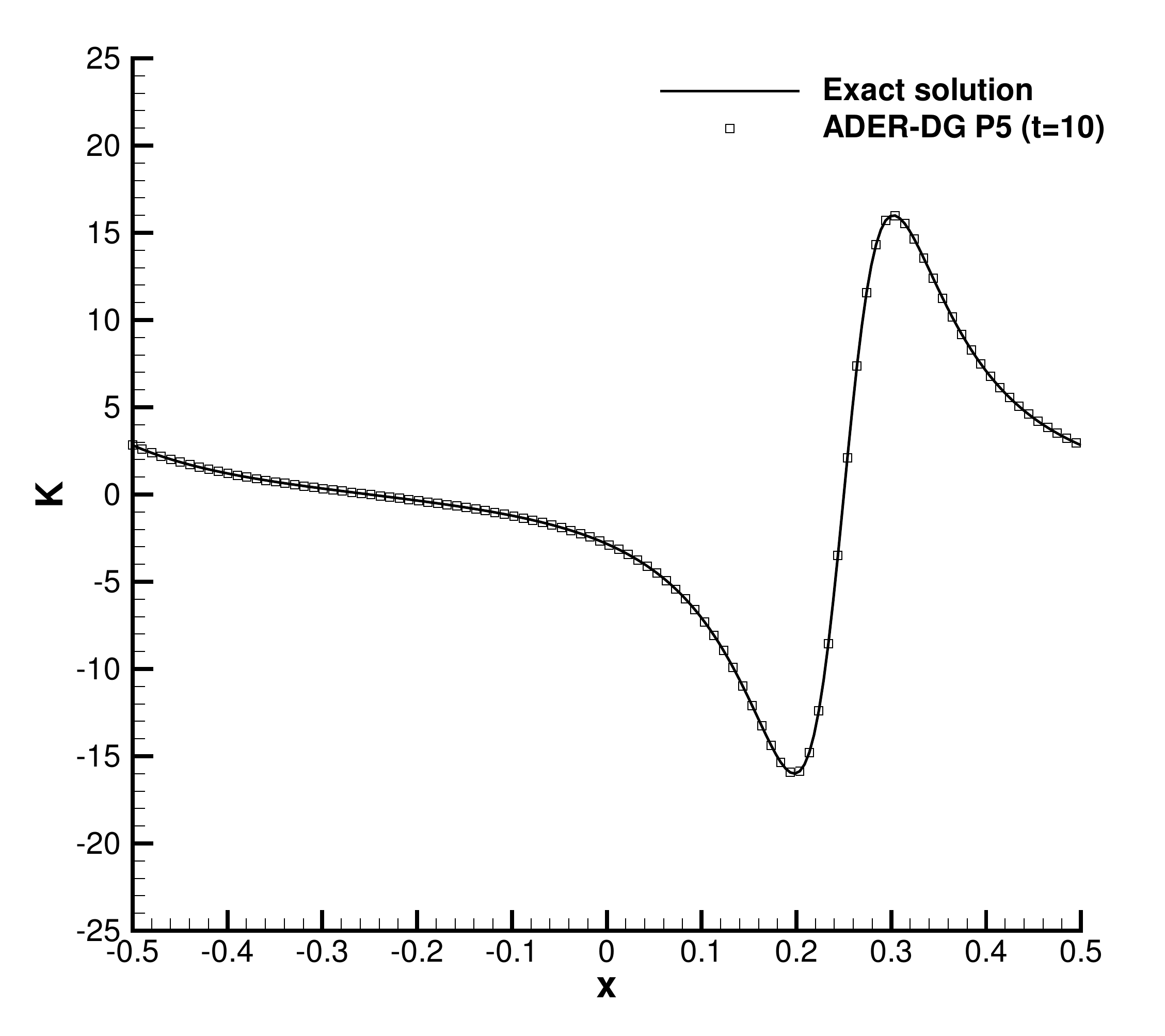}     &
	\includegraphics[width=0.40\textwidth]{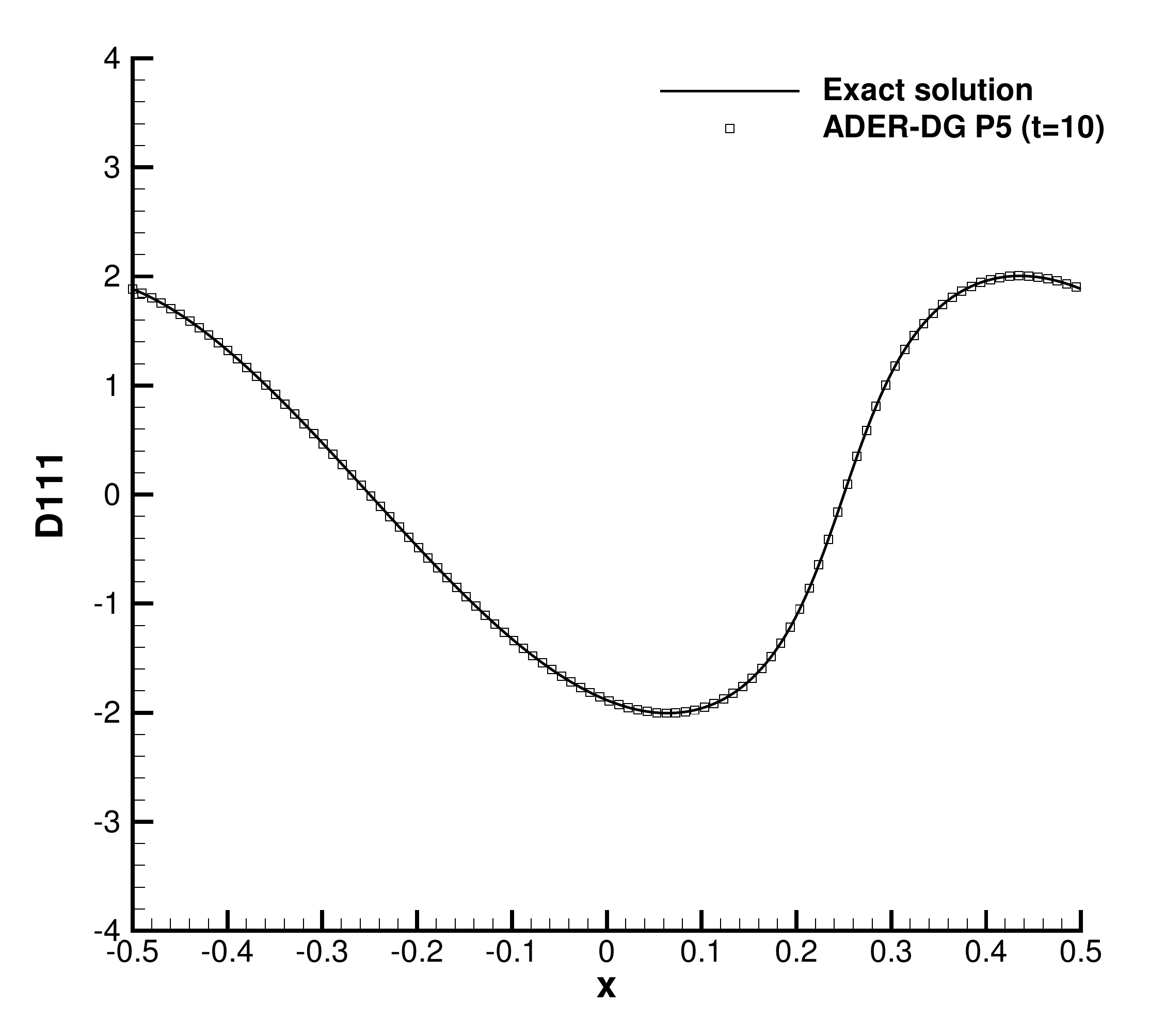}
    \end{tabular}
    \caption{Highly nonlinear gauge-wave test case with very large
      amplitude $A=0.9$. Comparison of the wave form with the exact
      solution at time $t=10$ for an ADER-DG $P_5$ scheme and $100 \times
      10$ elements.}
    \label{fig.gauge.xxl}
\end{center}
\end{figure}

The $L_2$ error norms of the conformal factor $\phi$, the lapse $\alpha$ and
the trace of the extrinsic curvature $K$, together with the observed
order of accuracy of the different ADER-DG schemes are reported in Table
\ref{tab.conv1}. We observe essentially the expected order of accuracy of
the scheme for $N=3$ and $N=4$, while a superconvergence is observed for
$N=5$ and $N=7$. We think that this is due to the strong nonlinearities
of the PDE system appearing in the regime in which we run this test case
with $A=0.9$ and that some leading errors may be dominated by quadratic
terms in the metric and the conformal factor, which can lead to a faster
error decay than $N+1$ for \textit{coarse} meshes. However, we expect
that this superconvergence will disappear on sufficiently refined meshes;
but since the absolute errors are already getting close to machine
accuracy on the meshes used here, it is not possible to refine the mesh
much more with double-precision arithmetics, at least in the $N=7$ case.
For the ADER-DG $P_5$ scheme using $100 \times 10$ elements a comparison
between numerical and exact solution of the nonlinear waveforms for
$\phi$, $\alpha$, $K$ and $D_{xxx}$ is provided in Fig.
\ref{fig.gauge.xxl} at $t=10$, where we can note again an excellent
agreement between exact and numerical solution.

\begin{table}
\caption{Numerical convergence results for the large amplitude gauge wave
    test problem with $A=0.9$ at a final time of $t=10$.  The $L_2$
    errors and corresponding observed convergence order are reported for
    the variables $\phi$, $\alpha$ and $K$.}
\begin{center}
\begin{tabular}{ccccccc}
\hline
  $N_x \times N_y$ & ${L_2}$ error $\phi$ & $\mathcal{O}(\phi)$  & ${L_2}$ error $\alpha$ & $\mathcal{O}(\alpha)$  & ${L_2}$ error $K$ & $\mathcal{O}(K)$  \\
\hline
  \multicolumn{7}{c}{$N=3$}   \\
\hline
$60 \times  6$  & 2.8663E-05 &     & 5.4876E-05 &      & 3.8469E-03 &      \\
$80 \times  8$  & 1.0574E-05 & 3.5 & 2.2314E-05 & 3.1  & 7.0357E-04 & 5.9  \\
$100\times 10$  & 3.8760E-06 & 4.5 & 8.0170E-06 & 4.6  & 2.3112E-04 & 5.0  \\
$120\times 12$  & 1.6311E-06 & 4.7 & 3.2521E-06 & 4.9  & 9.7392E-05 & 4.7  \\
\hline
  \multicolumn{7}{c}{$N=4$}   \\
\hline
$60 \times  6$  & 4.2966E-06 &     & 1.1408E-05 &      & 2.1910E-04 &      \\
$80 \times  8$  & 8.9473E-07 & 5.5 & 2.3725E-06 & 5.5  & 5.0194E-05 & 5.1  \\
$100\times 10$  & 2.5596E-07 & 5.6 & 6.8053E-07 & 5.6  & 1.5781E-05 & 5.2  \\
$120\times 12$  & 9.0039E-08 & 5.7 & 2.4064E-07 & 5.7  & 6.1004E-06 & 5.2  \\
\hline
  \multicolumn{7}{c}{$N=5$}   \\
\hline
$40 \times  4$  & 8.9305E-07 &     & 2.1971E-06 &      & 1.3614E-04 &      \\
$60 \times  6$  & 5.2103E-08 & 7.0 & 1.2756E-07 & 7.0  & 5.9568E-06 & 7.7  \\
$80 \times  8$  & 7.1947E-09 & 6.9 & 1.7348E-08 & 6.9  & 8.4259E-07 & 6.8  \\
$100\times 10$  & 1.5357E-09 & 6.9 & 3.6421E-09 & 7.0  & 1.8093E-07 & 6.9  \\
\hline
  \multicolumn{7}{c}{$N=7$}   \\
\hline
$30 \times 3$  & 1.7693E-08 &      & 3.9004E-08 &       & 6.3103E-06 &       \\
$40 \times 4$  & 1.8387E-09 & 7.9  & 4.1751E-09 &  7.8  & 5.5791E-07 &  8.4  \\
$60 \times 6$  & 6.2824E-11 & 8.3  & 1.4304E-10 &  8.3  & 2.1519E-08 &  8.0  \\
$80 \times 8$  & 5.6521E-12 & 8.4  & 1.3455E-11 &  8.2  & 1.7085E-09 &  8.8  \\
\hline
\end{tabular}
\end{center}
\label{tab.conv1}
\end{table}

\subsection{Robust stability test}

The so-called robust stability test is the last standard test problem
that we take from Ref. \cite{Alcubierre:2003pc}. While in the previous
test problems we have used a simple frozen shift condition $\partial_t
\beta^i = 0$ by setting $s=0$ in the FO-CCZ4 system, here we employ the
classical Gamma-driver shift condition. Furthermore, we employ the
$1+\log$ slicing condition, setting the slicing function to
$g(\alpha)=2/\alpha$ and the parameter $f$ of the Gamma driver to
$f=0.75$, which is also the typical value used for the BSSNOK system and
for the classical second-order CCZ4 system (see \cite{Alic:2011a} for
details). We further set $e=2$, $\kappa_1=\kappa_2=\kappa_3=0$, $K_0=0$,
$c=1$ and $\eta=0$.

As customary in this test, we start from the flat Minkowski metric
\begin{equation}
d s^2 = -d t^2 + d x^2 + d y^2 + d z^2.
\label{eqn.robstab.metric}
\end{equation}
We then add uniformly distributed \textit{random perturbations} to
\textit{all} quantities of the FO-CCZ4 system, \ie to all primary and
auxiliary variables and also to $\Theta$ and $\hat{\Gamma}^i$.  The
two-dimensional computational domain is $\Omega = [-0.5,0.5]^2$ and we
run different simulations with an unlimited ADER-DG $P_3$ scheme on four
successively refined meshes composed of $10 \rho \times 10 \rho$
elements, corresponding to $40 \rho \times 40 \rho$ degrees of freedom,
where $\rho \in \left\{ 1, 2, 4, 8 \right\}$ is the refinement factor.
The perturbation amplitude is $\epsilon = 10^{-7}/\rho^2$, which
corresponds to perturbation amplitudes that are three orders of magnitude
larger that those suggested in Ref. \cite{Alcubierre:2003pc}.

\begin{figure}[!htbp]
\begin{center}
    \begin{tabular}{cc}
        \includegraphics[width=0.4\textwidth]{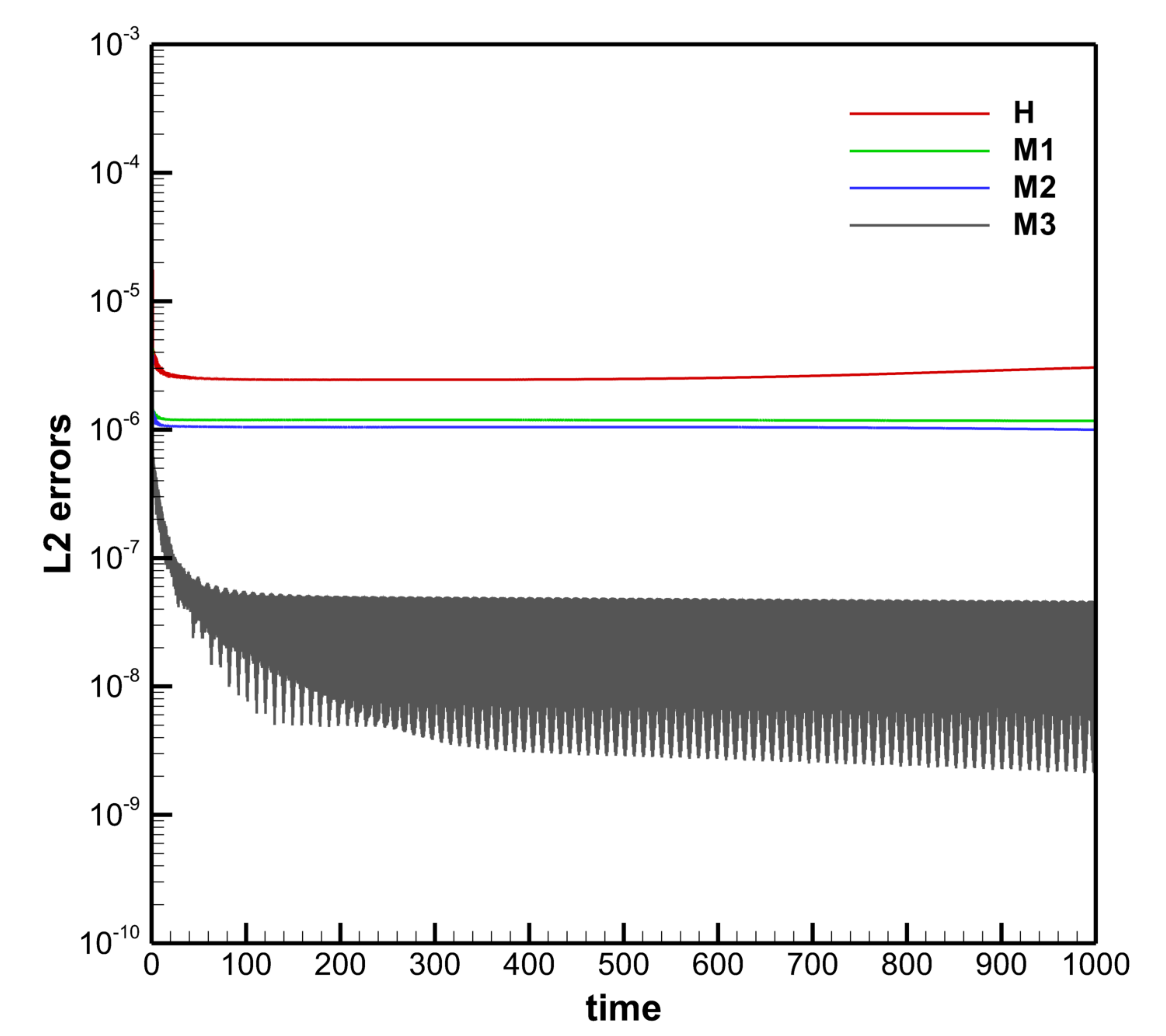} &
	\includegraphics[width=0.4\textwidth]{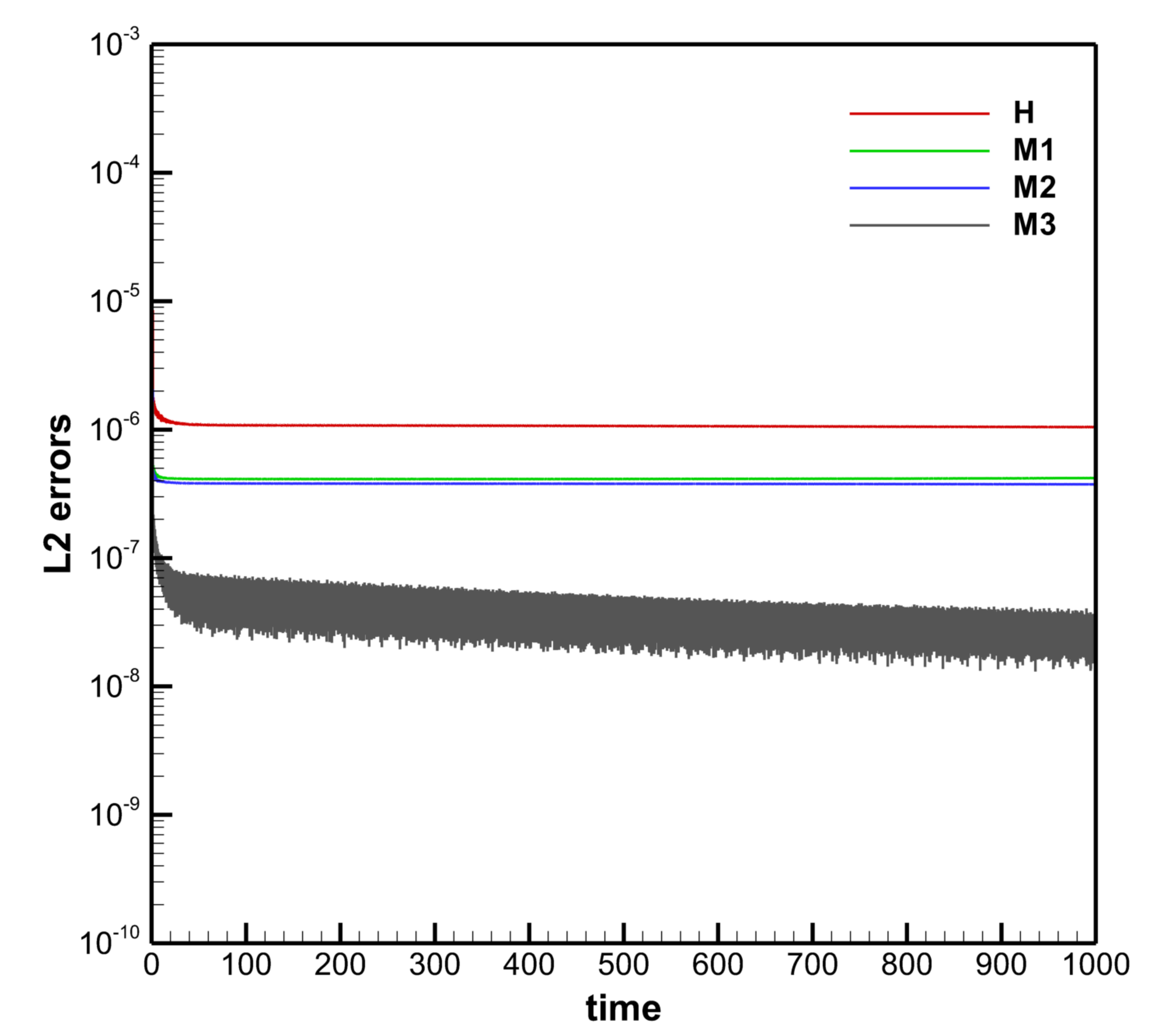} \\
	\includegraphics[width=0.4\textwidth]{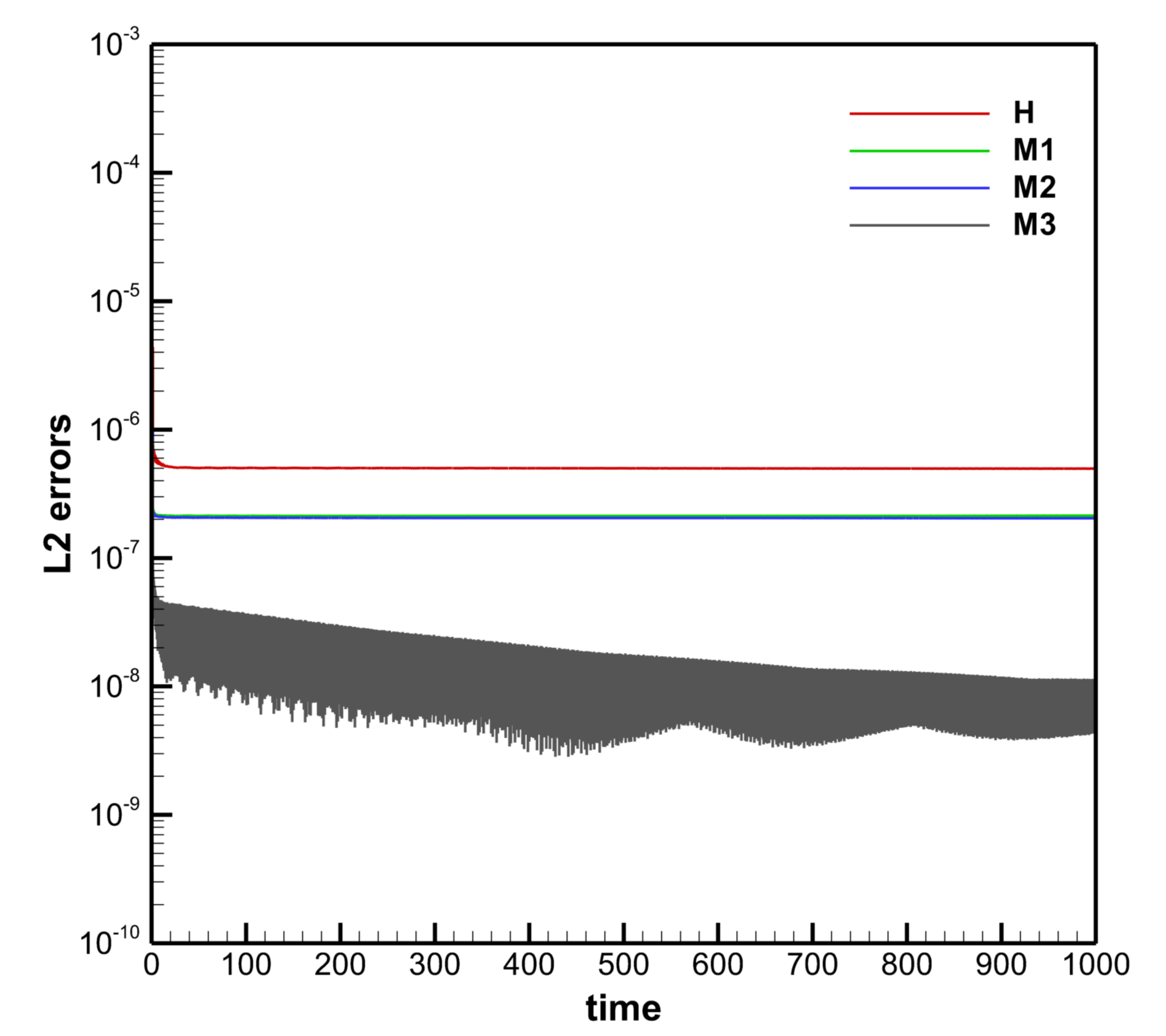} &
	\includegraphics[width=0.4\textwidth]{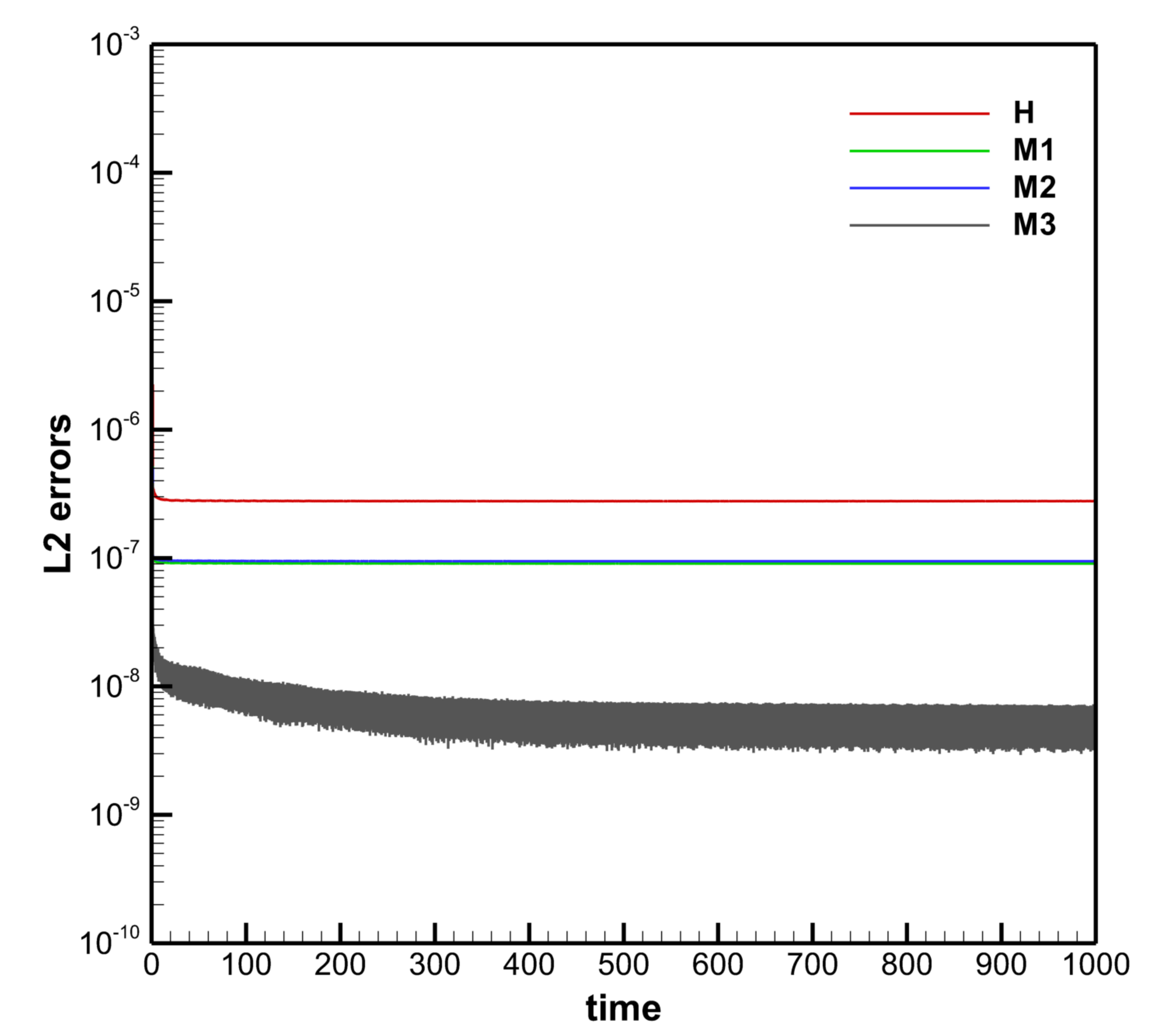}
    \end{tabular}
    \caption{Robust stability test case with Gamma-driver shift condition
      and $1+\log$ slicing with random initial perturbation of amplitude
      $10^{-7}/\rho^2$ in all quantities on a sequence of successively
      refined meshes on the unit square in 2D using an ADER-DG $P_3$
      scheme. Top left: $10\times10$ elements, corresponding to $40\times40$
      degrees of freedom ($\rho=1$). Top right: $20\times20$ elements,
      corresponding to $80\times80$ degrees of freedom ($\rho=2$). Bottom
      left: $40\times40$ elements, corresponding to $160\times160$ degrees of
      freedom ($\rho=4$). Bottom right: $80\times80$ elements, corresponding
      to $320\times320$ degrees of freedom ($\rho=8$). }
    \label{fig.robstab}
\end{center}
\end{figure}

The time evolution of the ADM constraints is reported in Fig.
\ref{fig.robstab} for all four simulations. One can observe that after an
initial decay the constraints remain essentially constant in time for all
different grid resolutions, indicating that our FO-CCZ4 system indeed
passes the robust stability test with the standard Gamma driver and
$1+\log$ gauge conditions (see \cite{Cao:2012} for similar tests with the
Z4c system).

\subsection{Convergence tests on three-dimensional black-hole spacetimes}

In this test we consider the evolution of isolated Schwarzschild and Kerr
black holes in 3D Cartesian Kerr-Schild coordinates, with $M=1$ the mass
of the black hole and $a$ the dimensionless spin. The metric in these
coordinates is known analytically and thus the primary variables of our
evolution system are given by
\begin{equation}
  \alpha = S^{-\halb}\,, \qquad
	\beta^i = \frac{2 H}{S} l_i\,, \qquad
	\gamma_{ij} = \left( \begin{array}{ccc}
	 1 + 2 H l_x^2 & 2 H l_x l_y & 2 H l_x l_z \\
	 2 H l_x l_y & 1 + 2 H l_y^2  & 2 H l_y l_z \\
	 2 H l_x l_z & 2 H l_y l_z & 1 + 2 H l_z^2
	\end{array} \right)\,,
	\label{eqn.kerr.metric}
\end{equation}
with
\begin{equation*}
	 H := M \frac{r^3}{r^4 + a^2 z^2}\,, \qquad
	 S := 1 + 2 H\,, \qquad
	 l_x := \frac{r x + a y}{r^2 + a^2}\,, \qquad
	 l_y := \frac{r y - a x}{r^2 + a^2}\,, \qquad
	 l_z := \frac{z}{r}\,,
\end{equation*}
and
\begin{equation*}
r := \sqrt{ (x^2 + y^2 + z^2 - a^2)/2 + \sqrt{((x^2 + y^2 + z^2 -
    a^2)/2)^2 + z^2 a^2} }\,.
\end{equation*}
We furthermore use the fact that the solution is stationary, \ie
$\partial_t \gamma_{ij}=0$, hence the extrinsic curvature $K_{ij}$ is
computed as follows (see \cite{Rezzolla_book:2013})
\begin{equation}
K_{ij} = \frac{1}{2\alpha} \left( \nabla_i \beta_j + \nabla_j \beta_i
\right)\,.
\end{equation}
The function $K_0$ is chosen as $K_0 = \left( K - \beta^k \partial_k
\alpha \right)/{\left(\alpha^2 g(\alpha)\right)}$, so that $ \partial_t
\alpha = 0$ and in this test the Gamma-driver shift condition is
simplified to $\partial_t \beta^i = f b^i$, $\partial_t B_k^i = f
\partial_k b^i$ and $\partial_t b^i = \partial_t \hat{\Gamma}^i$, with
the consequence that the above exact solution corresponds to a stationary
solution of the FO-CCZ4 system. In other words, we remove the advection
terms from the evolution equations of the shift $\beta^i$ and the
variable $b^i$ (see also \cite{Alcubierre:2008}). The conformal factor
$\phi$ and the auxiliary variables can be computed according to their
definition. The computational domain is chosen as $\Omega = [1,5]^3 \,
M^3$, and the exact solution given by the initial condition is imposed on
all boundaries in all variables at all times. Note that this choice of
boundary conditions is appropriate to study convergence since the exact
solution is also a stationary solution of our PDE system. Note also that
the black hole is centered at $x=y=z=0$, so that we evolve only a section
of the domain offset from the singularity, but encompassing regions both
inside and outside of the event horizon; this effectively amounts to
employing an excision of the black-hole interior.  We furthermore set
$e=2$, $c=1$, $\eta=0$, and consider the undamped CCZ4 system with the
$1+\log$ slicing, \ie we set $\kappa_1 = \kappa_2 = \kappa_3 = 0$,
$f=0.75$ and $g(\alpha)=2 / \alpha$.

The simulations were performed with different ADER-DG schemes on a
sequence of successively refined meshes until a final time of
$t=10\,M$. The Rusanov method is used as approximate Riemann solver at
the element interfaces. In the case of the Schwarzschild black hole we
use $a=0$, while for the Kerr black hole we set $a=0.9$. The
corresponding numerical convergence rates are reported for both cases in
Table \ref{tab.conv.bh}, where we observe that the designed order of
accuracy $N+1$ of our high-order fully-discrete one-step ADER-DG schemes
has been properly reached.

\begin{table}
\caption{Numerical convergence results of FO-CCZ4 with simplified Gamma
    driver for the Schwarzschild black hole (left) and the Kerr black
    hole (right) in 3D Cartesian Kerr-Schild coordinates at a final time
    of $t=10$.  The $L_2$ errors and corresponding observed convergence
    order are reported for the variables $\phi$.}
\begin{center}
\renewcommand{\arraystretch}{1.0}
\begin{tabular}{cccccccccccc}
\hline
    \multicolumn{6}{c}{\textbf{Schwarzschild black hole} ($a=0$)}   & \multicolumn{6}{c}{\textbf{Kerr black hole} ($a=0.9$)}   \\
\hline
  $N_x$ & ${L_2}$ error $\phi$ & $\mathcal{O}(\phi)$  & $N_x$ & ${L_2}$ error $\phi$ & $\mathcal{O}(\phi)$  &
  $N_x$ & ${L_2}$ error $\phi$ & $\mathcal{O}(\phi)$  & $N_x$ & ${L_2}$ error $\phi$ & $\mathcal{O}(\phi)$  \\
\hline
    \multicolumn{3}{c}{$N=3$}  &  \multicolumn{3}{c}{$N=5$}    &
		\multicolumn{3}{c}{$N=3$}  &  \multicolumn{3}{c}{$N=5$}    \\
\hline
10	& 9.9982E-06	&     &   5 & 2.1837E-06 &     & 10	& 1.4270E-05	&     &   5 & 2.6679E-06 &     \\
15	& 1.8439E-06	& 4.2	&  10 & 2.8327E-08 & 6.3 & 15	& 2.8279E-06	& 4.0 &  10 & 6.5136E-08 & 5.4    \\
20	& 5.8521E-07	& 4.0	&  15 & 2.3649E-09 & 6.1 & 20	& 8.9487E-07	& 4.0 &  15 & 6.0944E-09 & 5.8    \\
25	& 2.4322E-07	& 3.9	&  20 & 4.1176E-10 & 6.1 & 25	& 3.6468E-07	& 4.0 &  20 & 1.1087E-09 & 5.9    \\
\hline
\end{tabular}
\end{center}
\label{tab.conv.bh}
\end{table}

\subsection{Evolution of a single puncture black hole}
\label{sec.single.punctures}

We next have applied the FO-CCZ4 formulation to a single puncture black
hole \cite{Brandt97b} with mass $M=1$ and dimensionless spin $a=0$
located at the origin of a three-dimensional computational domain $\Omega
= [-150,150]^3 \, M^3$ with periodic boundary conditions everywhere. The
domain is discretized with an AMR mesh with grid spacing $\Delta x =
\Delta y = \Delta z = 2.5\,M$ within the inner box $\Omega_b = [-15,15]^3
\, M^3$, while $\Delta x = \Delta y = \Delta z = 7.5\,M$ is used in the
outer part of the domain. In the innermost zone $\Omega_l = [-3,3]^3 \,
M^3$ the third-order subcell ADER-WENO finite-volume limiter is activated
throughout the entire simulation. For details on the AMR framework and
the subcell finite-volume limiter we refer the interested reader again to
\cite{Dumbser2014,Zanotti2015c,Zanotti2015b}. We also stress that this
simulation can be run only after activating the finite-volume subcell
limiter, since a robust scheme is needed in order to deal with the
puncture singularity. Without such a limiter, \ie with a pure DG scheme,
the code crashes after a few timesteps since the high-order unlimited DG
scheme is \textit{not} robust enough to deal with the puncture metric. In
our simulation we use an ADER-DG $P_3$ scheme ($N=3$), which leads to
$2N+1 = 7$ finite-volume subcells per DG element, \ie the effective mesh
spacing in terms of points (cell averages) inside the domain $\Omega_l$
is $\Delta x = \Delta y = \Delta z = 0.357\,M$.  Note that we set up the
mesh so that the puncture is located at the boundary of the DG elements;
given the location of the degrees of freedom in the subcell grid (see
Fig. \ref{fig.subcellgrid}), no grid point coincides with the puncture.
We set the CCZ4 parameters to $\kappa_1 = 0.1$, $\kappa_2=0$, $\kappa_3 =
0.5$ and $\eta=0$. The constant $\mu$ accounting for the second-order
ordering constraints in the evolution of $B^i_k$ is set to $\mu=1/5$,
while for this test we use $c=1$, $f=0.75$ and $e=1$ to be as close as 
possible to a standard second-order CCZ4 formulation, where the cleaning 
of the Hamiltonian constraint is done at the speed of light.
The initial metric and lapse are provided by the \texttt{TwoPunctures}
initial data code \cite{Ansorg:2004ds} (part of the Einstein Toolkit
software \cite{loeffler_2011_et}). Explicitly, the lapse is set initially
to
\begin{equation}
    \alpha = \halb \left(\frac{1-\halb \left({M}/{r^*}\right)}{1+\halb
        \left({M}/{r^*}\right)}+1\right)\,,
\end{equation}
where $r^*:=(r^4+10^{-24})^{\frac{1}{4}}$ and $r$ is the coordinate
distance of a grid point from the puncture. The auxiliary quantities
(which are spatial derivatives of the primary quantities) are obtained
via a simple fourth order central finite difference applied to the
primary variables $\alpha$ and $\gamma_{ij}$. Initially the shift and the
extrinsic curvature are set to zero, \ie $\beta^i = 0$ and $K_{ij}=0$.

\begin{figure}[!htbp]
\begin{center}
    \begin{tabular}{ccc}
        \includegraphics[width=0.32\textwidth]{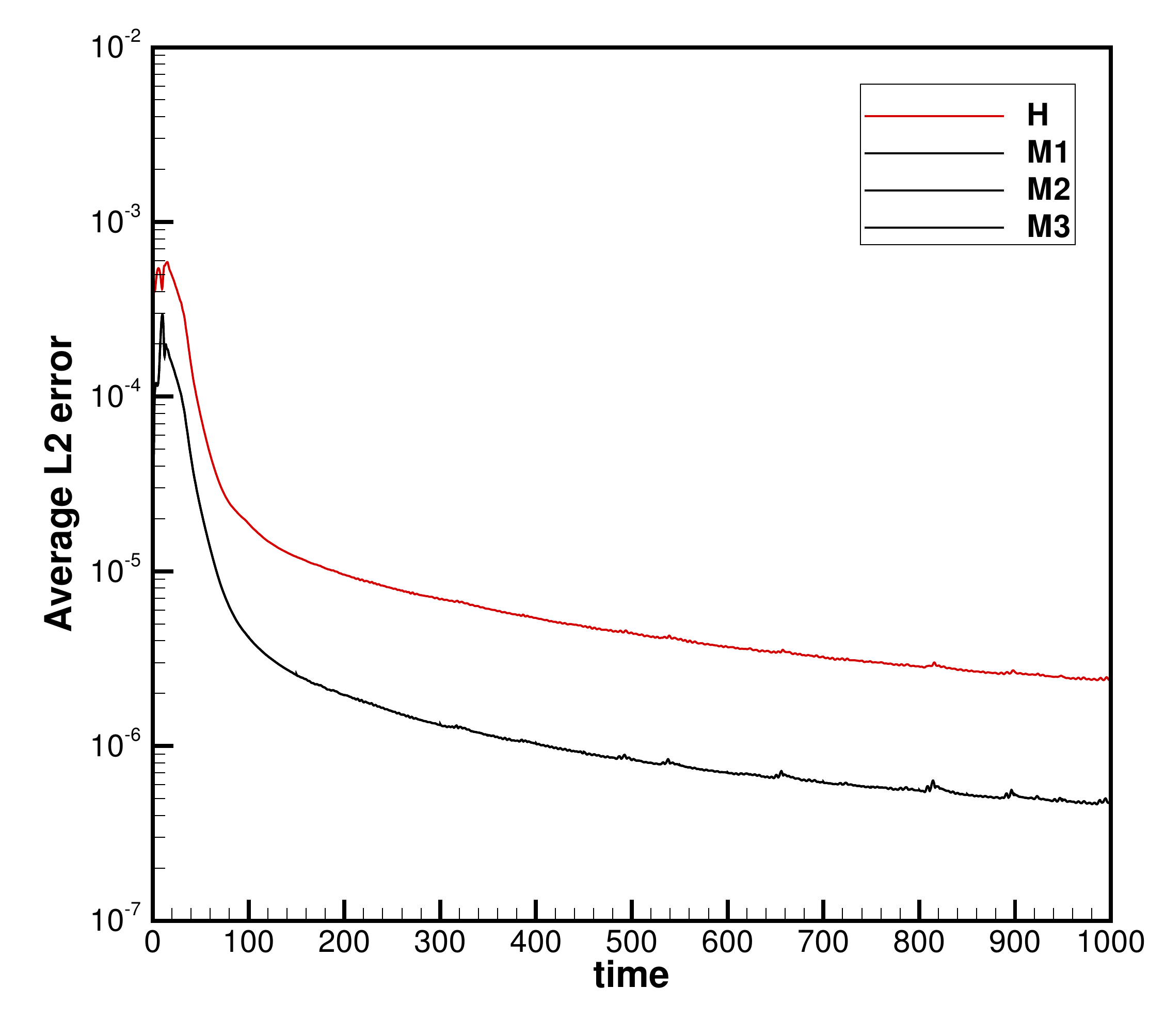} &
        \includegraphics[width=0.32\textwidth]{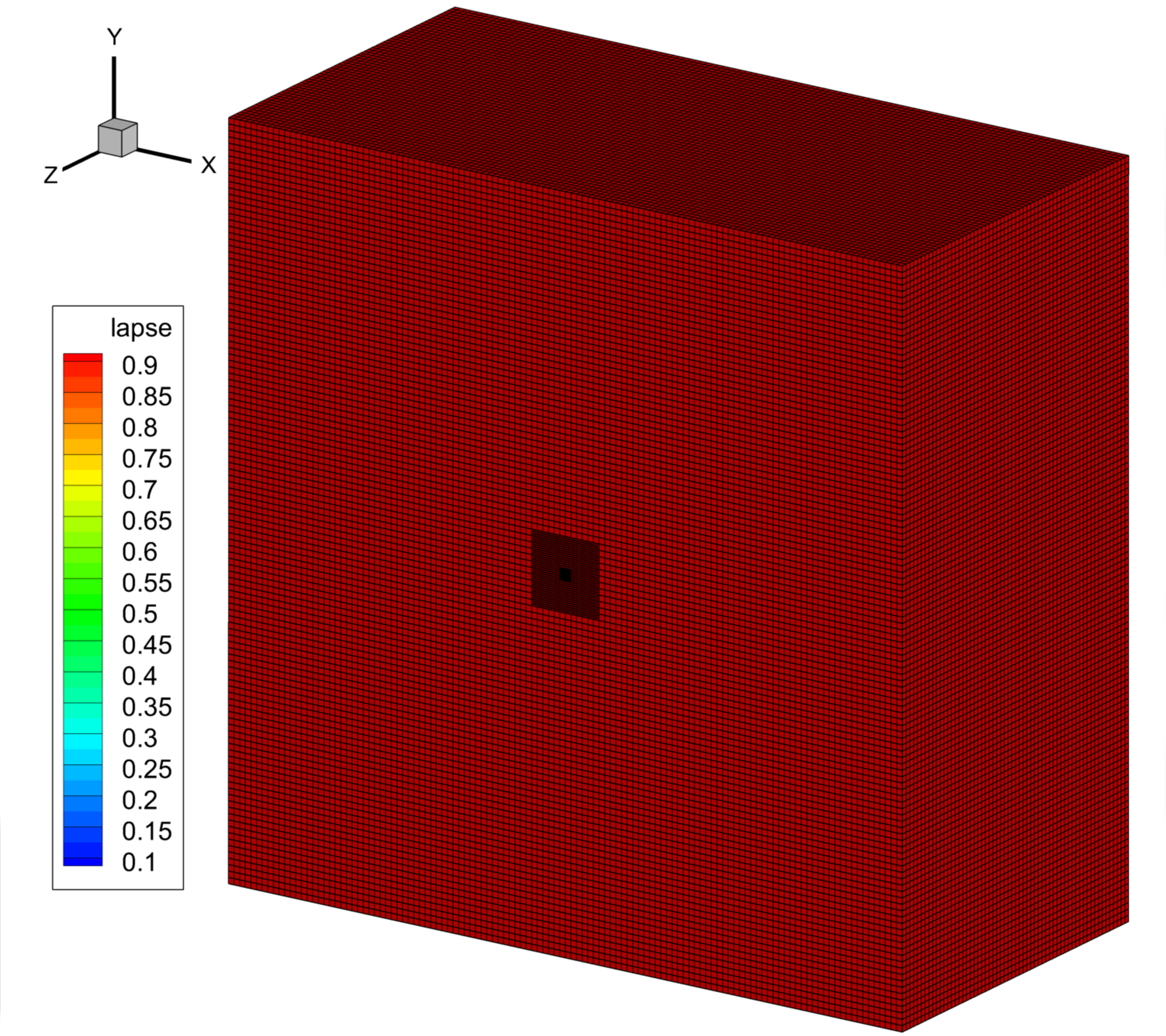} &
        \includegraphics[width=0.32\textwidth]{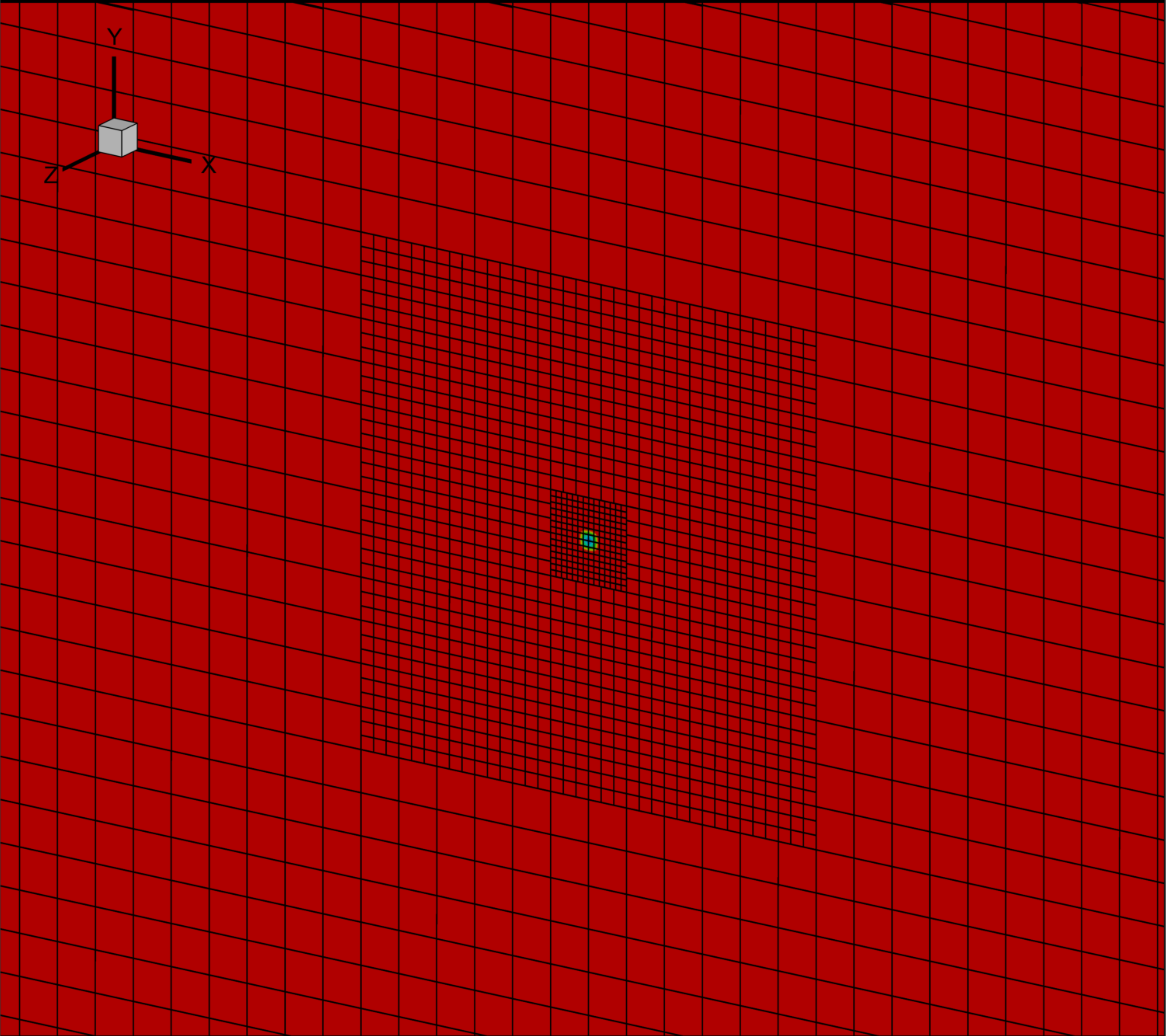}
    \end{tabular}
    \caption{Time evolution of the ADM constraints for the single
    puncture black hole using an ADER-DG $P_3$ scheme with AMR and
    ADER-WENO subcell finite-volume limiter until $t=1000$ (left). Color
    contours for the lapse at $t=200$ and grid setup showing the domain
    $\Omega$, the refined box $\Omega_b$ and the zone with active subcell
    finite-volume limiter $\Omega_l$ (center). Zoom into the center
    region at $t=200$ with color contours for $\alpha$ (right).}
    \label{fig.puncture}
\end{center}
\end{figure}

The evolution was carried out until a final time of $t=1000\,M$ and
Fig. \ref{fig.puncture} reports the evolution of the average $L_2$ error
of the ADM constraints, which we define as
\begin{equation*}
\overline{L}_2 = \sqrt{ \frac{\int_\Omega \epsilon^2
    \d\boldsymbol{x}}{\int_\Omega d\boldsymbol{x} } }\,,
\end{equation*}
where $\epsilon$ denotes the local error of each of the ADM quantities,
\ie Hamiltonian $H$ and momentum constraints $M_i$. In Fig. \ref{fig.puncture}
also a view of the 3D grid setup is shown together with a zoom into the center 
region with the contour colors of the lapse function at a time of $t=200\,M$.  

It is probably worth recalling that, to the best of our knowledge, these
are the first results obtained for a puncture black-hole spacetime using
a fully three-dimensional DG finite-element method with AMR and LTS.
Previous results obtained with high-order DG schemes for black-hole
spacetimes were essentially limited to the one-dimensional case (see, \eg
Refs. \cite{field10, Brown2012, Miller2016}).

\subsection{Preliminary results for moving punctures}\label{sec.movpunct}

The last test considered is a preliminary application of the FO-CCZ4
system to a binary system of two moving punctures. In particular, we
consider a head-on collision of two nonrotating black holes of equal mass
$M=1$ with zero linear momentum initially located at $\boldsymbol{x}^- =
(-1,0,0)$ and $\boldsymbol{x}^+ = (+1,0,0)$. The three-dimensional
computational domain is given by $\Omega = [-25,25]^3 \, M^3$ and flat
Minkowski spacetime is imposed as boundary condition everywhere.  The
CCZ4 parameters are set to $\kappa_1 = 0.1$, $\kappa_2=0$, $\kappa_3 =
0.5$, $\eta=0$ and furthermore we choose $c=1$, $e=1$, $f=1$ and $\mu=1/5$.
Again, the initial metric and the lapse are provided by the
\texttt{TwoPunctures} initial data code \cite{Ansorg:2004ds}, with the lapse
set initially to
\begin{equation}
    \alpha =
    \halb\left(
    \frac{1-\halb\left({m_-}/{r_-^*}\right)-\halb\left({m_+}/{r_+^*}\right)}
    {1+\halb\left({m_-}/{r_-^*}\right)+\halb\left({m_+}/{r_+^*}\right)}
    +1  \right)\,,
\end{equation}
where $r^*_-$ and $r^*_+$ are the coordinate distances of a grid point
from either puncture (defined analogously to the previous section) and
$m_-$ and $m_+$ are the so-called bare masses of the two black holes (see
\cite{Ansorg:2004ds}) and in this case they are equal. The auxiliary
quantities are computed from the primary variables via a fourth-order
central finite-difference method. We use the simple and robust Rusanov
method as approximate Riemann solver on the element boundaries.  The
shift and extrinsic curvature are initially set to $\beta^i = 0$ and
$K_{ij}=0$.

The domain is discretized with an AMR mesh of mesh spacing $\Delta x =
\Delta y = \Delta z = 5/12\,M$ within the inner box $\Omega_b =
       [-2.5,2.5]^3 \, M^3$, while $\Delta x = \Delta y = \Delta z =
       1.25\,M$ is used in the outer part of the domain. In the innermost
       zone $\Omega_l = [-5/3,5/3]^3 \, M^3$ the third-order subcell
       ADER-WENO finite-volume limiter is activated throughout the entire
       simulation. As for a single puncture, we use an ADER-DG $P_3$
       scheme ($N=3$), whose $2N+1 = 7$ finite-volume subcells lead to an
       effective mesh spacing inside the domain $\Omega_l$ of $\Delta x =
       \Delta y = \Delta z = 0.0595$. Once again we remark that the use
       of the finite-volume subcell limiter is essential in order to
       obtain a stable evolution.

\begin{figure}[!htbp]
  \begin{center}
	    \begin{tabular}{ccc}
      \includegraphics[width=0.31\textwidth]{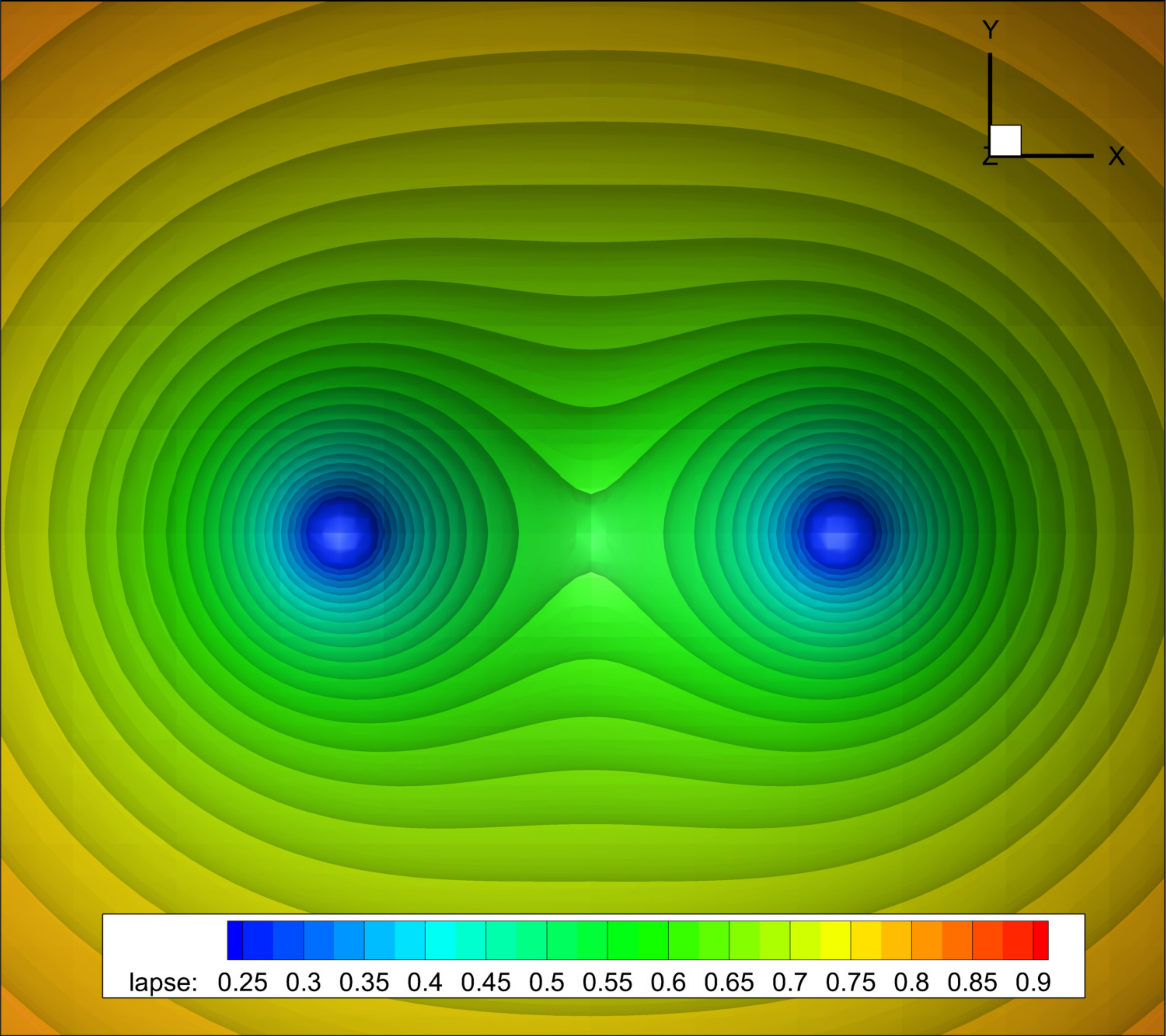} &
      \includegraphics[width=0.31\textwidth]{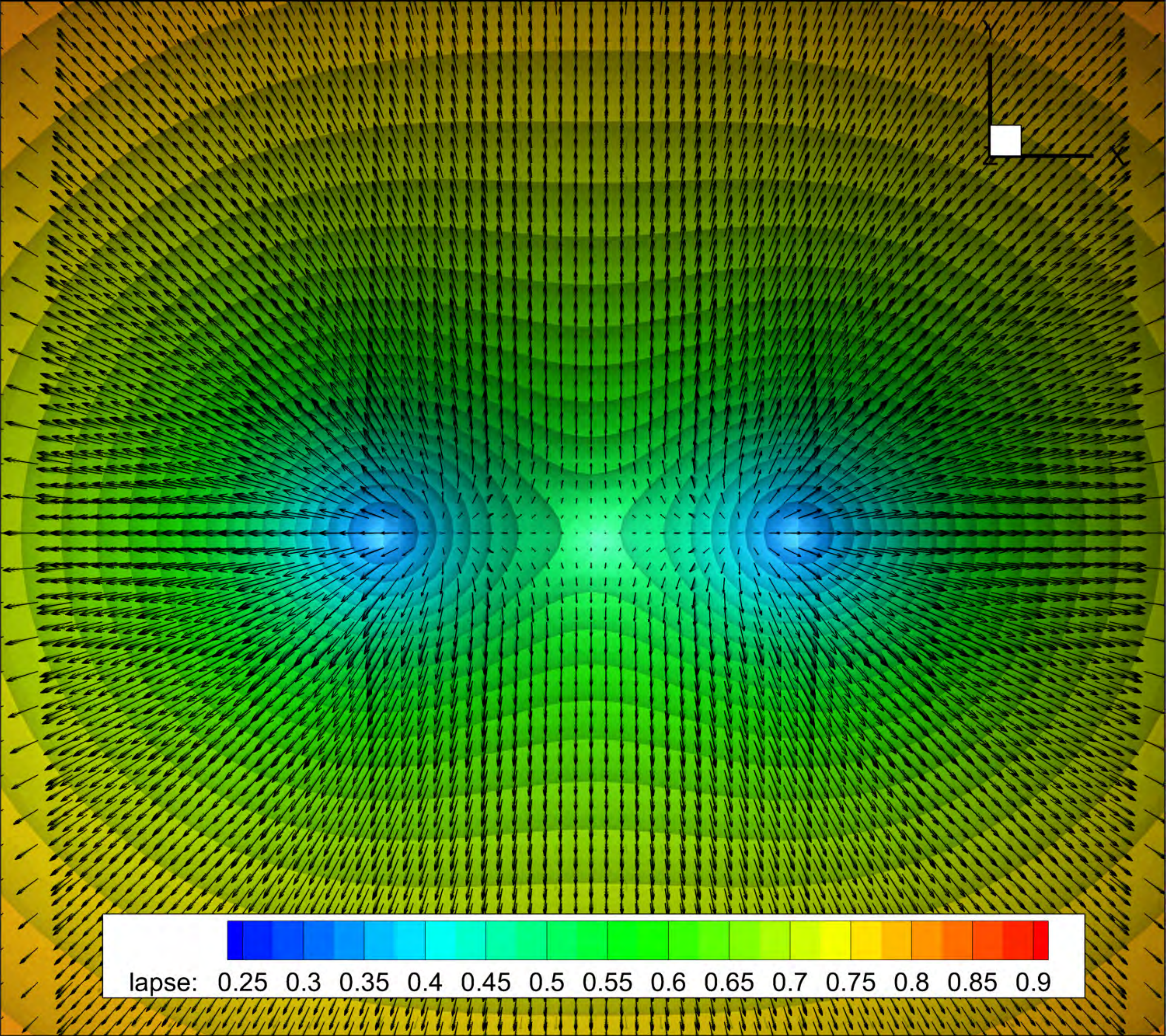} &
      \includegraphics[width=0.31\textwidth]{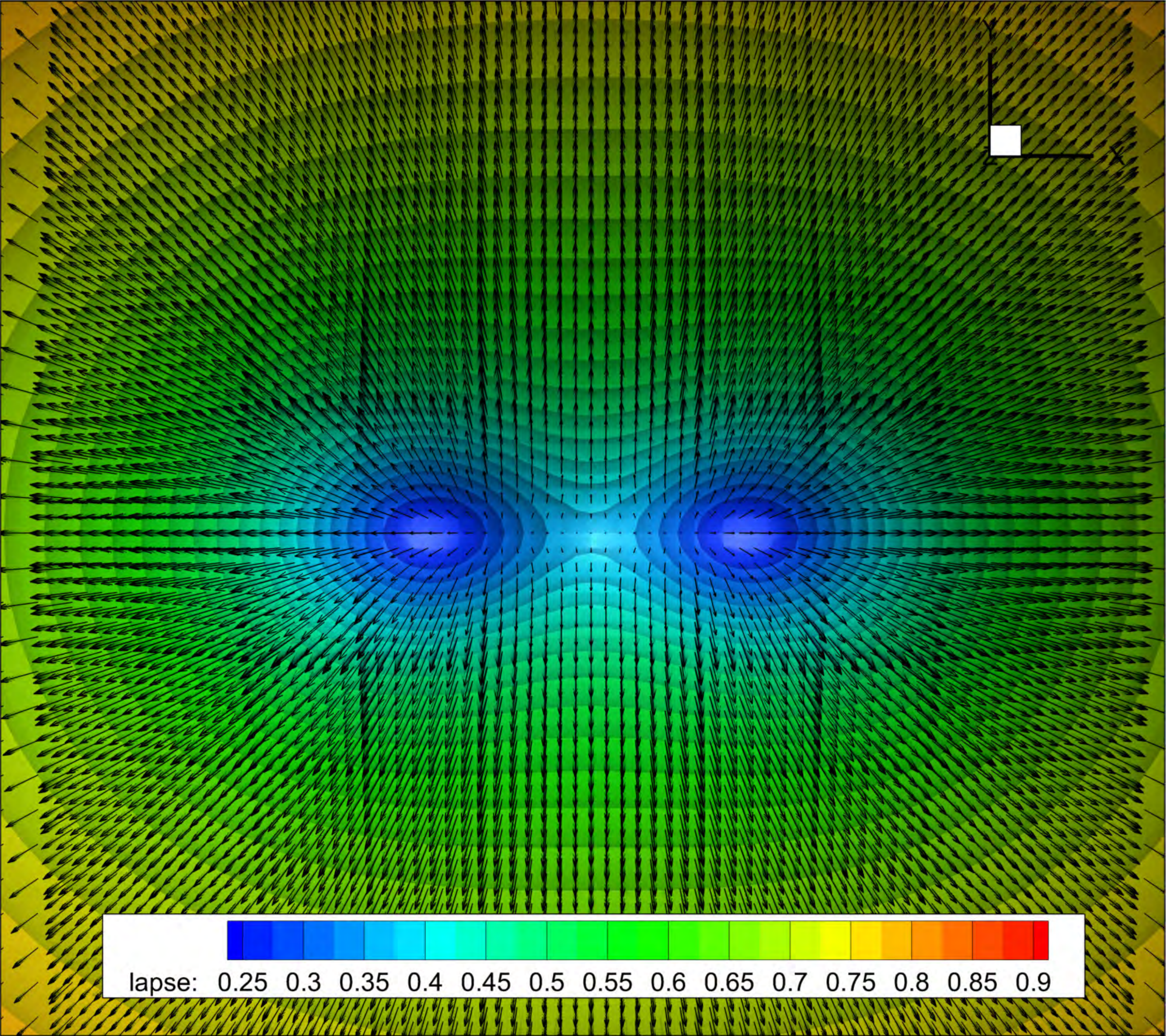} \\
      \includegraphics[width=0.31\textwidth]{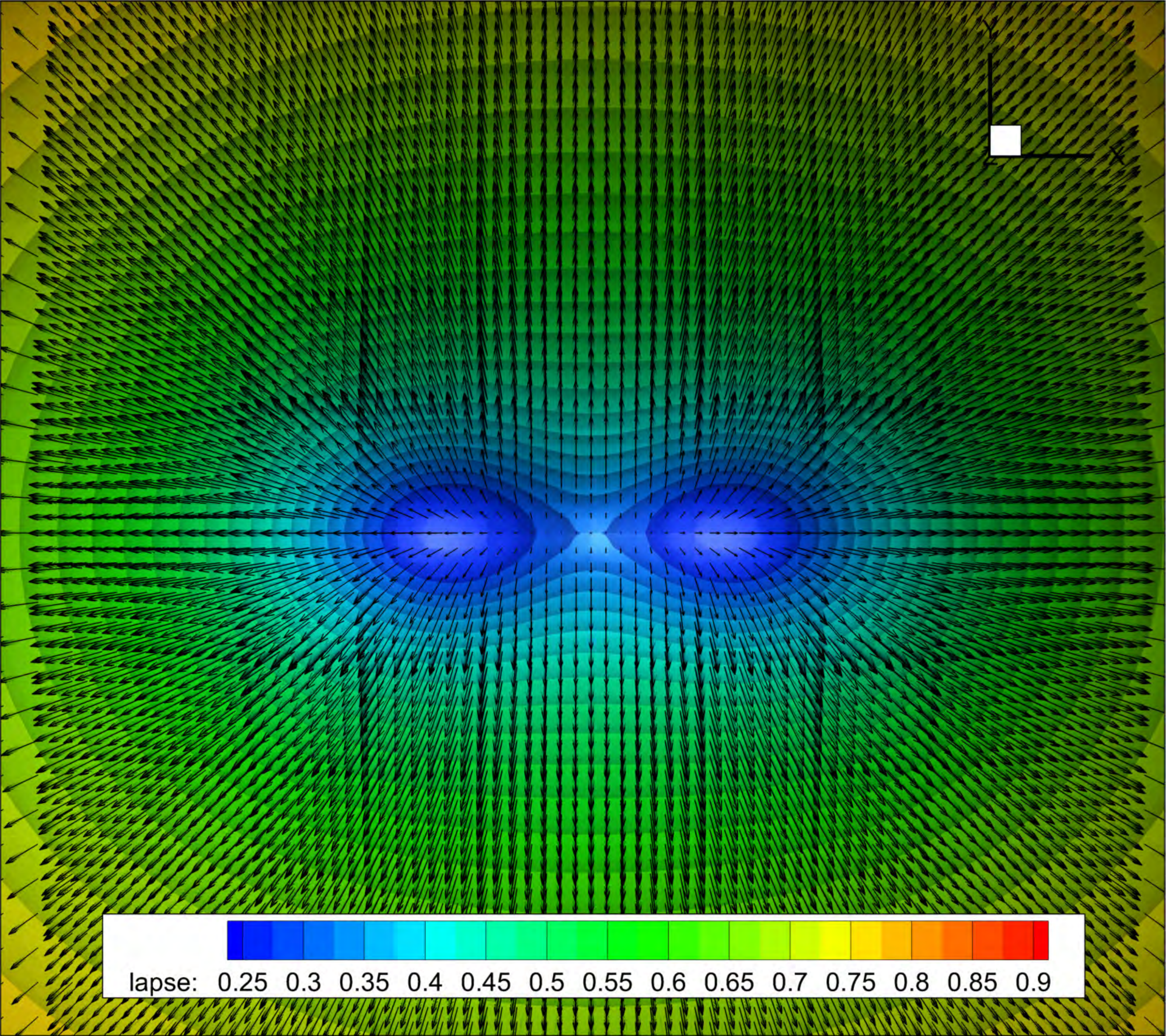} &
      \includegraphics[width=0.31\textwidth]{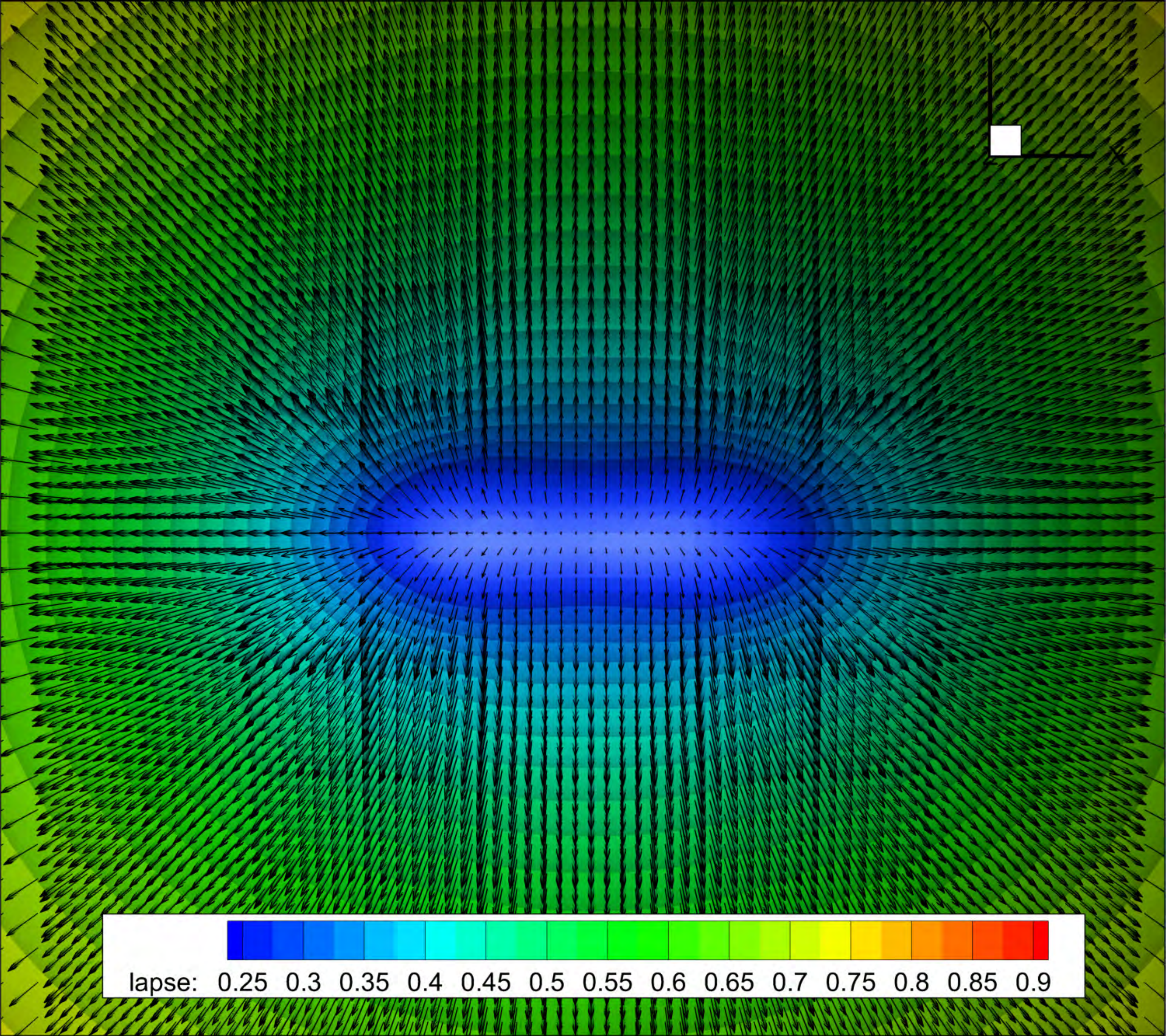} &
      \includegraphics[width=0.31\textwidth]{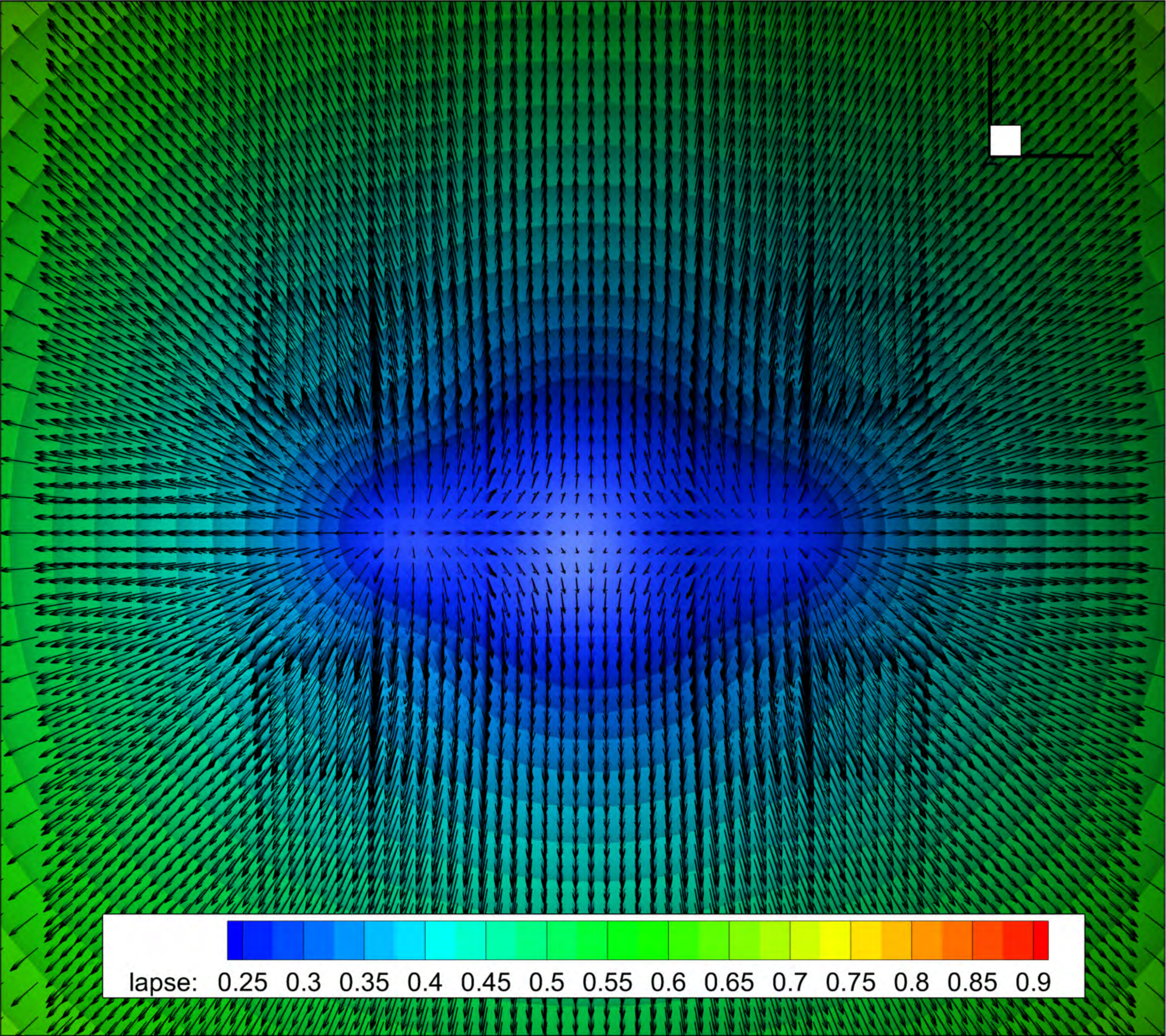} \\
			\end{tabular}
    \caption{Time evolution of the contour surfaces of the lapse $\alpha$
      and the shift vector $\beta^i$ for the head-on collision of two
      puncture black holes of equal mass $M=1$ at times
      $t=0,\,5,\,7,\,8,\,10\,M$ and $t=15\,M$, from top left to bottom
      right.}
    \label{fig.twopunctures}
	\end{center}
\end{figure}

\begin{figure}[!htbp]
  \begin{center}
	    \begin{tabular}{cc}
      \includegraphics[width=0.45\textwidth]{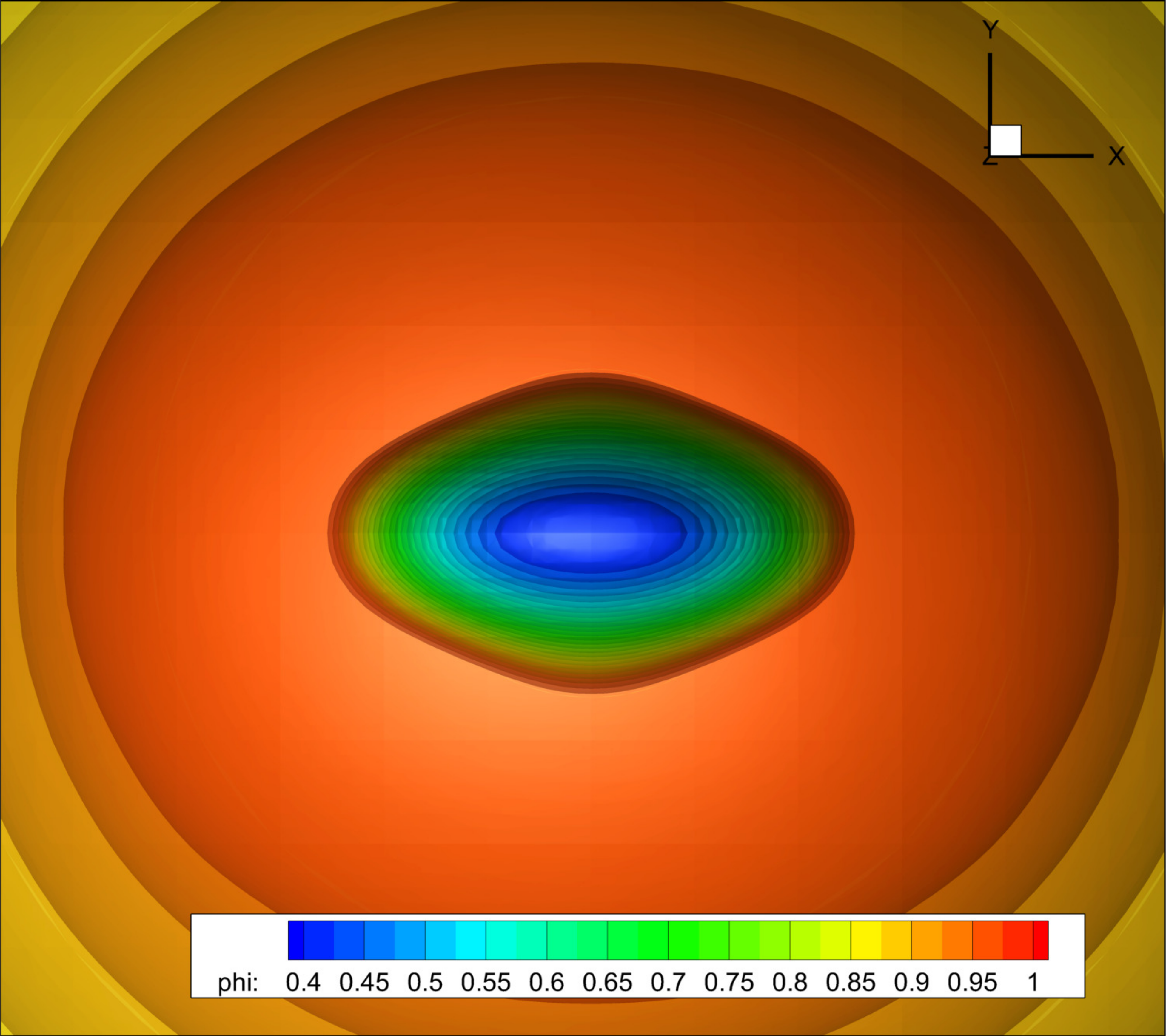} &
      \includegraphics[width=0.45\textwidth]{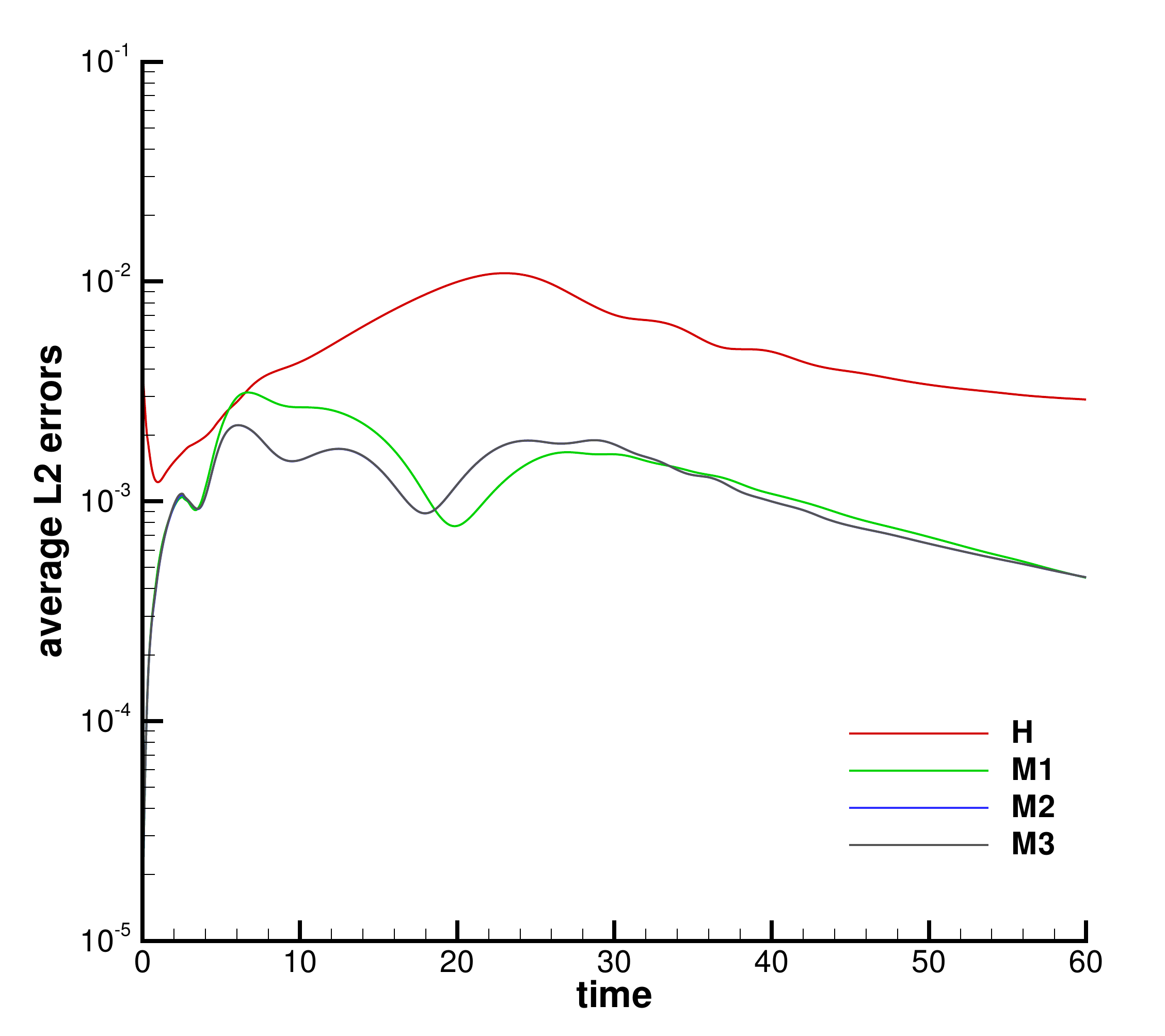}
			\end{tabular}
    \caption{Head-on collision of two puncture black holes: contour
      surfaces of the conformal factor $\phi$ at time $t=34\,M$ after the
      merger (left) and time evolution of the ADM constraints (right).
      The curves for the second and third momentum constraint almost
      coincide. }
    \label{fig.twopunctures.adm}
	\end{center}
\end{figure}

The simulation is run until a final time of $t=60\,M$ and the evolution
of the contour surfaces of the lapse and the shift vector are reported in
Fig. \ref{fig.twopunctures}. The contour surfaces of the conformal factor
at the final time as well as the evolution of the ADM constraints are
depicted in Fig. \ref{fig.twopunctures.adm}. Clearly, no sign of growth
in the violation of the constraints appears after the two punctures have
merged at $t\simeq 10\,M$.

Although these results are meant mostly as a proof-of-concept rather than
as a realistic modelling of the inspiral and merger on binary black-hole
systems, they provide convincing evidence that binary systems of puncture
black holes can be evolved stably with our path-conservative ADER-DG
scheme with ADER-WENO subcell finite-volume limiter on AMR grids based on
the FO-CCZ4 formulation proposed here. A more detailed and systematic
investigation, which includes the emission of gravitational waves from
binary systems of rotating black holes in quasi-circular orbits (see, \eg
\cite{Alic:2011a}), will be the subject of future work.

\section{Conclusions}
\label{sec.conclusions}

We have proposed a possible FO-CCZ4 formulation of the Einstein
equations, that is, a \textit{strongly hyperbolic} first-order
formulation of the conformal covariant Z4 (CCZ4) system of Alic et al.
\cite{Alic:2011a}. The system comprises 58 evolution equations for the
complete state vector given by ${\boldsymbol Q}^T := \left(
\tilde\gamma_{ij}, \ln{\alpha}, \beta^i, \ln{\phi}, \tilde A_{ij}, K,
\Theta, \hat\Gamma^i, b^i, A_k, B^i_k, D_{kij}, P_k \right)$. To the best
of our knowledge, this is the first time that a first-order hyperbolic
formulation of the CCZ4 system has been proposed and has been employed in 
a systematic series of numerical tests of increasing complexity in one, 
two and three spatial dimensions. 

The key idea for obtaining strong hyperbolicity in the new formulation is
the \textit{approximate symmetrization} of the sparsity pattern of the
system matrix, that is, the appropriate use of ordering constraints and
by using the fact that the trace of $\tilde{A}_{ij}$ is zero, in order to
avoid the appearance of Jordan blocks that cannot be diagonalized.
Another important idea employed to obtain the FO-CCZ4 formulation is the
use of first-order ordering constraints in a way that reduces the
evolution equations of the lapse $\alpha$, the shift $\beta^i$, the
conformal metric $\tilde{\gamma}_{ij}$ and the conformal factor $\phi$ to
a pure system of \textit{ordinary} differential equations
\cite{Alcubierre:2008}. In other words, whenever differential terms with
respect to $\alpha$, $\beta^i$, $\phi$ and $\tilde{\gamma}_{ij}$ appear,
they are replaced by the corresponding auxiliary variables $A_k$,
$B^i_k$, $P_k$ and $D_{kij}$ and thus become algebraic source terms. This
leads to a very particular split structure of the system that also
greatly simplifies the analysis of the resulting FO-CCZ4 system, since
the eigenvalues and eigenvectors associated with $\alpha$, $\beta^i$,
$\phi$ and $\tilde{\gamma}_{ij}$ become trivial, being zero and unit,
respectively. This has also the advantage that for the rest of the
analysis a reduced system of \textit{partial} differential equations
relative to only 47 dynamic variables $\boldsymbol{U}^T=(\tilde{A}_{ij}, K,
\Theta, \hat{\Gamma}^i, b^i, A_k, B_k^i, D_{kij}, P_k)$ can be
considered. Furthermore, the matrix of the reduced system in the dynamic
variables is only a function of $\alpha$, $\beta^i$, $\phi$ and
$\tilde{\gamma}_{ij}$, and not of the dynamic variables themselves, which
not only substantially simplifies the hyperbolicity analysis but which
also leads to the important result that \textit{all fields} of our
FO-CCZ4 system are \textit{linearly degenerate}.

When compared to the first-order Z4 system proposed in Refs. 
\cite{Bona97a, Alic:2009}, our entire FO-CCZ4 system is written in a 
fully non-conservative form, which is another key idea of our FO-CCZ4 
formulation. We stress that the previously mentioned simplifications are
\textit{not possible} if a \textit{conservative} formulation of the
system based on the divergence of fluxes is used, \eg like the one
proposed in \cite{Bona97a,Alic:2009}, since the Jacobian $\partial
\boldsymbol{F} / \partial{\boldsymbol{Q}}$ of the flux
$\boldsymbol{F}(\Q)$ will also depend on the dynamical variables and the
quasi-linear form of the system will also contain differential terms in
$\alpha$, $\beta^i$, $\phi$ and $\tilde{\gamma}_{ij}$, while in our
genuinely non-conservative formulation, no differential terms of the
latter quantities appear.

We have also provided a proof of \textit{strong hyperbolicity} of our
FO-CCZ4 system for two standard gauge choices, namely i) harmonic lapse
and zero shift and ii) 1+log slicing combined with the Gamma-driver.  In
both cases we have computed the analytical expressions of all eigenvalues
and all right eigenvectors of the system. For zero shift and harmonic
lapse it was also possible to provide the inverse of the right
eigenvector matrix, i.e. the so-called left eigenvectors of the
system. To the best knowledge of the authors, this is the first time that
the hyperbolicity of a \textit{first-order reduction} of the CCZ4 system
is analyzed, in particular including the Gamma-driver shift condition.

We have numerically solved the FO-CCZ4 system after discretizing it with
the aid of a family of high-order fully-discrete one-step ADER
discontinuous Galerkin (DG) schemes, supplemented with an ADER-WENO
finite-volume limiter in order to deal with the physical singularities
arising with black holes. The non-conservative formulation of the system
is naturally treated within the framework of path-conservative schemes,
first proposed by Castro and Par\'es in the finite-volume context
\cite{Castro2006,Pares2006} and later extended also to ADER-DG schemes in
\cite{Dumbser2009a,Dumbser2010}. Furthermore, in order to ensure
positivity of the numerical solution in terms of $\alpha$ and $\phi$, we
have decided to evolve the logarithms $\ln{\alpha}$ and $\ln{\phi}$ of
these quantities in time, rather than the quantities themselves.

As customary for novel formulations of the Einstein equations, we have
applied the strongly hyperbolic FO-CCZ4 system to a series of standard
test cases suggested in Ref. \cite{Alcubierre:2003pc}, such as the
gauge-wave test, the robust stability test and the linear-wave test bed.
Besides providing evidence that the new system is able to reproduce the
analytic solution accurately, we have carried out numerical convergence
studies of the method on the gauge-wave test in the highly nonlinear
regime of the equations, as well as on the Schwarzschild and a Kerr black
hole using 3D Cartesian Kerr-Schild coordinates. We have also provided
numerical evidence that our ADER-DG scheme with ADER-WENO finite-volume
subcell limiter is able to perform a long time integration of a single
puncture black hole with the usual Gamma driver and $1 + \log$ gauge
conditions that are typically used in simulations carried out with the
BSSNOK evolution system. Finally, we have also shown some first
preliminary results for two moving puncture black holes. To the best of
our knowledge, the numerical results shown in this paper represent the
first simulations of the 3+1 Einstein equations ever done with high-order
DG and WENO finite-volume schemes on three-dimensional adaptive grids.
All previous simulations of black-hole spacetimes with high-order DG
schemes, in fact, were limited to the one-dimensional case only.

Future research will concern the extension of the present algorithms to
dynamic AMR and the extraction of the gravitational waveforms generated
by binary black-hole mergers (see \cite{Centrella:2010,Bishop2016} for
reviews) and binary neutron-star mergers (see \cite{Baiotti2016} for a
review). For the latter case, the present FO-CCZ4 system will be properly 
coupled with the GRMHD equations, using the high order DG 
schemes on space-time adaptive AMR meshes with \textit{a posteriori} 
subcell finite volume limiter proposed in \cite{Zanotti2015b, 
  Zanotti2015c,Fambri2018}. We also plan to extend the unified 
formulation of Newtonian continuum mechanics recently proposed in 
\cite{Peshkov2014,Dumbser2015a,Dumbser2016} to the general relativistic
case and couple it with the FO-CCZ4 system presented in this paper.

\acknowledgments

It is a pleasure to thank D. Alic, C. Bona, and C. Palenzuela for helpful
and inspiring discussions. We also thank the referee very much for his 
very useful, insightful and constructive comments that helped to improve 
the quality of this paper substantially. 
This research was funded by the European Union's Horizon 2020 Research 
and Innovation Programme under the project \textit{ExaHyPE}, grant 
agreement number no. 671698 (call FETHPC-1-2014). 
It was also supported by the ERC synergy grant ``BlackHoleCam: Imaging 
the Event Horizon of Black Holes" (Grant No. 610058), by ``NewCompStar'', 
COST Action MP1304 and by the LOEWE-Program in the Helmholtz International 
Center (HIC) for FAIR. The simulations were performed on the SuperMUC 
cluster at the LRZ in Garching, on the LOEWE cluster in CSC in Frankfurt 
and on the HazelHen cluster at the HLRS in Stuttgart.

%%%%%%%%%%%%%%%%%%%%%%%%%%%%%%%%%%%%%%%%%%%%%%%%%%%%%%%%%%%%%%%%%%%%
%
%   B I B L I O G R A P H Y
%
%%%%%%%%%%%%%%%%%%%%%%%%%%%%%%%%%%%%%%%%%%%%%%%%%%%%%%%%%%%%%%%%%%%%

%

%merlin.mbs apsrev4-1.bst 2010-07-25 4.21a (PWD, AO, DPC) hacked
%Control: key (0)
%Control: author (72) initials jnrlst
%Control: editor formatted (1) identically to author
%Control: production of article title (-1) disabled
%Control: page (0) single
%Control: year (1) truncated
%Control: production of eprint (0) enabled
%

\appendix

\section{The eigenstructure of the FO-CCZ4 system}
\label{sec:eigenappendix}

The ordering of the 58 variables for the complete state vector $\Q$ used
in this paper is explicitly given below as
\begin{eqnarray}
\label{eqn.pde.Q}
{\boldsymbol Q}^T &=& \Big(
\tilde\gamma_{xx}, \tilde\gamma_{xy}, \tilde\gamma_{xz},
\tilde\gamma_{yy}, \tilde\gamma_{yz}, \tilde\gamma_{zz},
\ln{\alpha},
\beta^x, \beta^y, \beta^z,
\ln{\phi},
\tilde A_{xx}, \tilde A_{xy}, \tilde A_{xz},
\tilde A_{yy}, \tilde A_{yz}, \tilde A_{zz},
K, \Theta,
\hat\Gamma^x, \hat\Gamma^y, \hat\Gamma^z,
b^x, b^y, b^z,
\nonumber \\
&&
% 3 A quantities
A_x, A_y, A_x
% 9 B_k^i quantities
B^x_x, B^x_y, B^x_z,
B^y_x, B^y_y, B^y_z,
B^z_x, B^z_y, B^z_z,
% 18 D_kij quantities
D_{xxx}, D_{xxy}, D_{xxz},
D_{xyy}, D_{xyz}, D_{xzz},
D_{yxx}, D_{yxy}, D_{yxz},
\nonumber \\
&&
D_{yyy}, D_{yyz}, D_{yzz},
D_{zxx}, D_{zxy}, D_{zxz},
D_{zyy}, D_{zyz}, D_{zzz},
% 3 P_k quantities
P_x, P_y, P_z
\Big)
\,.
\end{eqnarray}

We emphasize again that for the hyperbolicity analysis of our FO-CCZ4
system it is sufficient to consider the reduced evolution system 
\eqref{eqn.pde.red} of the dynamic variables in the vector 
$\boldsymbol{U}$, while the quantities defining the 4-metric in the
vector $\boldsymbol{V}^T=(\tilde\gamma_{ij}, \ln{\alpha}, \beta^i,
\ln{\phi})$ are only evolved in time via pure ODEs, hence the associated 
eigenvalues are trivially zero and the eigenvectors are the unit vectors. 

\subsection{Zero shift with harmonic lapse} 

For a harmonic lapse ($g(\alpha)=1$) and zero shift ($\beta^i=0$, $s=0$)
and using the standard setting $c=1$ that is necessary for achieving
strong hyperbolicity in first and second order formulations of the Z4
system (with $e=1$), see \cite{Bona:2003qn, Bona:2004yp} for a detailed
analysis, the eigenvalues are given by
\begin{equation}
\lambda_{1,2,\cdots,21} = 0, \quad 
\lambda_{22,23} = \pm \sqrt{\tilde{\gamma}^{11}} \phi \, \alpha \,e, \quad 
\lambda_{24,25,\cdots,29} = +\sqrt{\tilde{\gamma}^{11}} \phi \, \alpha, \quad 
\lambda_{30,31,\cdots,35} = -\sqrt{\tilde{\gamma}^{11}} \phi \, \alpha.
\end{equation}
The associated 35 right eigenvectors of the reduced FO-CCZ4 system in  
the dynamic variables $\boldsymbol{U}^T=(\tilde{A}_{ij},K,\Theta, 
\hat{\Gamma}^i, A_k, D_{kij}, P_k)$, following the ordering chosen in 
equation \eqref{eqn.pde.Q}, read 
\begin{eqnarray}
\boldsymbol{r}_1^T & = &     
\left( \frac{\tilde{\gamma}_{11}}{\tilde{\gamma}_{33}},
       \frac{\tilde{\gamma}_{12}}{\tilde{\gamma}_{33}},
			 \frac{\tilde{\gamma}_{13}}{\tilde{\gamma}_{33}},
			 \frac{\tilde{\gamma}_{22}}{\tilde{\gamma}_{33}},
       \frac{\tilde{\gamma}_{23}}{\tilde{\gamma}_{33}},
		   \frac{\tilde{\gamma}_{33}}{\tilde{\gamma}_{33}},
			0,0,0,0,0,0,0,0,0,0,0,0,0,0,0,0,0,0,0,0,0,0,0,0,0,0,0,0,0 \right) \nonumber \\
\boldsymbol{r}_{2}^T & = &  \left(   
 0,0,0,0,0,0,0,0, \phantom{2} \tilde{\gamma}^{11} \tilde{\gamma}^{11}, \phantom{2} \tilde{\gamma}^{11} \tilde{\gamma}^{12}, \phantom{2} \tilde{\gamma}^{11} \tilde{\gamma}^{13},0,0,0,1,0,0,0,0
,0,0,0,0,0,0,0,0,0,0,0,0,0,0,0,0
\right) \nonumber \\
  \boldsymbol{r}_3^T & = &   
	\left( 0,0,0,0,0,0,0,0, 2 \tilde{\gamma}^{11} \tilde{\gamma}^{12}, 2 \tilde{\gamma}^{11} \tilde{\gamma}^{22}, 2 \tilde{\gamma}^{11} \tilde{\gamma}^{23},
	0,0,0,0,1,0,0,0,0,0,0,0,0,0,0,0,0,0,0,0,0,0,0,0 \right) \nonumber \\
\boldsymbol{r}_4^T & = &  
\left( 0,0,0,0,0,0,0,0,2 \tilde{\gamma}^{11} \tilde{\gamma}^{13}, 2 \tilde{\gamma}^{11} \tilde{\gamma}^{23}, 2 \tilde{\gamma}^{11} \tilde{\gamma}^{33},
0,0,0,0,0,1,0,0,0,0,0,0,0,0,0,0,0,0,0,0,0,0,0,0 \right) \nonumber \\
\boldsymbol{r}_{5}^T & = &   
\left( 0,0,0,0,0,0,0,0,0,\phantom{-} {\frac {{\tilde{\gamma}_{33}}}{ \tilde{\gamma}^{11} }},-{\frac {{\tilde{\gamma}_{23}}}{ \tilde{\gamma}^{11}}}, -\frac{\tilde{\gamma}^{12}}{\tilde{\gamma}^{11}},1,
 0,0,0,0,0,0,0,0,0,0,0,0,0,0,0,0,0,0,0,0,0,0,0 \right) \nonumber \\
% 0,0,0,0,\cdots,0,0,0 \right) \nonumber \\
%
\boldsymbol{r}_{6}^T & = &  \left(
0,0,0,0,0,0,0,0,0,-{\frac {{\tilde{\gamma}_{23}}}{ \tilde{\gamma}^{11} }},\phantom{-} {\frac {{\tilde{\gamma}_{22}}}{
 \tilde{\gamma}^{11} }},
-\frac{\tilde{\gamma}^{13}}{\tilde{\gamma}^{11}},
0,1,0,0,0,0,0,0,0,0,0,0,0,0,0,0,0,0,0,0,0,0,0,0
%0,1,0,0,0,\cdots,0,0,0
\right) \nonumber \\
\boldsymbol{r}_{7}^T & = &  \left(   
0,0,0,0,0,0,0,0,\phantom{2} \tilde{\gamma}^{12} \tilde{\gamma}^{11}, \phantom{2} \tilde{\gamma}^{12} \tilde{\gamma}^{12}, \phantom{2} \tilde{\gamma}^{12} \tilde{\gamma}^{13},
0,0,0,0,0,0,0,0,0,1,0,0,0,0,0,0,0,0,0,0,0,0,0,0
\right) \nonumber \\
\boldsymbol{r}_{8}^T & = &   
\left( 0,0,0,0,0,0,0,0,2 \tilde{\gamma}^{12} \tilde{\gamma}^{12}, 2 \tilde{\gamma}^{12} \tilde{\gamma}^{22}, 2 \tilde{\gamma}^{12} \tilde{\gamma}^{23},
0,0,0,0,0,0,0,0,0,0,1,0,0,0,0,0,0,0,0,0,0,0,0,0 \right) \nonumber \\
\boldsymbol{r}_{9}^T & = &  
\left( 0,0,0,0,0,0,0,0,2 \tilde{\gamma}^{12} \tilde{\gamma}^{13},2 \tilde{\gamma}^{12} \tilde{\gamma}^{23},2 \tilde{\gamma}^{12} \tilde{\gamma}^{33},
0,0,0,0,0,0,0,0,0,0,0,1,0,0,0,0,0,0,0,0,0,0,0,0 \right) \nonumber \\
\boldsymbol{r}_{10}^T & = &    
\left( 0,0,0,0,0,0,0,0,0,0,0,0,0,0,0,0,0,-\frac{\tilde{\gamma}^{12}}{\tilde{\gamma}^{11}},0,0,0,0,0,1,0,0,0
,0,0,0,0,0,0,0,0 \right) \nonumber \\
\boldsymbol{r}_{11}^T & = &  \left(   
0,0,0,0,0,0,0,0,0,0,0,0,0,0,0,0,0,0,-\frac{\tilde{\gamma}^{12}}{\tilde{\gamma}^{11}},0,0,0,0,0,1,0,0,0,0,0,0,0,0,0,0 \right) \nonumber \\
\boldsymbol{r}_{12}^T & = &  \left(   
0,0,0,0,0,0,0,0,0,0,0,0,0,0,0,0,0,0,0,-\frac{\tilde{\gamma}^{12}}{\tilde{\gamma}^{11}},0,0,0,0,0,1,0,0,0,0,0,0,0,0,0 \right) \nonumber \\
\boldsymbol{r}_{13}^T & = &     
\left( 0,0,0,0,0,0,0,0,\phantom{2} \tilde{\gamma}^{13} \tilde{\gamma}^{11}, \phantom{2} \tilde{\gamma}^{13} \tilde{\gamma}^{12}, \phantom{2} \tilde{\gamma}^{13} \tilde{\gamma}^{13},
0,0,0,0,0,0,0,0,0,0,0,0,0,0,0,1,0,0,0,0,0,0,0,0 \right) \nonumber \\
\boldsymbol{r}_{14}^T & = &  
\left( 0,0,0,0,0,0,0,0,2 \tilde{\gamma}^{13} \tilde{\gamma}^{12}, 2 \tilde{\gamma}^{13} \tilde{\gamma}^{22}, 2 \tilde{\gamma}^{13} \tilde{\gamma}^{23},
0,0,0,0,0,0,0,0,0,0,0,0,0,0,0,0
,1,0,0,0,0,0,0,0 \right) \nonumber \\
	\boldsymbol{r}_{15}^T & = &  
	\left( 0,0,0,0,0,0,0,0, 2 \tilde{\gamma}^{13} \tilde{\gamma}^{13}, 2 \tilde{\gamma}^{13} \tilde{\gamma}^{23}, 2 \tilde{\gamma}^{13} \tilde{\gamma}^{33},
	0,0,0,0,0,0,0,0,0,0,0,0,0,0,0,0,0,1,0,0,0,0,0,0 \right) \nonumber \\
\boldsymbol{r}_{16}^T & = &   
\left( 0,0,0,0,0,0,0,0,0,0,0,0,0,0,0,0,0,-\frac{\tilde{\gamma}^{13}}{\tilde{\gamma}^{11}},0,0,0,0,0,0,0,0
,0,0,0,1,0,0,0,0,0 \right) \nonumber \\
\boldsymbol{r}_{17}^T & = &   
\left( 0,0,0,0,0,0,0,0,0,0,0,0,0,0,0,0,0,0,-\frac{\tilde{\gamma}^{13}}{\tilde{\gamma}^{11}},0,0,0,0,0,0,0,0
,0,0,0,1,0,0,0,0 \right) \nonumber \\
\boldsymbol{r}_{18}^T & = &  \left(   
0,0,0,0,0,0,0,0,0,0,0,0,0,0,0,0,0,0,0,-\frac{\tilde{\gamma}^{13}}{\tilde{\gamma}^{11}},0,0,0,0,0,0,0,0,0,0,0,1,0,0,0
\right) \nonumber \\
\boldsymbol{r}_{19}^T & = &  \left(    
0,0,0,0,0,0,0,0,
 -\tilde{\gamma}^{11}(2\tilde{\gamma}_{11}\tilde{\gamma}^{11} +3\tilde{\gamma}_{12} \tilde{\gamma}^{12}+3\tilde{\gamma}_{13}\tilde{\gamma}^{13}),
 -\tilde{\gamma}^{12}(\tilde{\gamma}_{12} \tilde{\gamma}^{12} + \tilde{\gamma}_{22} \tilde{\gamma}^{22}+\tilde{\gamma}_{23} \tilde{\gamma}^{23} )+\tilde{\gamma}_{11} \tilde{\gamma}^{12}\tilde{\gamma}^{11},
\right. \nonumber \\
 && \left.
-\tilde{\gamma}^{13}(\tilde{\gamma}_{13} \tilde{\gamma}^{13} + \tilde{\gamma}_{23} \tilde{\gamma}^{23}+\tilde{\gamma}_{33} \tilde{\gamma}^{33} )+\tilde{\gamma}_{11} \tilde{\gamma}^{13}\tilde{\gamma}^{11},
0,0,0,0,0,0,\tilde{\gamma}_{22},\tilde{\gamma}_{23},\tilde{\gamma}_{33},0,0,0,0,0,0,0,0,0,0,0,0,1,0,0
\right) \nonumber \\
\boldsymbol{r}_{20}^T & = &   
\left(
0,0,0,0,0,0,0,0,
 -\tilde{\gamma}^{12}(2\tilde{\gamma}_{11}\tilde{\gamma}^{11} +3\tilde{\gamma}_{12} \tilde{\gamma}^{12}+3\tilde{\gamma}_{13}\tilde{\gamma}^{13}),
 -\tilde{\gamma}^{22}(\tilde{\gamma}_{12} \tilde{\gamma}^{12} + \tilde{\gamma}_{22} \tilde{\gamma}^{22}+\tilde{\gamma}_{23} \tilde{\gamma}^{23} )+\tilde{\gamma}_{11} \tilde{\gamma}^{12}\tilde{\gamma}^{12},
\right. \nonumber \\
 && \left.
-\tilde{\gamma}^{23}(\tilde{\gamma}_{13} \tilde{\gamma}^{13} + \tilde{\gamma}_{23} \tilde{\gamma}^{23}+\tilde{\gamma}_{33} \tilde{\gamma}^{33} )+\tilde{\gamma}_{11} \tilde{\gamma}^{13}\tilde{\gamma}^{12},
0,0,0,0,0,0,
  \tilde{\gamma}_{22} \frac{\tilde{\gamma}^{12}}{\tilde{\gamma}^{11}},
  \tilde{\gamma}_{23} \frac{\tilde{\gamma}^{12}}{\tilde{\gamma}^{11}},
  \tilde{\gamma}_{33} \frac{\tilde{\gamma}^{12}}{\tilde{\gamma}^{11}},
0,0,0,\cdots,0,0,0,1,0
\right) \nonumber \\
\boldsymbol{r}_{21}^T & = &    
\left( 0,0,0,0,0,0,0,0,
 -\tilde{\gamma}^{13}(2\tilde{\gamma}_{11}\tilde{\gamma}^{11} +3\tilde{\gamma}_{12} \tilde{\gamma}^{12}+3\tilde{\gamma}_{13}\tilde{\gamma}^{13}),
 -\tilde{\gamma}^{23}(\tilde{\gamma}_{12} \tilde{\gamma}^{12} + \tilde{\gamma}_{22} \tilde{\gamma}^{22}+\tilde{\gamma}_{23} \tilde{\gamma}^{23} )+\tilde{\gamma}_{11} \tilde{\gamma}^{12}\tilde{\gamma}^{13},
\right. \nonumber \\
 && \left.
-\tilde{\gamma}^{33}(\tilde{\gamma}_{13} \tilde{\gamma}^{13} + \tilde{\gamma}_{23} \tilde{\gamma}^{23}+\tilde{\gamma}_{33} \tilde{\gamma}^{33} )+\tilde{\gamma}_{11} \tilde{\gamma}^{13}\tilde{\gamma}^{13},
0,0,0,0,0,0,
  \tilde{\gamma}_{22} \frac{\tilde{\gamma}^{13}}{\tilde{\gamma}^{11}},
  \tilde{\gamma}_{23} \frac{\tilde{\gamma}^{13}}{\tilde{\gamma}^{11}},
  \tilde{\gamma}_{33} \frac{\tilde{\gamma}^{13}}{\tilde{\gamma}^{11}},
0,0,0,\cdots,0,0,0,1 \right) \nonumber \\
\boldsymbol{r}_{22,23}^T & = &    % ---22  
\left( 0,0,0,0,0,0,0,\mp \frac{1}{2} \sqrt{\tilde{\gamma}^{11}} e \phi, \tilde{\gamma}^{11} , \tilde{\gamma}^{12},
\tilde{\gamma}^{13},1,0,0,0,0,0,0,0,0,0,0,0,0,0,0,0,0,0,0,0,0,0,0,0 \right) \nonumber \\
\boldsymbol{r}_{24,30}^T & = &     
\left(
0,0,0,0,0,0,\mp 3 \phi \sqrt{\tilde{\gamma}^{11}},0,-4 \tilde{\gamma}^{11},-4 \tilde{\gamma}^{12},-4 \tilde{\gamma}^{13},-3,0,0,0,0,0,0,0,0,0,0,0,0,0,0,0,0,0,0,0,0,1,0,0
\right)  \nonumber \\
\boldsymbol{r}_{25,31}^T & = &    
\left( \mp
2 \phi \tilde{\gamma}^{12} / \sqrt{\tilde{\gamma}^{11}},
\pm \phi \sqrt{\tilde{\gamma}^{11}},0,0,0,0,0,0,0,0,0,0,0,0, -2 \frac{\tilde{\gamma}^{12}}{\tilde{\gamma}^{11}} ,1,0,0,0,0,0,0,0,0,0,0,0,0,0,0,0
,0,0,0,0
\right)  \nonumber \\
\boldsymbol{r}_{26,32}^T & = &     
\left(
\mp 2 \phi \tilde{\gamma}^{13} / \sqrt{\tilde{\gamma}^{11}},0,\pm \phi \sqrt{\tilde{\gamma}^{11}},0,0,0,0,0,0,0,0,0,0,0,-2 \frac{\tilde{\gamma}^{13}}{\tilde{\gamma}^{11}},0,1,0,0,0,0
,0,0,0,0,0,0,0,0,0,0,0,0,0,0
\right)    \nonumber \\
\boldsymbol{r}_{27,33}^T &=&      
\left(
\phantom{2} \mp \phi \tilde{\gamma}^{22}/\sqrt{\tilde{\gamma}^{11}},0,0,
\pm \phi \sqrt{\tilde{\gamma}^{11}},0,0,0,0
,0,0,0,0,0,0, -\frac{\tilde{\gamma}^{22}}{\tilde{\gamma}^{11}},0,0,1,0,0,0,0,0,0,0,0,0,0,0,0,0,0,0,0,0
\right) \nonumber \\
\boldsymbol{r}_{28,34}^T &=&      
\left( \mp
2 \phi \tilde{\gamma}^{23}/\sqrt{\tilde{\gamma}^{11}},0,0,0, \pm \phi \sqrt{\tilde{\gamma}^{11}},0,0
,0,0,0,0,0,0,0,-2\frac{\tilde{\gamma}^{23}}{\tilde{\gamma}^{11}},0,0,0,1,0,0,0,0,0,0,0,0,0,0,0,0
,0,0,0,0 \right) \nonumber \\
\boldsymbol{r}_{29,35}^T & = &    
\left( \phantom{2}
\mp \phi \tilde{\gamma}^{33} / \sqrt{\tilde{\gamma}^{11}},0,0,0,0, \pm \phi \sqrt{\tilde{\gamma}^{11}},0,0,0,0
,0,0,0,0,-\frac{\tilde{\gamma}^{33}}{\tilde{\gamma}^{11}},0,0,0,0,1,0,0,0,0,0,0,0,0,0,0,0,0,0,0,0
\right) \nonumber \\
\label{eqn.revc1}
\end{eqnarray}
The associated 35 left eigenvectors, which define the inverse of the
right eigenvector matrix ($\boldsymbol{L} = \boldsymbol{R}^{-1}$), read


For completeness, we also report the eigenvectors for the non-standard
case $c=0$, which in the original Z4 and CCZ4 framework is \textit{not}
strongly hyperbolic, see the analysis provided in \cite{Bona:2003qn,
  Bona:2004yp}.  However, in the case $c=0$ the FO-CCZ4 system can be
made strongly hyperbolic by choosing a faster cleaning speed $e > 1$.
For $c=0$ all eigenvectors are the same as in \eqref{eqn.revc1}, apart
from the pair $\boldsymbol{r}_{22,23}$, which in this case reads 
\begin{eqnarray} 
\boldsymbol{r}_{22,23}^T & = & \left(    %% 22, 23
\mp \frac{1}{\sqrt{\tilde{\gamma}^{11}}} \left( 2 \tilde{\gamma}_{11} \tilde{\gamma}^{11} + 3\tilde{\gamma}_{12} \tilde{\gamma}^{12} + 3\tilde{\gamma}_{13} \tilde{\gamma}^{13} \right) e \phi,
\pm \sqrt{\tilde{\gamma}^{11}} \tilde{\gamma}_{12} e \phi,
\pm \sqrt{\tilde{\gamma}^{11}} \tilde{\gamma}_{13} e \phi,
\pm \sqrt{\tilde{\gamma}^{11}} \tilde{\gamma}_{22} e \phi,
\pm \sqrt{\tilde{\gamma}^{11}} \tilde{\gamma}_{23} e \phi, \right. \nonumber \\
&& \left.
\pm \sqrt{\tilde{\gamma}^{11}} \tilde{\gamma}_{33} e \phi,
\mp 3 \sqrt{\tilde{\gamma}^{11}} e \phi,
\mp \frac{3}{2} \sqrt{\tilde{\gamma}^{11}} (e^2-1) e \phi,
(3e^2-7)\tilde{\gamma}^{11},
(3e^2-7)\tilde{\gamma}^{12},
(3e^2-7)\tilde{\gamma}^{13},
-3,0,0,
\right. \nonumber \\
&& \left.
-1/{\tilde{\gamma}^{11}} \left( 2\tilde{\gamma}_{11} \tilde{\gamma}^{11} + 3\tilde{\gamma}_{12} \tilde{\gamma}^{12} + 3 \tilde{\gamma}_{13} \tilde{\gamma}^{13} \right),
\tilde{\gamma}_{12},
\tilde{\gamma}_{13},
\tilde{\gamma}_{22},
\tilde{\gamma}_{23},
\tilde{\gamma}_{33},
0,0,0,0,0,0,0,0,0,0,0,0,1,0,0 \right) 
\end{eqnarray}
and which is linearly independent of the other right eigenvectors only
for $e \neq 1$. The associated left eigenvectors contain a factor $e^2-1$
in the denominator, which underlines the necessity of choosing $e \neq 1$
in the non-standard case $c=0$, but they are not reported here.

\subsection{Gamma driver shift condition with 1+log slicing} 

Here we present the 47 eigenvalues as well as the complete set of 47
linearly independent right eigenvectors of the reduced dynamic system
\eqref{eqn.pde.red} in the most general case when the gamma driver shift
condition is used, combined with the 1+log slicing for the lapse
($g(\alpha)=2/\alpha$). We recall that in this case the vector of dynamic
quantities is $\boldsymbol{U}^T :=\left( \tilde A_{ij}, K, \Theta,
\hat\Gamma^i, b^i, A_k, B^i_k, D_{kij}, P_k \right)$. Here, we have used
the standard setting $e=c=1$. In the final expressions we have used the
property $\textnormal{det}(\tilde{\gamma}_{ij})=1$.  Note that for
$\alpha=\phi=1$ the eigenvectors $\mathbf{r}_{16,22}$ are linearly
independent of the others only for $f \neq \frac{3}{4}$.  The eigenvalues
read
\begin{equation}
   \lambda_{1,2,3,\cdots,14,15} = -\beta^1, \qquad 
	 \lambda_{16,17,18,19,20,21}  = -\beta^1 + \sqrt{ \tilde{\gamma}^{11} } \phi \alpha, \qquad  
	 \lambda_{22,23,24,25,26,27}  = -\beta^1 - \sqrt{ \tilde{\gamma}^{11} } \phi \alpha, \qquad  
\end{equation} 
\begin{equation}
 \lambda_{28,29} = -\beta^1 \pm \sqrt{2} \sqrt{ \alpha \tilde{\gamma}^{11} } \phi, \qquad  
 \lambda_{30,31} = -\beta^1 \pm \sqrt{ \frac{4}{3} f \tilde{\gamma}^{11} }, \qquad 
\end{equation}
\begin{equation} 
 \lambda_{32,33} = -\beta^1 +  \frac{\sqrt{2}}{2} \sqrt{ \mu \tilde{\gamma}^{11} } \alpha, \qquad 
 \lambda_{34,35} = -\beta^1 -  \frac{\sqrt{2}}{2} \sqrt{ \mu \tilde{\gamma}^{11} } \alpha,
\end{equation}
\begin{equation} 
 \lambda_{36,37} = -\beta^1 + \sqrt{ f \tilde{\gamma}^{11} }, \qquad   
 \lambda_{38,39} = -\beta^1 - \sqrt{ f \tilde{\gamma}^{11} }, 
\end{equation}
\begin{equation} 
 \lambda_{40,41,42,43} = -\beta^1 + \frac{1}{2}  \sqrt{ \mu \tilde{\gamma}^{11} } \alpha,  \qquad 
 \lambda_{44,45,46,47} = -\beta^1 - \frac{1}{2}  \sqrt{ \mu \tilde{\gamma}^{11} } \alpha.  
\end{equation} 
The associated right eigenvectors are 
 
where we have used the abbreviation 
\begin{eqnarray}
 T &=& -64 f \phi^2 \left( \tilde{\gamma}^{12} \right)^2 + \mu f \left( 4 \tilde{\gamma}_{33} \left( -2 \tilde{\gamma}_{11} \tilde{\gamma}^{11} - \tilde{\gamma}_{13} \tilde{\gamma}^{13} \right) 
+ 4 \tilde{\gamma}^{12} \left( 8 \tilde{\gamma}^{12} - \tilde{\gamma}_{12}\tilde{\gamma}_{33} \right) \right) 
\nonumber \\
&& 
+ \alpha^2 \phi^2 \mu \left(  36 \left( \tilde{\gamma}^{12} \right)^2 + 
8 \tilde{\gamma}^{11} \left( -2 \tilde{\gamma}^{22} + \tilde{\gamma}_{13}^2 \right)  \right) 
- \mu^2 \alpha^2 \left( \tilde{\gamma}_{33} \tilde{\gamma}_{11} \tilde{\gamma}^{11}  - 6 \tilde{\gamma}^{11} \tilde{\gamma}^{22} + 12 \left( \tilde{\gamma}^{12} \right)^2 \right). 
\end{eqnarray}

\end{document}